\documentclass[11pt]{article}

\usepackage[titletoc,toc,title]{appendix}
\usepackage[english]{babel}

\usepackage{amssymb}
\usepackage{amsmath}

\numberwithin{equation}{section}

\usepackage{mathtools}
\usepackage{mathrsfs}
\usepackage{fullpage}
\usepackage{graphicx}
\usepackage[small,skip=0pt]{caption} 
\usepackage{subcaption}

\usepackage{authblk}

\usepackage{slashed} 

\usepackage{tikz}

\newcommand{\Z}{\mathbb{Z}}

\newcommand{\bd}{\begin{displaymath}}
\newcommand{\ed}{\end{displaymath}}
\newcommand{\be}{\begin{equation}}
\newcommand{\ee}{\end{equation}}
\newcommand{\bea}{\begin{eqnarray}}
\newcommand{\eea}{\end{eqnarray}}
\newcommand{\C}{\mathbb{C}}
\newcommand{\R}{\mathbb{R}}
\newcommand{\N}{\mathbb{N}}
\newcommand{\e}{\epsilon}

\newcommand{\nn}{\nonumber \\}
\newcommand{\ssum}[1]{\sum_{#1 =1}^{\infty}}

\newcommand{\ira}[1]{#1 \rightarrow i\infty}

\newcommand{\bfB}{\textbf{B}}
\newcommand{\bfH}{\textbf{H}}
\newcommand{\bff}{\textbf{f}}
\newcommand{\mK}{\mathcal{K}}
\newcommand{\slashconv}{\hat{\slashed{\star}}}

\newcommand{\rhs}{right hand side\space}
\newcommand{\uhp}{upper half-plane\space}

\newcommand{\ad}{c}
\newcommand{\s}{\mathbf{s}}
\newcommand{\Pf}{\mathbf{P}}
\newcommand{\Pft}{\tilde{\Pf}}
\newcommand{\Qf}{\mathbf{Q}}
\newcommand{\Qft}{\tilde{\Qf}}
\newcommand{\Qm}{\mathcal{Q}}

\newcommand{\mA}{\mathcal{A}}
\newcommand{\mAp}{\mathcal{A}^{\text{p}}}

\newcommand{\hbT}{\hat{\mathbb{T}}}
\newcommand{\mbT}{\mathbb{T}}
\newcommand{\bfT}{\textbf{T}}
\newcommand{\JT}{\mathscr{T}}
\newcommand{\mcT}{\mathcal{T}}
\newcommand{\hmcT}{\hat{\mathcal{T}}}
\newcommand{\mcTL}{\mcT_L}
\newcommand{\mcTR}{\mcT_R}

\newcommand{\mF}{\mathcal{F}}

\newcommand{\bO}[1]{\mathcal{O}\left(#1\right)}

\def \mut {\tilde{\mu}}
\def \omegat {\tilde{\omega}}
\def\ads{{\rm AdS}_5\times {\rm S}^5}

\DeclareMathOperator{\arcsinh}{arcsinh}

\newcommand{\factornumbering}[2]{\ensuremath{ \underbrace{ #1}_{\tiny
\begin{tikzpicture}[baseline=-2mm]
\node[circle, draw] {#2};
\end{tikzpicture}
} } }

\makeatletter
\DeclareRobustCommand\widecheck[1]{{\mathpalette\@widecheck{#1}}}
\def\@widecheck#1#2{%
    \setbox\z@\hbox{\m@th$#1#2$}%
    \setbox\tw@\hbox{\m@th$#1%
       \widehat{%
          \vrule\@width\z@\@height\ht\z@
          \vrule\@height\z@\@width\wd\z@}$}%
    \dp\tw@-\ht\z@
    \@tempdima\ht\z@ \advance\@tempdima2\ht\tw@ \divide\@tempdima\thr@@
    \setbox\tw@\hbox{%
       \raise\@tempdima\hbox{\scalebox{1}[-1]{\lower\@tempdima\box
\tw@}}}%
    {\ooalign{\box\tw@ \cr \box\z@}}}
\makeatother

\newcommand{\cD}{\widecheck{\Delta}}
\newcommand{\ewD}{\widecheck{\Delta}_e}

\usepackage{marginnote}

\newcommand{\defeq}{:=}

\usepackage[plainpages=false,hypertexnames=false,linktocpage=true]{hyperref}

\title{\bf Quantum Spectral Curve for the $\eta$-deformed $\ads$ superstring}
\date{}
\author[1]{Rob Klabbers \thanks{rob.klabbers@desy.de}}
\affil[1]{II. Institut f\"{u}r Theoretische Physik, Universit\"{a}t Hamburg, Luruper Chaussee 149, 22761 Hamburg, Germany \vspace{7pt}}
\author[2]{Stijn J. van Tongeren\thanks{svantongeren@physik.hu-berlin.de}}
\affil[2]{Institut f\"ur Mathematik und Institut f\"ur Physik, Humboldt-Universit\"at zu Berlin, IRIS Geb\"aude, Zum Grossen Windkanal 6, 12489 Berlin, Germany}

\begin{document}

\begin{flushright}
    \footnotesize
   ZMP-HH-17-25\\ HU-EP-17-21
\end{flushright}
{\let\newpage\relax\maketitle} 
\thispagestyle{empty}


\begin{abstract}
The spectral problem for the $\ads$ superstring and its dual planar maximally supersymmetric Yang-Mills theory can be efficiently solved through a set of functional equations known as the quantum spectral curve. We discuss how the same concepts apply to the $\eta$-deformed $\ads$ superstring, an integrable deformation of the $\ads$ superstring with quantum group symmetry. This model can be viewed as a trigonometric version of the $\ads$ superstring, like the relation between the XXZ and XXX spin chains, or the sausage and the $\mathrm{S}^2$ sigma models for instance. We derive the quantum spectral curve for the $\eta$-deformed string by reformulating the corresponding ground-state thermodynamic Bethe ansatz equations as an analytic $Y$ system, and map this to an analytic $T$ system which upon suitable gauge fixing leads to a $\Pf\mu$ system -- the quantum spectral curve. We then discuss constraints on the asymptotics of this system to single out particular excited states. At the spectral level the $\eta$-deformed string and its quantum spectral curve interpolate between the $\ads$ superstring and a superstring on ``mirror'' $\ads$, reflecting a more general relationship between the spectral and thermodynamic data of the $\eta$-deformed string. In particular, the spectral problem of the mirror $\ads$ string, and the thermodynamics of the undeformed $\ads$ string, are described by a second rational limit of our trigonometric quantum spectral curve, distinct from the regular undeformed limit.
\end{abstract}
\newpage
\setcounter{tocdepth}{2}
\tableofcontents
\section{Introduction}
The discovery of integrable models in the planar limit of the AdS/CFT correspondence has led to remarkable advances in this area \cite{Arutyunov:2009ga,Beisert:2010jr}. Using well-known tools of integrability,\footnote{A series of pedagogical reviews in this context can be found in \cite{Bombardelli:2016rwb}.} it is for instance possible to find a closed set of functional equations that nonperturbatively describe the spectrum of scaling dimensions in planar $\mathcal{N}=4$ supersymmetric Yang-Mills theory, or equivalently the energy spectrum of a superstring moving in $\ads$. This set of equations is known as the quantum spectral curve (QSC) \cite{Gromov:2013pga}. Here we present a quantum deformation of this quantum spectral curve, describing the spectrum of the light-cone gauge-fixed, quantum deformed $\ads$ superstring.

Conceptually, the solution of the $\ads$ spectral problem involves fixing a light-cone gauge, doing a double Wick rotation to arrive at the so-called mirror model \cite{Arutyunov:2007tc}, and using its exact S matrix as input for the thermodynamic Bethe ansatz (TBA) \cite{Arutyunov:2009zu,Bombardelli:2009ns,Arutyunov:2009ur,Gromov:2009bc}. This results in an involved set of infinitely many coupled integral equations, encoding the spectrum. Fortunately, these equations can be simplified significantly \cite{Arutyunov:2009ux,Cavaglia:2010nm,Balog:2011nm,Gromov:2011cx,Balog:2012zt}. The end result is the quantum spectral curve, a set of algebraic equations for only a handful of functions, taking the form of a natural quantisation of the classical spectral curve for the $\ads$ superstring \cite{Gromov:2013pga}. Similar QSCs have also been found for strings on $\mathrm{AdS}_4 \times \mathbb{CP}^3$ \cite{Bombardelli:2017vhk}, and for the Hubbard model \cite{Cavaglia:2015nta}. In this setting, different states of the string correspond to different solutions of the quantum spectral curve, with the charges labelling a state appearing in specific power-law asymptotics of the functions defined on a complex plane with cuts. Loosening these analytic constraints by allowing additional exponential asymptotics for instance, the quantum spectral curve can also describe the spectrum of Cartan-twisted strings \cite{Gromov:2015dfa,Kazakov:2015efa}, such as strings on the real-$\beta$ Lunin-Maldacena background \cite{Lunin:2005jy,Frolov:2005ty,Frolov:2005dj}. In this light, it is interesting to ask which theories and observables the quantum spectral curve can describe, and what the corresponding analytic constraints are. In this paper we will show how the quantum spectral curve can be naturally ``trigonometrised'' by defining it on a cylinder -- providing interesting new perspectives on some of its analytic structure -- such that it describes a quantum deformation of the spectral problem of the $\ads$ superstring.\footnote{A related quantum deformation of the $\mathrm{S}^2$ sigma model -- the sausage model -- has recently been investigated in a related fashion \cite{Ahn:2017mff}. Unlike our integrable strings and the Hubbard model, however, this model has a relativistic S matrix and no complicating branch cuts.}

The quantum deformed spectrum thus described is associated to the $\eta$-deformed superstring \cite{Delduc:2013qra}, a model with quantum deformed $\mathfrak{psu}(2,2|4)$ symmetry with real deformation parameter $q$ \cite{Arutyunov:2013ega,Delduc:2014kha}. This model is an example of a Yang-Baxter sigma model \cite{Klimcik:2002zj,Klimcik:2008eq}, a class of models including the Lunin-Maldacena case mentioned above \cite{Matsumoto:2014nra}. While the AdS/CFT interpretation of twisted string models like the Lunin-Maldacena one is generically understood \cite{vanTongeren:2015uha,vanTongeren:2016eeb}, the same does not apply to this quantum deformed model. In fact, the background of the $\eta$-deformed model does not solve the supergravity equations of motion \cite{Arutyunov:2015qva}, meaning that the $\eta$-deformed sigma model is not conformal at the quantum level. Interestingly, it is classically equivalent (T dual) to a conformal but non-unitary type IIB$^*$ string model \cite{Hoare:2015wia}.\footnote{Yang-Baxter models preserve $\kappa$ symmetry by construction \cite{Delduc:2013qra}, see also \cite{Wulff:2016tju,Borsato:2016ose}. However, this only implies that the background solves a set of generalised supergravity equations \cite{Arutyunov:2015mqj,Wulff:2016tju}, which are not sufficient to guarantee Weyl invariance in the standard sense. Still, solutions of the generalised supergravity equations are always formally T dual to solutions of the standard ones, except possibly in specific ``null'' cases. See \cite{Baguet:2016prz,Sakamoto:2017wor} for related discussions in the context of exceptional and double field theory. Let us also mention that the $\lambda$ model \cite{Sfetsos:2013wia,Hollowood:2014qma,Demulder:2015lva}, which can be viewed as a deformation of the non-abelian T dual of the $\ads$ string, has quantum group symmetry with $q$ a root of unity, and is Weyl invariant \cite{Borsato:2016ose}.} Moreover, the maximal deformation limit of the $\eta$ model appears to be equivalent to the $\ads$ mirror model at the quantum (S-matrix) level \cite{Arutynov:2014ota}, which is a conformal and unitary model \cite{Arutyunov:2014cra,Arutyunov:2014jfa}. In this light, it is important to better understand the status of the $\eta$-deformed model at the quantum level, starting from its spectrum.\footnote{Various classical solutions for this model have been investigated in e.g. \cite{Arutynov:2014ota,Arutyunov:2014cda,Banerjee:2014bca,Kameyama:2014vma,Khouchen:2015jfa,Roychowdhury:2016bsv,Hernandez:2017raj}.} Our equations describe the spectrum of the classically light-cone gauge-fixed $\eta$-deformed sigma model, a quantum deformation of the $\ads$ superstring spectrum interpolating between $\ads$ and its mirror version.\footnote{Strictly speaking, our quantum spectral curve equations depend on two coupling constants whose precise identification in terms of the two coupling constants in the Lagrangian remains to be determined. The identification one finds by expanding the exact S matrix at tree level \cite{Arutyunov:2013ega} cannot apply at finite coupling for reasons of unitarity \cite{Arutynov:2014ota}, providing further motivation to investigate the quantum model in detail.} This interpolation reflects a curious property of the full $\eta$-deformed model, dubbed mirror duality \cite{Arutynov:2014ota}: performing a double Wick rotation in the light-cone gauge, is equivalent to inverting the deformation parameter in a suitable parametrisation. In other words, this family of models is self-similar under the mirror transformation. As such, our quantum spectral curve also carries information on the thermodynamics of these models in the decompactification limit. In particular, a suitable limit of our equations should have interesting applications in the integrability-based computation of the Hagedorn temperature of \cite{Harmark:2017yrv}.

To arrive at our trigonometric quantum spectral curve, we will follow the route taken for the undeformed string \cite{Cavaglia:2010nm,Balog:2011nm,Gromov:2011cx,Gromov:2013pga}. We will start from the thermodynamic Bethe ansatz equations for the $\eta$-deformed model \cite{Arutynov:2014ota}, and determine the analytic properties of the associated $Y$ functions. This allows us to recast the TBA equations in the form of an analytic $Y$ system: a standard $Y$ system, together with analyticity data reflecting the underlying $\eta$-deformed model. From here we translate the analytic $Y$ system to an analytic $T$ system with gauge freedom, by the usual Hirota map. With suitable gauge choices, these $T$ functions can be parametrised in an elegant fashion, where the analyticity constraints result in a $\Pf\mu$ system, i.e. the quantum spectral curve.

As much of the discussion is quite technical,  we will start in Section \ref{sec:overviewandresult} with a basic overview of the deformed model followed by a summary of our main result: the quantum spectral curve for the $\eta$-deformed string. The main line of the derivation is given in Section \ref{sec:mainderivation}, while we have moved lengthy but important technical derivations to various appendices. In Section \ref{sec:mirrordualityandlimit} we discuss the mirror duality of the $\eta$-deformed string and its QSC in more detail, including related interesting limits and applications. In the conclusions we discuss various interesting open questions and possible future directions.

\section{Overview of the model and its quantum spectral curve}
\label{sec:overviewandresult}
\subsection{Basic parametrisation of the $\eta$-deformed string}

\vspace{-5pt}
\paragraph{Parameters.} The spectrum of the undeformed $\ads$ string is arranged in superconformal multiplets with corresponding highest weight states. Only the energy of highest weight states depends non-trivially on the one coupling constant of the string: the effective string tension. These highest weight states can be labelled through six quantum numbers
\begin{equation}
\{\Delta,S_1,S_1,J_1,J_2,J_3\},
\end{equation}
where the $S_i$ are the spins of the conformal group, the $J_i$ are the weights of $\mathfrak{so}(6)$ -- angular momenta on the sphere -- and $\Delta=E-J_1$ denotes the scaling dimension, or just the target space energy of the string. For the $\eta$-deformed string, the spectrum is still similarly organised, as the representation theory of the quantum deformed superconformal group is essentially unchanged, at least for real $q$. As such, we will use the same quantum numbers to label our states, with $\Delta$ now viewed purely as the target space energy. The spectrum itself, however, now depends on the Lagrangian deformation parameter $\varkappa$ as well as the effective string tension $T$. The exact S-matrix of the model can be taken to depend on two parameters $h$ and $q$ instead. As in other integrable models, the precise dependence of this $h$ and $q$ on the Lagrangian parameters $\varkappa$ and $T$ is not a priori clear, see Appendix \ref{App:parametereta} for further discussion. We work with $h$ and $q$ which we parametrise as in \cite{Arutynov:2014ota}
\begin{equation}
\label{eq:parameters}
q = e^{-\ad}\qquad \mbox{ and } \qquad h \sinh \ad = \sin \frac{\theta}{2},
\end{equation}
where $\ad\geq 0$ and $\theta \in [0,\pi]$, covering all possible inequivalent models.\footnote{Note that we denote $-\log q$ by $\ad$ instead of the $a$ used in \cite{Arutynov:2014ota}.} We will see that the parameter $\theta$ is a natural deformed analogue of the undeformed coupling constant $g$. The undeformed string arises in the limit $q\rightarrow1$, i.e. $\ad \rightarrow 0$, while $h$ remains finite and becomes the effective $\ads$ string tension $g$. To describe a deformation of the $\ads$ string, we will enforce the level matching condition of vanishing total world-sheet momentum ($P=0$) in our integrable model. The dispersion relation of our model is given by
\begin{equation}
\label{eq:dispersionsolved}
\mathcal{E}(p) = \frac{2}{\ad} \arcsinh \sqrt{\sec^2\frac{\theta}{2}\sinh^2 \frac{\ad}{2} + \tan^2\frac{\theta}{2} \sin^2 \frac{p}{2}}.
\end{equation}
In short, we are dealing with a one parameter deformation of the integrable model describing the undeformed $\ads$ string, and our goal is to account for this deformation in the quantum spectral curve.

\begin{figure}[!t]
\centering
\begin{subfigure}{7.5cm}
\includegraphics[width=7.5cm]{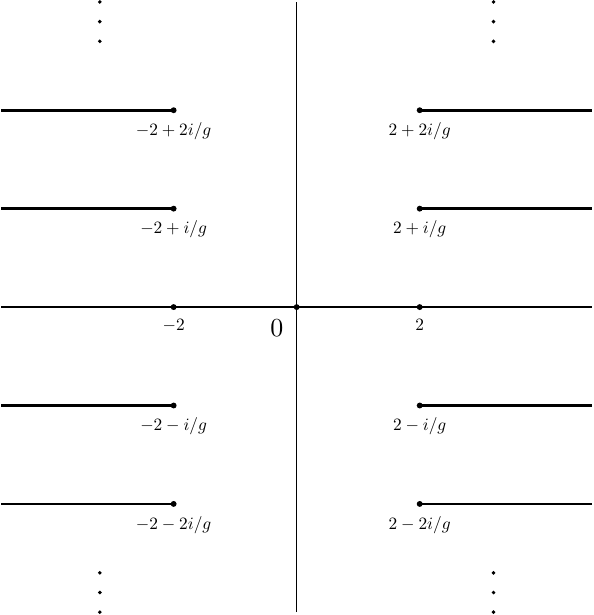}
\caption{}
\end{subfigure}
\qquad \qquad
\begin{subfigure}{6cm}
\includegraphics[width=6cm]{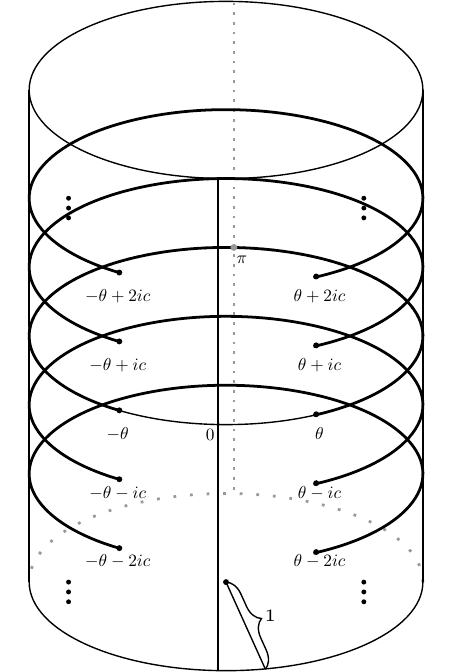}
\caption{}
\end{subfigure}
\caption{The cut structures of the undeformed (a) and deformed (b) models. Possible branch cuts are indicated by thick lines.}
\label{fig:cutstructures}
\end{figure}

\vspace{-5pt}
\paragraph{Basic analytic structure.} The S matrix, Bethe ansatz, TBA and QSC of the undeformed string all have a discrete representation theoretic structure that would appear in any superconformal model, e.g. a superconformal spin chain, while the true nature of the string is captured by the analytic structure of the functions entering the game. Namely, at the most basic level, the natural rapidity variable of the undeformed string lives on a plane with branch cuts, see Fig. \ref{fig:cutstructures}.
These branch cuts run in the real direction, and there can be multiple, shifted by discrete steps in the imaginary direction. The string tension enters by defining the ratio of the distance between the branch points and the elementary shift distance, which is an invariant under rescaling the rapidity. Discontinuity relations of suitable variables across these branch cuts ultimately provide the detailed characterisation of the model. In our deformed case, the rapidity plane becomes a cylinder, introducing an extra scale which allows us to account for the deformation parameter. Concretely, we choose to take a cylinder of radius one, i.e.
\begin{equation}
\mbox{Re}(u) \in\, (-\pi,\pi],
\end{equation}
making our branch points lie at $\mbox{Re}(u) = \pm \theta$, with unit shifts implemented as $u \rightarrow u \pm i \ad$, see Fig. \ref{fig:cutstructures}. For shifted functions we use the notation
\begin{align}
\label{eq:shift}
f^{[n]}(u) \defeq f(u+i\ad n), \quad f^{\pm} = f^{[\pm 1]},\quad f^{\pm\pm}= f^{[\pm2]}.
\end{align}
As we will see in detail below, the TBA equations for the deformed string determine discontinuity relations across these various branch cuts, which, together with discrete group theoretic data, can be recast in the form of the quantum spectral curve. In general terms the deformation amounts to putting the quantum spectral curve on a suitable cylinder.

For the undeformed string we distinguish between ``short'' branch cuts that run through zero and ``long'' branch cuts that run through infinity. The presence of these cuts is an indication of the non-relativistic nature of the light-cone gauge-fixed string. Indeed, the gauge-fixed string is not invariant under the double Wick rotation underlying the application of the \emph{thermodynamic} Bethe ansatz to the \emph{spectral} problem, but instead becomes an inequivalent model known as the \emph{mirror model}. At the level of analytic structure, broadly speaking the double Wick rotation exchanges long and short cuts. For the deformed string however, we see that there is no natural notion of long and short cuts, as this identification flips when we cross $\theta=\pi/2$. This is a reflection of the mirror duality mentioned in the introduction, which we will come back to in Section \ref{sec:mirrordualityandlimit}. To keep with the terminology of the undeformed model, we will refer to cuts running from $-\theta$ to $\theta$ as short, and ones from $\theta$ to $-\theta$ through $\pi$ as long. Let us also define
\begin{equation}
\label{eq:Zs}
\begin{aligned}
Z_N &= \left\{ u \in \C \, | \, u=v+i\ad N, \, v\in (-\pi,\pi]\right\},\\
\hat{Z}_N &= \left\{ u \in \C \, | \, u=v+i\ad N, \, v\in [-\theta,\theta]\right\},\\ \check{Z}_N &= \left\{ u \in \C \, | \, u=v+i\ad N, \, v\in [-\pi,-\theta]\cup [\theta,\pi]\right\}.
\end{aligned}
\end{equation}
The part of the cylinder with $|$Im$(u)|<\ad$ is the \emph{physical strip}.

\vspace{-5pt}
\paragraph{$x$ functions.} Most differences in the analytic structure between the undeformed and the deformed models can already be seen from the $x$ functions, which govern the rapidity-variable dependence of the S-matrix data and dispersion relation that enters the Bethe ansatz. For our deformed model, the \emph{string} and \emph{mirror} $x$ functions are defined as the $2\pi$-periodic functions
\begin{align}
x_s(u)& =-i \csc \theta \left(e^{iu}-\cos \theta - (1-e^{iu}) \sqrt{\frac{\cos u-\cos \theta}{\cos u - 1}} \right),\label{eq:xs} \text{ with a } \hat{Z}_0 \text{ branch cut},\\
x_m(u)& =-i \csc \theta \left(e^{iu}-\cos \theta + (1+e^{iu}) \sqrt{\frac{\cos u-\cos \theta}{\cos u + 1}} \right), \text{ with a } \check{Z}_0 \text{ branch cut}. \label{eq:xm}
\end{align}
In this parametrisation, the undeformed limit means decompactifying the cylinder to the plane, while keeping the branch points at finite distance from each other. Concretely, with $\theta = \theta(h,a)$ via Eqn. \eqref{eq:parameters}, and $h=g$, our rapidity cylinder and $x$ functions become the rapidity plane and $x$ functions of \cite{Bombardelli:2009ns,Arutyunov:2009ur} if we rescale
\begin{equation}
\label{eq:undefrescale1}
u \rightarrow \ad g u,
\end{equation}
and consider the limit $\ad\rightarrow 0$. Taking instead $h=2g$,
\begin{equation}
\label{eq:undefrescale2}
u \rightarrow 2 \ad u
\end{equation}
we arrive at the parametrisation of \cite{Gromov:2009bc,Gromov:2013pga} in the limit $\ad\rightarrow 0$.

\subsection{Quantum spectral curve for the $\eta$-deformed string}
The main purpose of this paper is to derive the $\eta$-deformed quantum spectral curve from the TBA equations of the $\eta$-deformed model, as we will discuss in detail below.  The functional form of its basic $\Pf\mu$ system coincides with that of the undeformed model, which can be seen as a reflection of the similarities in the representation theory of the symmetry algebras of both models. It reads
\be
\mut_{ab}-\mu_{ab} = \Pf_a \Pft_b-\Pf_b \Pft_a, \quad \Pft_a = \mu_{ab} \Pf^b,\quad \Pf_a\Pf^a = 0,\quad  \text{Pf}(\mu) = 1,
\ee
where we adopt the notation from \cite{Gromov:2014caa}: the unknown functions $\Pf_a,\Pf^a$ all have one $\hat{Z}_0$ cut, whereas the coefficients $\mu_{ab}$ of the $4\times 4$-matrix $\mu$ have a ladder of short cuts. $\text{Pf}$ denotes the Pfaffian. The notation $\tilde{f}$ indicates the continuation of $f$ around the branch point $\theta$.
\begin{figure}[!t]
\centering
\begin{subfigure}{7cm}
\includegraphics[width=7cm]{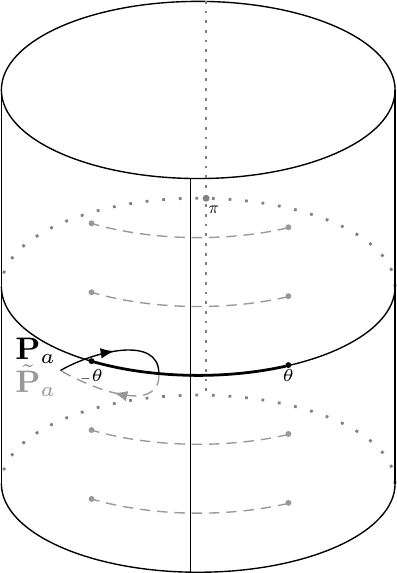}
\caption{}
\end{subfigure}
\qquad \qquad
\begin{subfigure}{7cm}
\includegraphics[width=7cm]{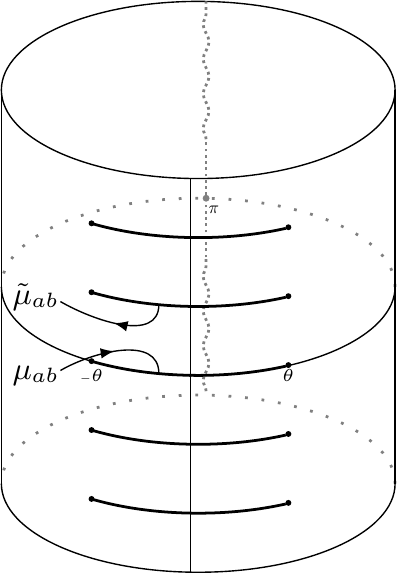}
\caption{}
\end{subfigure}
\caption{The analytic structure of the $\Pf_a$ and $\mu_{ab}$. The thick lines indicate branch cuts on the first sheet, dashed lines branch cut on the second sheet after continuation through the cut on the real axis. The dotted line through $\pi$ in (a) indicates the possible discontinuity of the $\Pf_a$ due to anti-periodicity. The squiggly line in (b) indicates that for generic $\theta$ outside the physical strip the $\mu_{ab}$ cannot be put on (a finite cover of) the cylinder.}
\label{fig:PmuQomega}
\end{figure}
Therefore, these equations relate first and second sheet evaluations of several functions.

From the $\Pf\mu$ system we can derive the associated $\Qf\omega$ system: a set of functions $\Qf_i$, $\Qf^i$ and an anti-symmetric matrix $\omega_{ij}$ with $i,j=1,\cdots,4$ satisfying the equations
\be
\omegat_{ij}-\omega_{ij} = \Qf_i \Qft_j-\Qf_j \Qft_i, \quad \Qft_i = \omega_{ij} \Qf^j,\quad \Qf_i\Qf^i = 0,\quad  \text{Pf}(\omega) = 1.
\ee
The $\Qf_j$ have one $\check{Z}_0$ cut and the $\omega_{ij}$ have a ladder of long cuts. The effect of the deformation is seen in the fact that the $\Pf$ and $\Qf$ functions are defined as (anti-)periodic functions on a cylinder as in Fig. \ref{fig:cutstructures}.\footnote{Our starting point -- the ground-state TBA equations -- is periodic, but the $\Pf$ and $\Qf$ functions enter quadratically in that picture, allowing for anti-periodicity.} The situation for $\mu_{ab}$ is more involved: we will show that $\mu_{ab}$ at zero coupling $\theta$ is (anti-)periodic, but for finite coupling this property is generically lost, depending on the choice of branch cuts. A non-trivial property of $\mu_{ab}$ that survives deformation is that when viewed as a function with long cuts it is $2i \ad$-periodic:\footnote{When necessary, we use a hat to indicate the short-cutted version of a function ($\hat{f}$) and a check for the long-cutted version ($\check{f}$). Here the long-cutted version of $\mu_{ab}$ coincides with the short-cutted one in the strip with $0<$Im$(u)< 2 i\ad$.}
\be
\check{\mu}_{ab}^{++}=\check{\mu}_{ab},
\ee
where we use the shift notation introduced in Eqn. \eqref{eq:shift}. By definition, the $\check{\mu}_{ab}$ are (anti-)periodic everywhere on the cylinder. When viewed as a function with short cuts, however, $\mu_{ab}$ generically are only (anti-)periodic on the upper half of the physical strip, where they coincide with the long-cutted $\check{\mu}_{ab}$. Outside of this region the $\mu_{ab}$ lose their periodicity properties at non-zero coupling to reflect the fact that the deformed string energy $\Delta$ is not integer. This can be interpreted as the trigonometric version of the phenomenon in the undeformed case that $\mu_{ab}$ develops a branch point at infinity for non-zero coupling.

In the QSC for the undeformed string, excited states are described by particular large $u$ asymptotics for the $\Pf$ functions. As discussed in more detail in Sections \ref{sec:mufromTBA} and \ref{sec:asymptotics}, in our case this direction is not available, and we consider large imaginary $u$ instead. Concretely, with $z= e^{-i u/2}$ we have
\be
\Pf_a \simeq A_a z^{-\tilde{M}_a}, \quad \Pf^a \simeq A^a z^{\tilde{M}_a}, \quad \Qf_i \simeq B_i z^{\hat{M}_i}, \quad \Qf^i \simeq B^i z^{-\hat{M}_i},
\ee
where $f \simeq g$ indicates that $\lim_{z\rightarrow \infty} f/g=1$. The powers are given in terms of the quantum numbers of a deformed string state as
\begin{align}
\tilde{M} &= \frac{1}{2}\left\{ J_1+J_2-J_3+2,J_1-J_2+J_3,-J_1+J_2+J_3,-J_1-J_2-J_3-2\right\}, \nn
\hat{M} &= \frac{1}{2}\left\{ \Delta-S_1 -S_2 +2 ,\Delta + S_1 +S_2,-\Delta-S_1 +S_2 ,-\Delta+S_1 -S_2-2\right\}.
\end{align}
The coefficients $A,B$ are constrained as
\be
A_{a_0}A^{a_0} =2\frac{\prod_{j}\sinh\left( \ad \frac{\tilde{M}_{a_0}- \hat{M}_{j}}{2}\right)}{\prod_{b\neq a_0} \sinh\left( \ad \frac{ \tilde{M}_{a_0}- \tilde{M}_{b}}{2}\right)}, \quad B^{j_0} B_{j_0}  =-2\frac{\prod_{a}\sinh\left(\ad  \frac{\hat{M}_{j_0}- \tilde{M}_{a}}{2}\right)}{\prod_{j\neq j_0} \sinh\left(\ad \frac{ \hat{M}_{j_0}- \hat{M}_{j}}{2}\right)},
\ee
for given $a_0,j_0$. The form of the asymptotics for $\Pf$ and $\Qf$ functions is very similar to those for the undeformed case, the crucial difference sits in the constants appearing in the $\tilde{M}$\&$\hat{M}$s. These constants come about naturally through the analysis of consistency of the QSC and the presence of trigonometric functions as opposed to polynomials, combined with a careful analysis of the undeformed limit, discussed in detail in Section \ref{sec:asymptotics}. This completes our overview of the QSC.

\section{Deriving the QSC}
\label{sec:mainderivation}
In our derivation of the QSC for the $\eta$-deformed model we will follow the same steps historically taken for the original $\ads$ QSC:
\begin{align}
\text{TBA equations} \rightarrow \text{Analytic $Y$ system}\rightarrow \text{Analytic $T$ system}\rightarrow \text{Quantum Spectral Curve} \nonumber
\end{align}
In this way we will see how this entire procedure can be trigonometrised, finding some nice structures and subtle differences to the undeformed case along the way.

\subsection{From TBA to analytic $Y$ system}
As discussed in \cite{Arutynov:2014ota} the ground-state TBA equations for the $\eta$-deformed model take the same form as those of the undeformed model, up to taking the $Y$ functions to be periodic functions on a cylinder rather than a plane -- both with branch cuts -- and replacing the various integration kernels by their deformed counterparts. As such, we define the standard complete, short, and long convolutions as
\begin{align}
f\star h(u,v)=\,&\int_{-\pi}^{\pi}\, dt\, f(u,t)h(t,v) &= \int_{Z_0}\, dt\, f(u,t)h(t,v) \, , \nonumber \\
f\, \hat{\star}\,  h(u,v)=\,&\int_{-\theta}^{\theta}\, dt\, f(u,t)h(t,v) &=  \int_{\hat{Z}_0}\, dt\, f(u,t)h(t,v) \, , \\
f\, \check{\star}\, h(u,v)=\,&\int_{-\pi}^{-\theta}\, dt\, f(u,t)h(t,v)+\int_{\theta}^{\pi}\, dt\, f(u,t)h(t,v) \hspace{-25pt}&= \int_{\check{Z}_0}\, dt\, f(u,t)h(t,v) \, ,\nonumber
\end{align}
respectively. The TBA equations are then
\begin{equation}
\label{TBAeqns}
\begin{aligned}
\log Y_{Q} =&\, -L \tilde{\mathcal{E}}_Q +\Lambda_{P} \star K^{PQ}_{\mathfrak{sl}(2)} + \sum_\alpha \Lambda^{(\alpha)}_{M+1|vw}\star K^{MQ}_{vwx} + \sum_\alpha L^{(\alpha)}_{\beta}\, \hat{\star} \,K_{\beta}^{yQ},  \\
\log Y^{(\alpha)}_{\beta} =& \,- \Lambda_P \star K^{Py}_{\beta} + \left(L^{(\alpha)}_{M+1|vw}-L^{(\alpha)}_{M+1|w}\right)\star K_M\\
\log Y^{(\alpha)}_{M|w} =&\, L^{(\alpha)}_{N|w} \star K_{NM}  + \left(L^{(\alpha)}_{-}-L^{(\alpha)}_{+}\right)\,\hat{\star}\, K_M \\
\log Y^{(\alpha)}_{M|vw} =&\, L^{(\alpha)}_{N|vw} \star K_{NM}  +  \left(L^{(\alpha)}_{-}-L^{(\alpha)}_{+}\right)\,\hat{\star}\, K_M  -\Lambda_{Q} \star K_{xv}^{QM},
\end{aligned}
\end{equation}
where for generic $Y$ functions
\begin{equation}
L_\chi = \,\log(1+1/Y_\chi), \qquad \Lambda_\chi =  \,\log(1+Y_\chi),
\end{equation}
with the exception of $L_{\pm}$ and $\Lambda_{\pm}$, defined as
\begin{equation}
L_\pm = \,\log(1-1/Y_\pm), \qquad \Lambda_\pm = \,\log(1-Y_\pm).
\end{equation}
Repeated indices are summed over, $M,N,\ldots \in \mathbb{N}^+$, $\beta=\pm$, and $\alpha = l,r$ distinguishes a so-called left and right set of $Y$ functions. The energy corresponding to a solution of these equations is given by
\bea
\label{eq:Energy} E(J) &=&-\int_{Z_0} du\, \sum_Q \frac{1}{2\pi}\frac{d\tilde{p}^Q}{du}\log\left(1+Y_Q\right).
\eea
The deformed kernels appearing above are defined in Appendix \ref{App:kernelsandenergies}. All $Y$ functions are periodic with period $2\pi$, and have branch cuts of square-root type on some $\check{Z}_N$s.

\subsubsection{Simplification and $Y$ system}

The above equations can be simplified by combining them in particular ways, namely by acting with
\begin{equation}
\label{eq:K+1invdef}
(K+1)^{-1}_{PQ} = \delta_{P,Q} - (\delta_{P,Q+1}+ \delta_{P,Q-1}) \s,
\end{equation}
where $\s$ is the doubly-periodic analogue of the the standard kernel $s(u) = (4 \cosh{\frac{\pi u}{2}})^{-1}$
\begin{equation}
\label{eq:sigmakerneldef}
\mathbf{s}(u)  \equiv \sum_{n \in \mathbb{Z}} \tfrac{1}{\ad} s\left(\tfrac{u + 2\pi n}{\ad}\right) =  \sum_{n\in \mathbb{Z}} \frac{1}{4 \ad \cosh \frac{\pi (u+ 2\pi n)}{2 \ad}} = \tfrac{K(m^\prime)}{2\pi \ad} \mbox{dn}(u)\, ,
\end{equation}
where dn is the corresponding Jacobi elliptic function with real period $2\pi$ and imaginary period $4 \ad$, and $K(m^\prime)=\sum_{l \in \mathbb{Z}} \frac{\pi}{2\cosh\pi^2 l/\ad}$ is the elliptic integral of the complementary elliptic modulus. This gives the set of simplified TBA equations presented in \cite{Arutynov:2014ota}, cf. Appendix \ref{App:simpTBA}. To get to the $Y$ system we define the usual operator $s^{-1}$
\begin{equation}
\label{eq:sinvdef}
 f \circ s^{-1}(u) = \lim_{\epsilon\rightarrow 0} f(u + i \ad - i\epsilon) +  f(u - i \ad + i\epsilon)\, ,
\end{equation}
so that
\begin{equation}
(f \star \s) \circ s^{-1}(u) = f(u)\, , \, \, \, \, \mbox{for} \, \, u \in (-\pi,\pi] \, ,
\end{equation}
just as for the undeformed $s(u)$. Note that $s^{-1}$ has a non-trivial kernel. Applying $s^{-1}$ to the simplified TBA equations we get the following $Y$-system equations, for real $u$.
\paragraph{Q particles}
\begin{align}
\label{Ysys:Q}
\frac{Y_1^+ Y_1^-}{Y_2} & = \frac{\prod_{\alpha}
\left(1-\frac{1}{Y^{(\alpha)}_{-}}\right)}{1+Y_2}\, , \, \, \, \, \, \mbox{for} \, \,
u\in \hat{Z}_0\, ,\\
\frac{Y_Q^+ Y_Q^-}{Y_{Q+1}Y_{Q-1}} & = \frac{\prod_{\alpha} \Bigg(
1+\frac{1}{Y_{Q-1|vw}^{(\alpha)}}\Bigg)}{(1+Y_{Q-1})(1+Y_{Q+1})} \, .
\end{align}
\paragraph{$w$ strings}
\begin{align}
\label{Ysys:w}
Y_{1|w}^+ Y_{1|w}^- & =(1+Y_{2|w})\left(\frac{1-Y_-^{-1}}{1-Y_+^{-1}}\right)^{\vartheta(\theta-|u|)},  \\
Y_{M|w}^+ Y_{M|w}^- & =(1+Y_{M-1|w})(1+Y_{M+1|w})\,,
\end{align}
\paragraph{$vw$ strings}
\begin{align}
\label{Ysys:vw}
Y_{1|vw}^+ Y_{1|vw}^- & =\frac{1+Y_{2|vw}}{1+Y_2}\left(\frac{1-Y_-}{1-Y_+}\right)^{\vartheta(\theta-|u|)},  \\
Y_{M|vw}^+ Y_{M|vw}^- & =\frac{(1+Y_{M-1|vw})(1+Y_{M+1|vw})}{1+Y_{M+1}} ,
\end{align}
\paragraph{$y$ particles}
\begin{equation}
\label{Ysys:y}
Y_{-}^{+}Y_{-}^{-} = \frac{1+Y_{1|vw}}{1+Y_{1|w}}\frac{1}{1+Y_1}\, .
\end{equation}
As for the undeformed string, there is no $Y$ system equation for $Y_+$, and to get this equation for $Y_-$ we used the identities $K^{Qy}_- \circ s^{-1} =  K_{xv}^{Q1} + \delta_{Q,1}$ and $K_M \circ s^{-1} = K_{M1} + \delta_{M,1}$. In the equations for $(v)w$ strings and $y$ particles we have suppressed the $\alpha$ index. $\vartheta$ is the usual Heaviside function.
 	
\subsubsection{Deriving the analyticity conditions}
\label{subsub:Deriving the Y}
The above $Y$ system is of the exact same form as the undeformed $Y$ system, even though the TBA equations of the two models contain different kernels, and solutions to the TBA equations have different analytic properties. This shows that the two models are distinguished at this level by additional requirements on the solution of the $Y$ system only. This extra analytic data is given in the form of functional equations on the discontinuities of the $Y$ functions. In the undeformed case these were found in  \cite{Arutyunov:2009ax,Cavaglia:2010nm,Balog:2011nm}. Given the right set of data it is  possible to rederive the ground-state TBA equations from the $Y$ system,  showing that the solutions of the former are encompassed in the solution set of the latter. Indeed, one can show that any solution of the $Y$ system that (1) obeys the additional discontinuity relations, (2) is pole free except possibly at branch points, and (3) has the correct boundary behaviour at $\pm i \infty$, also satisfies the ground-state TBA equations. Solutions with extra poles then correspond to excited states and lead to excited state TBA equations, which differ from the ground-state TBA equations by the presence of driving terms \cite{Dorey:1996re}. In this way, the $Y$ system supplemented with discontinuity relations -- known as the \emph{analytic $Y$ system} -- extends the range of the ground-state TBA equations to include excited states without changing the equations themselves.

\begin{figure}[!t]
\centering
\includegraphics[width=7cm]{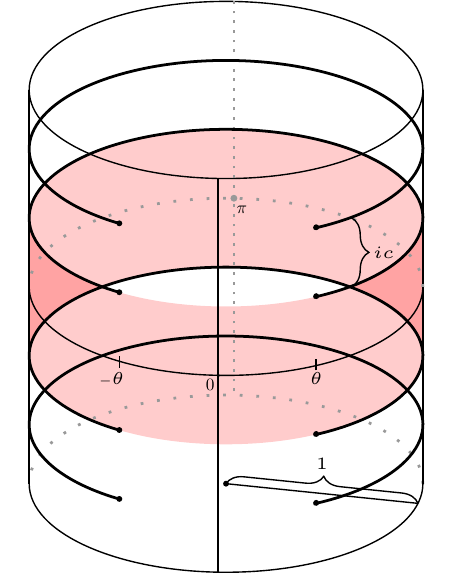}\\[2mm]
\caption{The analytic structure of the $Y$ functions on a cylinder with radius one. The thick lines indicate branch cuts on the first sheet, the analyticity strip is indicated in red}
\label{fig:Yanalytics}
\end{figure}

The extra analyticity data is analogous to that of the undeformed case, and can be found by following the analysis of the undeformed TBA equations of \cite{Cavaglia:2010nm}. We find that
\begin{itemize}
\item $Y_-^{(\alpha)}$ and $Y_+^{(\alpha)}$ are each others analytic continuation when continued through their cut on the real axis.
\item All other $Y$ functions $Y_{M}$ and $Y_{M|(v)w}^{(\alpha)}$ are analytic in the strip
\be
\left\{ u \in \C\, | \, |\text{Im}(u)| < M\ad\right\},
\ee
possibly with the exception of a finite number of poles, as one can verify by locating the branch points in the \rhs of the TBA equations. This is illustrated in Fig. \ref{fig:Yanalytics}.
\end{itemize}
All other required analytic information one needs to specify a solution comes in the form of discontinuity relations. We wil use the notation
\be
\left[f \right]_N(u) = f^{[N]}(u) - \widetilde{f^{[N]}}(u)
\ee
for the discontinuity of $f$ across a long cut. These discontinuities occur in the TBA equations because of the existence of the finite integration interval $[-\theta,\theta]$ and poles in the kernels: whereas the integration contours spanning the entire periodicity interval $(-\pi,\pi]$ can be deformed in such a way that while analytically continuing around one of the branch points the poles do not cross the contour, the integral contour $[-\theta,\theta]$ does not allow for such a deformation. Therefore, one typically picks up a Cauchy-type integral that represents the difference between the two different evaluations of our $Y$ functions (see \cite{Cavaglia:2010nm} for more details). The fact that our model is defined on a cylinder makes the analysis even cleaner, since the cancellation of integration contours no longer depends on the $Y$ functions falling off sufficiently fast. Instead, periodicity allows for  exact cancellation.

Written in appropriate notation the resulting discontinuity relations look the same as for the undeformed case.
\vspace{-10pt}
\paragraph{Discontinuity relations}
\begin{align}
\label{discs1}
\left[\log Y_{1|w}^{(\alpha)}\right]_{\pm1}(u) &= \left[L_{-}^{(\alpha)}\right]_0(u), \nn
\left[\log Y_{1|vw}^{(\alpha)}\right]_{\pm1}(u) &= \left[\Lambda_{-}^{(\alpha)}\right]_0(u),\nn
\left[\log \frac{Y_-}{Y_+} \right]_{\pm2N}(u) &= -\sum_{P=1}^{N} \left[\Lambda_P\right]_{\pm(2N-P)}(u) \text{ for } N\geq 1,\nn
\left[\Delta\right]_{\pm 2N} (u) &= \pm \sum_{\alpha} \left( \left[L^{(\alpha)}_{\mp}\right]_{\pm 2N}(u) + \sum_{M=1}^N \left[   L_{M|vw}^{(\alpha)}\right]_{\pm(2N-M)}(u) + \left[ \log Y_-^{(\alpha)}\right]_{0}(u) \right),
\end{align}
where
\begin{align}
\Delta(u) &=
\left\{
	\begin{array}{ll}
		\cD(u)  & \mbox{if Im}(u)>0  \\
		\cD(u_*)  & \mbox{if Im}(u)<0
	\end{array}\right. ,\nn
\cD (u) &= \left[\log Y_1 \right](u), \nn
\Delta(i u +\e) - \Delta(i u -\e) &= 2\pi L i, \text{ for } u\in \R.
\end{align}
This set of relations together with the $Y$ system equations (\ref{Ysys:Q}-\ref{Ysys:y}) forms the \emph{analytic $Y$ system}. Let us stress that at this stage the only information on the deformed model's parameters lies in the branch point location $\theta$, the shift distance $\ad$ and the periodicity of the $Y$ functions. As in the undeformed case, when supplemented with the assumption that the $Y$ functions do not have ``extra'' poles and have the right asymptotics we can rederive the ground-state TBA equations from the analytic $Y$ system. We will not treat the entire derivation of the discontinuity relations here, nor will we show the complete rederivation of the TBA equations. Instead, in the next two sections we will demonstrate how one derives the discontinuity relation for $Y_{1|w}$ and how to derive the TBA equation for $Y_-$ from the analytic $Y$ system, to explain the strategy and highlight the differences with the undeformed case. We also discuss the treatment of the $Y_Q$ functions, as these are the most complicated and require a careful analysis of the deformed dressing kernel. The derivation of the other discontinuity relations and the derivation of the TBA equations from the analytic $Y$ system follows the same strategy as our explicit examples.

\subsubsection{Deriving the $Y_{1|w}^{(\alpha)}$ discontinuity relation}
\label{sec:derivingY1wdisc}
The only term generating a discontinuity on $Z_{\pm 1}$ on the \rhs of the $Y_{1|w}$ TBA equation is
\be
\left(L_{-}^{(\alpha)}-L_{+}^{(\alpha)}\right)\hat{\star} K_1,
\ee
where the kernel $K_1$ has poles at $\pm i\ad$ in the physical strip. Continuing the function $H_{\pm}$ defined by
\be
H_{\pm}(u) = \int_{\hat{Z}_0} dv \left(L_{-}^{(\alpha)}-L_{+}^{(\alpha)}\right)(v) K_1(v-u\mp \ad i),
\ee
around the branch point at $\theta$ we find that
\be
\left[H \right]_{\pm1} (u) = L_{-}^{(\alpha)}(u)-L_{+}^{(\alpha)}(u).
\ee
Since $Y_+$ and $Y_-$ are related by analytic continuation, the discontinuity of $\log Y_{1|w}^{(\alpha)}$ can be written as
\be
\left[\log Y_{1|w}^{(\alpha)}\right]_{\pm1}(u) =\left[L_{-}^{(\alpha)}\right]_0(u).
\ee
\subsubsection{Deriving the $Y_-^{(\alpha)}$ TBA equation from the analytic $Y$ system}
\begin{figure}[!t]
\centering
\begin{subfigure}{5cm}
\includegraphics[width=5cm]{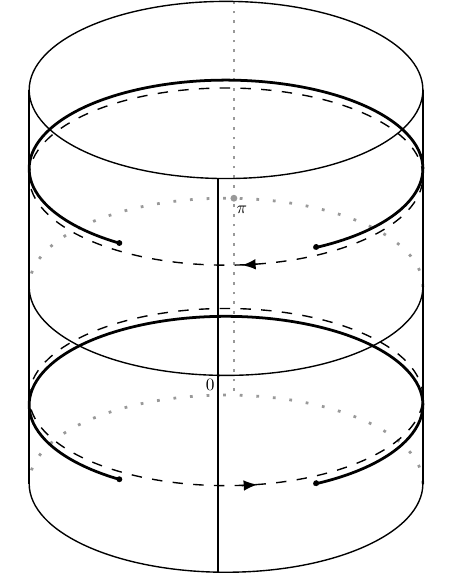}
\caption{}
\end{subfigure}
\quad
\begin{subfigure}{7cm}
\includegraphics[width=7cm]{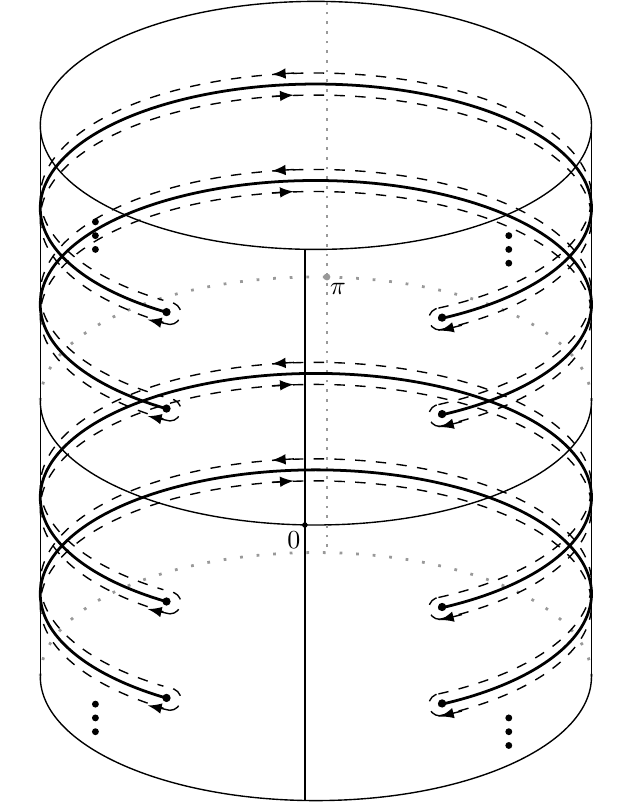}
\caption{}
\end{subfigure}
\caption{The integration contours $\gamma$ (in (a)) and $\Gamma$ (in (b)) are indicated by the dashed lines. The dots in Fig. (b) indicate that the contour continues to $\pm\infty$ along the cylinder in both directions. The solid lines are the branch cuts connecting $\theta\pm i\ad N $ to $-\theta\pm i\ad N$ for $N\in \Z$.}
\label{fig:gammas}
\end{figure}

To derive the TBA equation \eqref{TBAeqns} for $Y_-$ we write\footnote{We drop the index $\alpha$ since the derivation is the same for both values of $\alpha$.}
\be
\log Y_- = 1/2 \left( \log Y_-Y_+ + \log Y_-/Y_+ \right)
\ee
and derive expressions for the two terms on the right-hand side. As shown in Appendix \ref{App:discYmin}, the discontinuity relations for $\log Y_-/Y_+$ of Eqn. \eqref{discs1} are equivalent to the equation
\be
\label{ymin0}
\log \frac{Y_-}{Y_+} (u) = -\Lambda_P \star K_{Py}.
\ee
So to derive the TBA equation it remains to prove
\be
\label{ymin1}
\log Y_-Y_+ = (2L_{M|vw}-2L_{M|w}-\Lambda_M)\star K_M,
\ee
where both sides are branch-cut free on the real axis. One can derive this expression by writing
\be
\log Y_-Y_+(u) = \oint_{\gamma} \frac{dz}{2\pi i}\log Y_-Y_+(z)H(z-u),
\ee
which is valid for $u$ in the physical strip and where $\gamma$ is the positively oriented set of circles that run on the insides of the lines with Im$(u) = \pm i \ad$ as shown in Fig. \ref{fig:gammas}.
The function
\be
H(u) = \frac{1}{2}\cot \frac{u}{2}
\ee
is the natural trigonometric version of $1/u$ as commonly used in the rational case. We can deform the contour $\gamma$ immediately to $\Gamma$, see Fig. \ref{fig:gammas}, using the boundary condition that $\log Y_-Y_+$ vanishes as $u\rightarrow \pm i \infty$, consistent with the TBA equations \eqref{TBAeqns}. This gives
\be
\label{yplusmins}
\log Y_-Y_+(u) =
\sum_{N=1}^{\infty}\sum_{\tau=\pm}  \int_{Z_0} \frac{dz}{2\pi i}H(z+2i \tau N \ad-u) \left[\log Y_-Y_+ \right]_{2\tau N}(z).
\ee
Using the $Y$ system alone one derives as in the undeformed case that
\be
\label{res2}
\left[\log Y_- Y_+  \right]_{\pm 2N} = 2\sum_{J=1}^N \left[L_{J|vw} - L_{J|w}\right]_{\pm(2N-J)}-\sum_{Q=1}^{N}\left[\Lambda_Q \right]_{\pm(2N-Q)}.
\ee
Plugging this into Eqn. \eqref{yplusmins} and cancelling integrals whose contours run in opposite directions and can be deformed to overlap we find
\begin{align}
\log Y_-Y_+(u)  &= \sum_{J=1}^{\infty} \int_{Z_0} \frac{dz}{2\pi i}  \left(2L_{J|vw}(z) - 2L_{J|w}(z) -\Lambda_J(z) \right)\left(H(z-i J \ad-u)-H(z+i J \ad-u) \right),
\end{align}
which produces Eqn. \eqref{ymin1} after recognising the kernel $K_J$ in the difference of $H$s. Adding up Eqs. \eqref{ymin0} and \eqref{ymin1} then produces the TBA equation \eqref{TBAeqns} for $Y_-$.

\subsubsection{Treating the $Y_Q$s}
The TBA equations for the $Y_Q$s contain the notoriously complicated dressing-phase kernel $K^{PQ}_{\mathfrak{sl}(2)}$. This complicates the derivation of the relevant discontinuity equations \eqref{discs1} and rederivation of the TBA equations. We treat this derivation in Appendix \ref{App:discYQ}. Let us just note one technical difference between the deformed and undeformed analysis: in the undeformed case all $Y$ functions obeying the TBA equations vanish as $u\rightarrow \pm i\infty$. In the deformed case, the $Y_Q$ functions do not vanish as $u\rightarrow \pm i\infty$ due to the driving term $\mathcal{E}_Q$, but instead we see that
\be
\log Y_Q \rightarrow \mp \ad QL \text{ as } u \rightarrow \pm i \infty.
\ee
In the deformation of $\gamma$ into $\Gamma$ this leads to an additional constant next to the infinite sum of discontinuities as in Eqn. \eqref{yplusmins}, initially leading to an extra contribution to the $Y_Q$ TBA equations. However, this extra contribution gets cancelled by a constant appearing in the rederivation of the driving term $\mathcal{E}_Q$, see Eqn. \eqref{eq:rederiveenergy}. Namely, the presence of an extra branch cut on the imaginary axis leads to an integral over $K_Q$, which in the undeformed case vanishes identically. In the deformed case the contribution of this integral cancels the constant mentioned above, thereby correctly reproducing the $Y_Q$ TBA equations.

\subsection{From analytic $Y$ system to analytic $T$ system}
\label{sec:T-system}
\begin{figure}[!t]
\centering
\includegraphics[width=10cm]{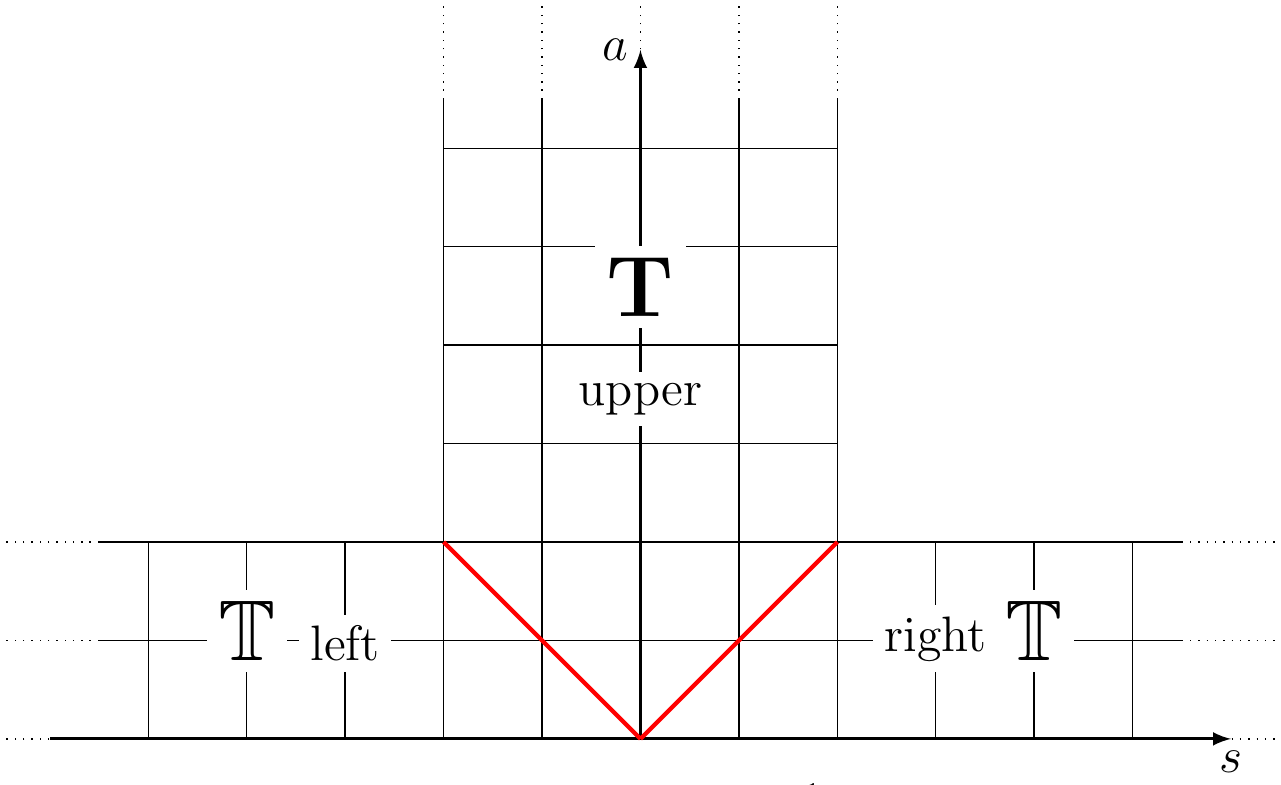}
\caption{The $T$ hook organising all the $T$ functions. It is naturally partioned into \emph{upper}, \emph{left} and \emph{right} band.}
\label{fig:thook}
\end{figure}

Now that we have established equivalence of the analytic $Y$ system to the ground-state TBA equations, we can try to simplify further by introducing $T$ functions to parametrise the $Y$ functions. On an algebraic level this works the same as in the undeformed case, matching the underlying representation theory.\footnote{For contrast, when the deformation parameter $q$ is a root of unity the TBA equations have a more intricate, truncated, structure \cite{Arutyunov:2012zt}, reflecting itself in the parametrisation in terms of $T$ functions \cite{Arutyunov:2012ai}.} Indeed, we have seen in the previous section that the main difference between the analytic $Y$ system for the $\eta$-deformed and the undeformed case is the periodicity of the $Y$ functions. With the standard Hirota map we have
\begin{align}
\label{YsinTs}
Y^{(\pm)}_{M|w} &= \frac{ T_{1,\pm(M+2)} T_{1,\pm M} }{T_{0,\pm(M+1)} T_{2,\pm (M+1)}}, \quad &Y^{(\pm)}_{M|vw} &= \frac{ T_{M+2,\pm 1} T_{M,\pm 1} }{  T_{M+1,0} T_{M+1,\pm 2} }, \quad
Y_{M} = \frac{ T_{M,1} T_{M,-1} }{T_{M-1,0} T_{M+1,0}} \nn
Y_{+}^{(\pm)} &= -\frac{ T_{2,\pm1} T_{2,\pm 3} }{T_{1,\pm 2} T_{3,\pm2}}, \quad &Y_{-}^{(\pm)} &= -\frac{ T_{0,\pm1} T_{2,\pm 1} }{T_{1,0} T_{1,\pm2}},
\end{align}
where the $T$ functions $T_{a,s}$ live on a \emph{$T$ hook}, see Fig \ref{fig:thook}.
$T$ functions with indices that fall outside of this hook are zero by definition. The $T$ functions on the $T$ hook homogenise the $Y$ system equations (\ref{Ysys:Q}-\ref{Ysys:y}) in the sense that if the $T$s satisfy the \emph{Hirota equation}
\be
\label{hirota}
T_{a,s}^+T_{a,s}^- = T_{a-1,s}T_{a+1,s}+ T_{a,s-1}T_{a,s+1},
\ee
then the $Y$s parametrised by these $T$s satisfy the $Y$ system equations (\ref{Ysys:Q}-\ref{Ysys:y}). We are of course interested in special solutions of Eqs. \eqref{hirota} that parametrise periodic $Y$ functions that also satisfy the additional analyticity properties listed in Section \ref{subsub:Deriving the Y}. Transferring these properties is not straightforward, however, due to gauge freedom: the parametrisation of $Y$s in terms of $T$s and the Hirota equation itself are left invariant under redefining
\be
\label{eq:Tgaugetransformation}
T_{a,s} \rightarrow g_1^{[a+s]}g_2^{[a-s]}g_3^{[-a+s]}g_4^{[-a-s]}T_{a,s},
\ee
where the $g_i$ are arbitrary (anti-)periodic functions. The $g_i$ can be anti-periodic as the parametrisation \eqref{YsinTs} is even in the $T$s.
 \begin{table}
\begin{center}
\begin{tabular}[t]{ l | l}
For $a\geq |s|$ & For $s\geq a$ \\
\hline
$\bfT_{a,0} \in \mathcal{A}_{a+1}$ & $\mathbb{T}_{0, \pm s } = 1$ \\
$\bfT_{a,\pm 1} \in \mathcal{A}_{a}$ & $\mathbb{T}_{1, \pm s } \in \mathcal{A}_s$ \\
$\bfT_{a,\pm 2} \in \mathcal{A}_{a-1}$ & $\mathbb{T}_{2, \pm s } \in \mathcal{A}_{s-1}$
\end{tabular}
\end{center}
\caption{Analyticity strips of the two $T$ gauges.}
\label{Tsanalyticity}
\end{table}

\subsubsection{Constructing the $T$ gauges}

We now follow the undeformed case \cite{Gromov:2011cx} in constructing the real gauges $\bfT$ and $\mathbb{T}$ for the upper band and left/right bands of the $T$ hook respectively out of a solution of the analytic $Y$ system. In these gauges the extra analyticity properties take a particular nice form and are listed in Table \ref{Tsanalyticity}. The construction of these gauges is fairly technical and we treat the details in Appendix \ref{App:Tgauges}. Here we will give an overview of the derivation.

We derive a gauge $\JT$ with the right analyticity strips using a construction first discussed in \cite{Balog:2011nm}.
We use the discontinuity relation for $Y_{\pm}^{(\alpha)}$ to construct a gauge transformation that defines the $\bfT$ gauge, which in addition to having the right analyticity strips also satisfies the ``group-theoretical'' properties
\be
\bfT_{3,\pm2} = \bfT_{2,\pm3}, \quad \bfT_{0,0}^+ =\bfT_{0,0}^-, \quad \bfT_{0,s} = \bfT_{0,0}^{[+s]}.
\ee
From the $\bfT$ gauge we directly construct the $\mbT$ gauge through the transformation
\be
\label{hbTinbfT}
\mathbb{T}_{a,s} = (-1)^{a s} \bfT_{a,s} \left( \bfT_{0,0}^{[a+s]}\right)^{\frac{a-2}{2}}.
\ee
The analytic properties of this gauge in the left and right band are not a priori clear. To investigate this we use that the following combination of $Y_{\pm}^{(R)}$ functions can be rewritten as
\be
\frac{1+1/Y_{+}^{(R)}}{1+Y_{-}^{(R)}} = \frac{T_{2,2}^+T_{2,2}^-T_{0,1}}{T_{1,1}^+T_{1,1}^-T_{2,3}},
\ee
i.e. only depend on $\mcT$s from the right wing of the $T$ hook. Under some technical assumptions this allows us to define a gauge $\mcT$ which does satisfy the properties listed in Table \ref{Tsanalyticity} in the right band. In particular we find that this gauge is \emph{$\Z_4$-symmetric}, ie. it satisfies
\be
\hat{\mcT}_{a,s}= (-1)^a \hat{\mcT}_{a,-s}
\ee
on the $T$ hook. The hatted version $\hat{\mcT}$ of the $\mcT$ gauge is defined as follows: on the right band ($s\geq a$) it coincides with the $\mcT$s on the strip with $-\ad<$Im$(u)<0$, but is defined with short cuts. Since the explicit parametrisation of the $\hat{\mcT}$s depends analytically on $s$, one can continue the $\hat{\mcT}$s outside the right band for any $s\in \Z$. This results in a gauge on an infinite horizontal band satisfying the Hirota equation.

Note that our convention of identifying in the lower half-plane is different from the one used in \cite{Gromov:2013pga}, where long- and short-cutted versions of functions are identified in the upper half-plane. Ultimately, this different convention is due to the fact that we have introduced our $x$ functions in Eqs. \eqref{eq:xs} and \eqref{eq:xm} in such a way that they coincide in the lower half-plane, whereas the $x$ functions used in \cite{Gromov:2013pga} coincide in the upper half-plane. We will be able to reconcile this after extracting the needed information from our TBA equations.

The final part of the construction is using the discontinuity relation for the $Y_Q$s to analyse the gauge transformation relating $\mcT$s and $\mbT$s. From this we conclude that the $\mbT$s must have the right analyticity strips and are in fact also $\Z_4$-symmetric, also giving rise to a solution of the Hirota equation on the infinite horizontal band. A byproduct of this derivation is that we find that $\hbT_{1,s}$ only has two (short) cuts at $Z_{\pm s}$. This last fact combined with the $\Z_4$ symmetry of the $\mbT$ gauge allows for the introduction of the $\Pf\mu$ system.
\subsection{Introducing the $\Pf\mu$ system}
\label{sec:pmusystem}
The final reparametrisation consists in introducing the $\Pf\mu$ system.
Starting from the assertion that every infinite band of $T$ functions satisfying the Hirota equation can be parametrised as a Wronskian determinant solution (see \cite{Krichever:1996qd,Gromov:2011cx}), we can greatly simplify the $\hbT$ gauge. Indeed, a general solution of the horizontal band on which the $\hbT$ gauge is defined consists of six independent functions and after enforcing $\Z_4$ symmetry and $\hbT_{0,s}=1$ we find that only two functions remain independent. We can adapt an argument in \cite{Gromov:2011cx} to conclude that these two functions each have only one short $Z_0$ cut, see Appendix \ref{App:hbT-gauge}. The result is a parametrisation in four functions $\Pf_1,\Pf_2,\Pf^3,\Pf^4$ each with only one short $\hat{Z}_0$ cut, as follows
\begin{align}
\label{eq:TsinPs}
\hbT_{0,s} &= 1 &\text{for $s\in \Z$}, \nn
\hbT_{1, s} &= \Pf_1^{[+s]}\Pf_2^{[-s]}-\Pf_1^{[-s]}\Pf_2^{[+s]} &\text{for $s>0$}, \nn
\hbT_{1, s} &= \Pf^{4[+s]}\Pf^{3[-s]}-\Pf^{4[-s]}\Pf^{3[+s]} &\text{for $s<0$}, \nn
\hbT_{2,\pm s} &= \hbT_{1,\pm 1}^{[+s]}\hbT_{1,\pm 1}^{[-s]} &\,\,\text{for $|s|>1$}.
\end{align}
The construction of the entire $\Pf\mu$ system from here is straightforward, since it depends solely on properties of the $T$ gauges that are the same as in the undeformed case: one can use the relation \eqref{hbTinbfT} and the Hirota equation \eqref{hirota} (valid for the $\mathbb{T}$s) to derive expressions for the upper band $\bfT$s in terms of the introduced $\Pf$ functions, and the additional function
\be
\label{defmu}
\mu_{12} = \left(\bfT_{0,1}\right)^{1/2}.
\ee
$\mu_{12}$ is most naturally defined with long cuts, since the right-hand side of Eqn. \eqref{defmu} has long cuts. However, in the remainder $\mu_{12}$ will be the short-cutted version of this function, whereas we will write $\check{\mu}_{12}$ for the long-cutted function. By demanding the analytic and group-theoretical properties of the $\bfT$ gauge one derives, as in \cite{Gromov:2013pga}, a set of relations between these $5$ quantities which can be economically packaged in the \emph{$\Pf\mu$ system}:
\be
\label{Pmusystem}
\mut_{ab}-\mu_{ab} = \Pf_a \Pft_b-\Pf_b \Pft_a, \quad \Pft_a = \mu_{ab} \Pf^b,\quad \Pf_a\Pf^a = 0,\quad  \text{Pf}(\mu) = 1,
\ee
where $\text{Pf}$ is the Pfaffian of the matrix $\mu$. To arrive at this form we introduced the auxiliary objects $\Pf^1,\Pf^2,\Pf_3,\Pf_4$ and $\mu_{ab}$, where the $\mu_{ab}$ are components of an antisymmetric $4\times 4$ matrix. This set of equations for the unknown $\Pf$s and $\mu$ is exactly the same as in the undeformed case. The auxiliary $\mu_{ab}$ are introduced as solutions to the Riemann-Hilbert problems
\be
\frac{\mut_{a 4}}{\mut_{12}}-\frac{\mu_{a 4}}{\mu_{12}} = \frac{\Pft_a \Pft^3-\Pf_a \Pf^3}{\mu_{12}\mut_{12}} \quad \frac{\mut_{a 3}}{\mut_{12}}-\frac{\mu_{a 3}}{\mu_{12}} = \frac{\Pft_a \Pft^4-\Pf_a \Pf^4}{\mu_{12}\mut_{12}},
\ee
where we impose that the (long-cutted) $\check{\mu}_{ab}$ are $2 i \ad $-periodic for consistency with the $T$ system. These equations are of the form
\be
\tilde{f}-f= g
\ee
for (anti-)periodic $g$ and always has a solution
\be
f(u) = \frac{1}{2\pi i} \int_{Z_0} K_{\ad}(u-v) g(v) dv,
\ee
modulo a regular doubly-periodic function, i.e. a constant. Here the kernel $K_{\ad}$ is defined as
\be
\label{zetakernel}
K_{\ad}(u) = \zeta(u) -\zeta(u+2\pi) +\zeta(2\pi),
\ee
with $\zeta$ the Weierstra\ss\, function with quasi half-periods $(2\pi,i\ad)$. This kernel satisfies
\begin{align}
K_{\ad}(u+2\pi) &= - K_{\ad}(u), \nn
\lim_{\ad\rightarrow 0} \ad \cdot K_{\ad} (\ad u) &= \pi \coth (\pi u),
\end{align}
i.e. $K_{\ad}$ is anti-periodic and has the expected undeformed limit. Solutions defined in this way are always anti-periodic in the real direction with period $2\pi$ and periodic in the imaginary direction with period $2 i\ad$, from which one can deduce that the newly introduced $\mu_{ab}$ will be anti-periodic when $\mu_{12}$ is periodic and vice versa. So we see that as in the undeformed case $\check{\mu}_{ab}^{++} = \check{\mu}_{ab}$ for all $a,b$ as a function with long cuts. Quite importantly, in our current conventions the property $\tilde{\hat{\mu}}_{ab} = \hat{\mu}_{ab}^{++}$ does not hold. Since we identify short- and long-cutted functions in the lower half-plane our short-cutted $\mu$ functions satisfy
\be
\tilde{\hat{\mu}}_{ab} = \hat{\mu}_{ab}^{--},
\ee
as a simple inspection shows. We can learn more about the analytic properties of $\mu_{12}$ from the TBA equations.
\subsection{Deducing analytic properties of $\mu$ from TBA}
\label{sec:mufromTBA}
From an algebraic point of view it is natural that the $\Pf\mu$ system defined in the previous section is exactly the same as the original system derived in \cite{Gromov:2013pga}, as the representation theory of $q$-deformed algebras for $q$ real is essentially the same as that of undeformed algebras. The crucial difference between the two cases sits in the particular solution one has to pick up in order for the QSC to describe a given state of a particular model. We distinguish between the undeformed and deformed model by considering solutions on a cylinder rather than a plane. Then, as in the undeformed case, we can try to pick out individual states by imposing extra conditions in the form of asymptotics. A natural starting point to find these asymptotics is a particularly simple constraint on $\mu_{12}$ following from the TBA equations.\footnote{Compare with Section 3.7 in \cite{Gromov:2011cx} or Section 5.3.4 in \cite{Gromov:2013pga}, where the upper-half-plane conventions where used.} Namely, with short cuts we have
\begin{equation}
\label{mufromTBA}
\frac{\mu_{12}^{--}}{\mu_{12}} =  \frac{Y_+}{Y_-} = \exp\left( \Lambda_P \star K^{Py}\right),
\end{equation}
where the dependence on the second argument of the kernel $K^{Py}$ is written in terms of the $x_s$ function to accommodate the short cuts. Expanding the right-hand side as $u \rightarrow -i \infty$ for which $x_s$ diverges we find
\begin{equation}
K^{Py}(v,u) = \frac{1}{2\pi i}  \partial_v \left(  -i \ad \tilde{p}^Q - \tilde{E}_Q + \mathcal{O}\left(x_s^{-1}(u)\right) \right),
\end{equation}
where the mirror energy $\tilde{E}_Q$ and mirror momentum $\tilde{p}_Q$ are defined in Eqn. \eqref{eq:mirrormomentum}. Using formula (6.6) in \cite{Arutynov:2014ota} we now find that
\begin{align}
\label{KPyexpansion}
\Lambda_P \star K^{Py} (u)= -i P + \ad E +\mathcal{O}\left(x_s^{-1}(u)\right) =  \ad E +\mathcal{O}\left(x_s^{-1}(u)\right),
\end{align}
where $E$ and $P$ here are the deformed string energy and momentum respectively and we used the level-matching condition $P=0$. Assume now that $\mu_{12}$ behaves as
\be
\mu_{12} \simeq e^{i u A},
\ee
with $A$ having positive real part\footnote{Assuming $A$ is imaginary or with negative real part leads to a contradiction} and $\simeq$ meaning that $\lim_{u\rightarrow -i \infty} e^{-i u A} \mu_{12} = 1$ where the limit has to be taken avoiding branch cuts, i.e. here $|\text{Re}(u)| > \theta$. We find
\begin{align}
\log \frac{\mu_{12}^{--}}{\mu_{12}} \simeq 2\ad A.
\end{align}
Comparing the two sides in Eqn. \eqref{mufromTBA} yields $A=E/2$, so that
\begin{equation}
\label{muasymp}
\mu_{12} \simeq e^{i u E/2 } = e^{i u \frac{\Delta-J}{2}}.
\end{equation}
Note that this expression is very similar to its undeformed counterpart $\mu_{12} \sim u^{E}$, although the way we obtain this expression is subtly different.  In the deformed case, we do not find energy and momentum as separate coefficients in the expansion of $K^{Py}$, but instead they come together already at the lowest order. This can be interpreted as a mixing of the conserved charges of the deformed theory as a result of the deformation. If we continue the expansion \eqref{KPyexpansion}, at second order we find an expression which manifestly gives the energy in the undeformed limit:
\begin{equation}
\frac{1}{x^{[-Q]}} - x^{[-Q]}-\frac{1}{x^{[+Q]}}+x^{[+Q]},
\end{equation}
even though in the deformed model this expression has no immediate physical interpretation.
\subsubsection{Periodicity properties of $\mu_{12}$}
\label{sec:muperiodicity}
It is important to note the factor of $1/2$ in the exponent in Eqn. \eqref{muasymp}, as it implies that for odd integer $E$ the $\mu_{12}$ function is anti-periodic instead of periodic. This is exactly a possible situation at lowest order in perturbation theory, where quantum corrections play no role yet. Indeed, at lowest order the branch cuts of $\mu_{12}$ dissapear as we send $\theta \rightarrow 0$. Using the regularity requirement for physical states as in the undeformed case, the only possible locations for poles are at the locations where the branch cuts dissapear, at $u=2i \ad N$ for integer $N$, but an analysis along the lines of \cite{Marboe:2014gma} shows that at lowest order there are no poles at these points. Therefore $\mu_{12}$ is analytic at lowest order. Actually, we know that $\mu_{12}^+$ is real analytic, as we show in Eqn. \eqref{mureality} below. As $\mu_{12}$ and $\check{\mu}_{12}$ coincide on the strip just below the real axis and $\check{\mu}_{12}$ is defined as the square root of a $2\pi$-periodic function, see Eqn. \eqref{defmu}, both are at least $4\pi$-periodic just below the real axis. This allows us to write a Fourier expansion
\be
\mu_{12}(u) = \sum_{k\in \mathbb{Z}} a_k e^{ik u/2}
\ee
valid in the strip and using analyticity we can continue this expression anywhere in the complex plane. Combining this expansion with the known asymptotic behaviour around $u=-i \infty$ and the fact that $\mu_{12}^+$ is real implies that $\mu_{12}$ is a (complex) trigonometric polynomial\footnote{Here a \emph{trigonometric polynomial} is a function $f$ of the form
\begin{displaymath}
f(u) = \sum_{k=0}^N \left( a_k \sin( k u/2) + b_k \cos( k u/2) \right).
\end{displaymath}
If the coefficients are real (complex), $f$ is a real (complex) trigonometric polynomial. The \emph{order} of $f$ is denoted $N\in \N$.
}. The asymptotic behaviour directly enforces the periodicity properties: in the asymptotic region $\mu_{12}$ is $2\pi$-(anti-)periodic depending on the parity of $E$ and since $\mu_{12}$ is a trigonometric polynomial this property must hold everywhere. Interestingly, this means that for odd $E$ $\mu_{12}$ is not continuous on the cylinder. In short, we find that for $\theta=0$
\be
\label{eq:muperiod}
\mu_{12}(u) = (-1)^E \mu_{12}(u+2\pi).
\ee
This property can be interpreted as the deformed version of odd-degree polynomials, which have a different limit depending on whether we take $u \rightarrow \pm \infty$. Note that this property does not survive once we go to finite $\theta$: $\check{\mu}_{12}$ retains its periodicity from the lowest order, as it is manifestly a square root of a $2\pi$-periodic function. $\mu_{12}$, however, will lose its periodicity outside of the physical strip in favour of obeying its asymptotics, which for finite $\theta$ are generically no longer integer. Instead, $E$ becomes a measure of the jump of $\mu_{12}$ over the line Re$(u)=\pi$ very far in the lower half-plane.

\subsection{Switching conventions}
\label{sec:conventions}
At this point we have extracted from the TBA equations what we need. This allows us to reconcile the convention issue we raised in Section \ref{sec:pmusystem} by flipping the sign of $u$, $u\rightarrow -u$. This changes only a couple of our current results and conventions as follows:
\begin{align}
\label{newconv}
\check{f} = \hat{f} \text{ in the strip } 0> \text{Im}(u)>-\ad \quad &\Longrightarrow \quad \check{f} = \hat{f} \text{ in the strip } 0< \text{Im}(u) < \ad, \nn
\tilde{\hat{\mu}}_{ab} = \hat{\mu}_{ab}^{--} \quad &\Longrightarrow \quad \tilde{\hat{\mu}}_{ab} = \hat{\mu}_{ab}^{++}, \nn
x(u) \rightarrow \infty \text{ as } u \rightarrow - i \infty &\Longrightarrow  x(u) \rightarrow \infty \text{ as } \ira{u}, \nn
\mu_{12} \simeq e^{i u \frac{\Delta-J}{2}} \quad &\Longrightarrow \mu_{12} \simeq e^{-i u \frac{\Delta-J}{2}},
\end{align}
where now $\simeq$ means we consider the dominant term as $\ira{u}$ avoiding branch cuts. We will use these new conventions in the remaining derivation of the quantum spectral curve.

\subsection{The $\Qf\omega$ system}

To complete the quantum spectral curve we will derive the associated $\Qf\omega$ system. As in the undeformed case, the $\Qf$s are defined as functions with long cuts by
\be
\label{defQ}
\Qf_i = -\Pf^a \Qm_{a|i}^+ \text{ for Im}(u)>0,
\ee
where the $\Qm_{a|i}$ are four independent solutions to the finite-difference equation
\be
X_a^- = \left( \delta_a^b + P_a P^b\right) X_b^+
\ee
labelled by $i=1,\cdots,4$ and analytic in the upper half-plane. The anti-symmetric $4\times 4$ matrix $\omega$ plays the role of $\mu$ in the $\Qf\omega$ system and is defined with long cuts as
\be
\label{defomega}
\omega_{ji}=\Qm_{a|j}^{-}\mu_{ab}\Qm_{b|i}^{-}.
\ee
We go through the derivation in Appendix \ref{App:qomega}. Ultimately one finds that the functions $\Qf_i,\Qf^i$ and $\omega_{ab}$ satisfy the equations
\be
\label{Qomegasystem}
\omegat_{ij}-\omega_{ij} = \Qf_i \Qft_j-\Qf_j \Qft_i, \quad \Qft_i = \omega_{ij} \Qf^j,\quad \Qf_i\Qf^i = 0,\quad  \text{Pf}(\omega) = 1,
\ee
which is the $\Qf\omega$ system as we know it already from the undeformed case.  From the objects defined so far we can now in fact define the entire $QQ$ system containing $256$ functions $\Qm_{A|I}$ with multi-indices $A$ and $I$: define the basic functions\footnote{The unimodularity constraint $\Qm_{1234|1234}=1$ follows from the $2i\ad$ periodicity of $\bfT_{1,0}$.}
\begin{align}
\Qm_{a|\emptyset} := \Pf_a, \quad \Qm_{\emptyset|i} := \Qf_i,\quad \Qm_{\emptyset|\emptyset}=\Qm_{1234|1234}=1
\end{align}
and generate the other $\Qm$s using the finite-difference equations known as the Pl\"{u}cker relations
\begin{align}
\Qm_{A|I}\Qm_{Aab|I} &= \Qm^+_{Aa|I}\Qm^-_{Ab|I}-\Qm^-_{Aa|I}\Qm^+_{Ab|I}, \nn
\Qm_{A|I}\Qm_{A|Iij} &= \Qm^+_{A|Ii}\Qm^-_{A|Ij}-\Qm^-_{A|Ii}\Qm^+_{A|Ij}, \nn
\Qm_{Aa|I}\Qm_{A|Ii} &= \Qm^+_{Aa|Ii}\Qm^-_{A|I}-\Qm^-_{Aa|Ii}\Qm^+_{A|I}.
\end{align}
The resulting $QQ$ system has the same algebraic structure as the one for the undeformed string.

\subsection{$H$ symmetry and reality}
The $QQ$ system derived above has a residual GL$(4)\times$GL$(4)$ gauge freedom dubbed \emph{$H$ symmetry}: all the equations are invariant under a transformation of the form
\be
Q_{A|I} \rightarrow \sum_{\substack{|B|=|A| \\ |J|=|I|}} \left(H_b^{\left[|A|-|I|\right]}\right)_A^B\left(H_f^{\left[|A|-|I|\right]}\right)_I^J Q_{B|J},
\ee
where $M_A^B = M_{a_1}^{b_1}M_{a_2}^{b_2}\cdots M_{a_{|A|}}^{b_{|B|}}$ and $H_b,H_f$ are arbitrary $u$-dependent GL$(4)$-matrices which are $2i\ad$-periodic and $2\pi$-(anti-)periodic. Note that due to the strict periodicity properties regular $H$-matrices are constant by Liouville's theorem, restricting the freedom of $H$ symmetry severely compared to the undeformed case. However, we will see that for most applications this amount of $H$ symmetry will suffice.

To analyse this a bit further, let us consider the conjugation properties of our basic functions $\Pf_a$,$\Qf_i$ and $\mu_{ab}$: it is not clear from our construction that we can ensure nice conjugation properties, but we do know that the energy extracted from the QSC should be real. As was argued in the undeformed case, this suggests that complex conjugation should be a symmetry of the QSC. Assuming this, we can pick up $\Pf_a$ and $\mu$ with nice reality properties using $H$ symmetry: for $\Pf_a$ the conjugation transformation then reads
\be
\bar{\Pf}_a = H_a^b \Pf_b,
\ee
where $H$ should be a lower-triangular constant matrix to not undo the ordering of $\Pf_a$ in the asymptotics as discussed in Section \ref{sec:asymptotics} or its analytic structure. The newly defined $\Pf' = H^{1/2}\Pf$ will now be real functions. By multiplying the relevant $\Pf_a$ by $i$ we can actually pick the following conjugation rule for $\Pf_a$:
\be
\bar{\Pf}_a = (-1)^a \Pf_a,
\ee
which lead to real $\mathbb{T}$, consistent with our original parametrisation \eqref{eq:TsinPs}. We will nevertheless choose our $\Pf$ functions to be real.

The conjugation properties of $\mu$ can now be derived as well: assuming that all $\Pf$ functions can be expressed as a real function times a convergent series in our $x$ function -- \emph{not} real analytic in our case -- we find that conjugation and continuation of these functions commute, since for $u$ in the lower/upper half-plane
\begin{align}
\widetilde{\overline{x(u)}} = \widetilde{\left( \frac{x(\bar{u})\xi+1}{x(\bar{u})+\xi} \right)} =  \frac{(x(\bar{u}))^{\pm1}\xi+1}{(x(\bar{u}))^{\pm1}+\xi}  =\left( \frac{x(\bar{u})\xi+1}{x(\bar{u})+\xi}\right)^{\pm1} =(\overline{x(u)})^{\pm1} = \overline{\tilde{x}(u)},
\end{align}
showing this somewhat surprising property for the $x$ functions. Now we can use the $\Pf\mu$ equations \eqref{Pmusystem} to derive that
\be
\mu_{ab} = - (-1)^{a+b} \mu_{ab}.
\ee
For short cuts we find that $\bar{\hat{\mu}}_{ab} = -(-1)^{a+b}\hat{\mu}_{ab}$ or
\be
\label{mureality}
\overline{\hat{\mu}^{+}}_{ab} = -(-1)^{a+b}\hat{\mu}_{ab}^{+}.
\ee
\subsection{Selecting the right solution}
\label{sec:asymptotics}
Thus far we have derived the QSC from the ground-state TBA equations of our $\eta$-deformed model. As in the undeformed case, we can conjecture that the QSC can be extended to describe any state in the system, provided we pick up the right ``boundary conditions'', in the form of asymptotics of the $\Pf$ and $\Qf$ functions. In constrast to the undeformed case we need to impose these asymptotics in a different part of the complex plane. We conjecture that a state in the $\eta$-deformed model with given charges can be described in the QSC through the following exponential large $iu$ asymptotics. Namely, with $z= e^{-i u/2}$, as we will shortly argue
\be
\Pf_a \simeq A_a z^{-\tilde{M}_a}, \quad \Qf_i \simeq B_i z^{\hat{M}_i}\quad \Pf^a \simeq A^a z^{\tilde{M}_a}, \quad \Qf^i \simeq B^i z^{-\hat{M}_i},
\ee
where
\begin{align}
\label{charges}
\tilde{M} &= \frac{1}{2}\left\{ J_1+J_2-J_3+2,J_1-J_2+J_3,-J_1+J_2+J_3,-J_1-J_2-J_3-2\right\}, \nn
\hat{M} &= \frac{1}{2}\left\{ \Delta-S_1 -S_2 +2 ,\Delta + S_1 +S_2,-\Delta-S_1 +S_2 ,-\Delta+S_1 -S_2-2
\right\},
\end{align}
which follows the distribution of global charges of the classical spectral curve of the $\ads$ superstring. Note that the constant shifts in the powers of the asymptotics of $\Qf_i$ and $\Pf^a$ differ from those of the undeformed case.

While our present choice of $z$ is the simplest function with asymptotic behaviour $e^{-i u/2}$, it is of course not unique. Moreover, it is not real, while this would be desirable given the reality of the $\Pf$ and $\Qf$ functions. It is in fact natural to consider $z=\sin u/2$. Although the undeformed limit of our asymptotics is ambiguous as it lives near the origin, we will see that formally taking  the undeformed limit of $z=\sin u/2 \rightarrow \ad u/2 +\mathcal{O}\left(\ad^3\right)$ gives the correct undeformed power law asymptotics.

\subsubsection{Deriving the asymptotics}
\label{sec:derivingasymptotics}
We can deduce the above asymptotics knowing the $\mu_{12}$ asymptotics from TBA and using the fact that our model should limit to the undeformed quantum spectral curve.

Let us start by noting that we found the canonical quantum numbers $\Delta,J$ in the asymptotics of $\mu_{12}$ as we sent $\ira{u}$ outside the branch cut strip. This motivates us to postulate that all the $\Pf_a$ and $\Qf_i$ have interesting asymptotics in this limit:
\be
\Pf_a \simeq A_a z^{-\tilde{M}_a}, \quad \Qf_i \simeq B_i z^{\hat{M}_i}\quad \Pf^a \simeq A^a z^{\tilde{M}^a}, \quad \Qf^i \simeq B^i z^{-\hat{M}^i},
\ee
where we refrain from interpreting the powers for now and do not impose a relation between the four sets of powers. Now, cf. Eqn. \eqref{QPF1}, we have
\be
\label{QPQ2}
\mathcal{Q}_{a|i}^+-\mathcal{Q}_{a|i}^- = \Pf_a \Qf_i
\ee
to deduce asymptotics for $\Qm_{a|i}$. Writing the asymptotics as
\be
\mathcal{Q}_{a|i} \simeq C_{a|i}z^{d_{a|i}}
\ee
we find that
\be
C_{a|i}\left(e^{\ad d_{a|i}/2}- e^{- \ad d_{a|i}/2} \right)z^{d_{a|i}} \simeq A_a z^{-\tilde{M}_a}B_i z^{\hat{M}_i},
\ee
which when comparing powers leads to $d_{a|i} = -\tilde{M}_a + \hat{M}_i$. We also find
\be
C_{a|i} = \frac{A_a B_i }{2\sinh \ad \frac{d_{a|i}}{2}}.
\ee
Performing the same analysis on Eqn. \eqref{defQ} we find by comparing the powers that $\tilde{M}^a = \tilde{M}_a$. Comparing coefficients in Eqn. \eqref{defQ} we find
\be
\label{eq:Acoefficients}
\frac{A_a A^ae^{ \ad \frac{d_{a|i}}{2}} }{2\sinh \left(\ad \frac{d_{a|i}}{2}\right)} =-1 \qquad \text{ for every } i.
\ee
Note that due to Eqn. \eqref{QPQ2} we could have written Eqn. \eqref{defQ} with a minus shift and comparing the coefficients of the asymptotics of that equation leads to a slightly different formula:
\be
\label{eq:Acoefficientsminus}
\frac{A_a A^ae^{-\ad \frac{d_{a|i}}{2}} }{2\sinh \left(\ad \frac{d_{a|i}}{2}\right)} =-1 \qquad \text{ for every } i.
\ee
The solutions of these equations coincide only when
\be
\label{eq:sumconstraint}
\sum_{a}\tilde{M}_a = \sum_{i}\hat{M}_i.
\ee
We then have
\be
\label{eq:A0A0}
A_{a_0}A^{a_0} =2\frac{\prod_{j}\sinh\left( \ad \frac{\tilde{M}_{a_0}- \hat{M}_{j}}{2}\right)}{\prod_{b\neq a_0} \sinh\left( \ad \frac{ \tilde{M}_{a_0}- \tilde{M}_{b}}{2}\right)} \qquad \text{ for } a_0=1,\cdots,4.
\ee
Performing a similar analysis on
\be
\label{derB}
\Pf_a = -\Qf^i \Qm_{a|i}^{\pm},
\ee
we find that $\hat{M}^i = \hat{M}_i$ and that
\be
\label{eq:B0B0}
B^{j_0} B_{j_0}  =-2\frac{\prod_{a}\sinh\left(\ad \frac{\hat{M}_{j_0}- \tilde{M}_{a}}{2}\right)}{\prod_{j\neq j_0} \sinh\left(\ad  \frac{ \hat{M}_{j_0}- \hat{M}_{j}}{2}\right)}\qquad\text{ for } j_0=1,\cdots,4.
\ee
Continuing to analysis along the same lines, we can also find asymptotics for $\omega$ and $\mu$: Since $\omega^{++} = \omega$ we find that $\omega\simeq 1$, as in the undeformed case. Choosing the right $\Qf_i$ basis we can choose it to be such that it is anti-symmetric with $\omega_{12}=1=\omega_{34}$ and all other elements vanishing. To derive $\mu$ asymptotics we analyse Eqn. \eqref{defomega}. Using the definition of $\omega^{ij}$ and $\Qm^{a|i} = -\Qm_{a|i}^{-t}$ we can invert these relations to obtain
\be
\mu^{ab} = (\Qm^{a|i})^-(\Qm^{b|j})^-\omega_{ij},
\ee
which we can use directly to find the asymptotics of $\mu$ in terms of $\tilde{M}_a$ and $\hat{M}_i$.

Our remaining task now is to find the dependency of the $M$s on the global charges. In order to deduce this we should consider the following set of constraints on the asymptotics:
\begin{itemize}
\item The $\mu_{12}$ asymptotics have a specified form as follows from TBA, see Eqn. \eqref{newconv}.
\item The sum of $\tilde{M}_a$ and the $\hat{M}_i$ should be equal, see Eqn. \eqref{eq:sumconstraint}.
\item We should find the appropriate Bethe equations at weak coupling, which we analyse in Section \ref{sec:weakcoupling} for the $\mathfrak{sl}_2$ sector.
\item In the left-right symmetric sector where $J_2=J_3=S_2=0$,\footnote{This is the sector where the left and right $Y$ functions and associated $T$ functions are (assumed to be) equal, not to be confused with the left and right \emph{bands} of the $T$ hook.} we get constraints on the powers. Namely, $ \Pf^4 = \Pf_1$ and $\Pf^3=\Pf_2$ implies that
\begin{align}
\tilde{M}_1=-\tilde{M}_4,\quad \tilde{M}_2=-\tilde{M}_3,\quad \hat{M}_1=-\hat{M}_4,\quad \hat{M}_2=-\hat{M}_3,
\end{align}
thus constraining their dependence on the other charges and on constants.
\item We can use $H$ symmetry of the $QQ$ system to make sure that all the $\Pf$ and $\Qf$ have different asymptotics and we can order them such that asymptotically
\be
|\Pf_1|<|\Pf_2|<|\Pf_3|<|\Pf_4| \text{ and } |\Qf_2|> |\Qf_1| > |\Qf_4|>|\Qf_3|,
\ee
as in the undeformed case, implying bounds on the asymptotics.
\item In the undeformed limit we should obtain the known asymptotics from the undeformed case.
\end{itemize}
We additionally impose that the dependence on the charges is linear and that the $\Pf$ and the $Q$ asymptotics, depend only on the $J$s, and only on $\Delta$ and the $S_i$, respectively. We use the $\mu_{12}$ asymptotics to find the dependence on $J_1$ and $\Delta$ and  constrain the dependence on $S_1$. From the ordering of $\Qf$ functions and the sum constraint we then find that also the $S$ dependence is exactly as in the undeformed case\footnote{We follow the undeformed case and use the ABA diagram, see Appendix C in \cite{Gromov:2014caa}.}. At this point it seems clear that the only way to ensure consistency with the undeformed asymptotics is to let the dependence on the other charges be the same as in the undeformed case. This leaves only a freedom to add constants, and at this point the asymptotics can be written using four independent constants $\alpha,\beta,\gamma,\delta$:
\begin{align}
\tilde{M} &= \frac{1}{2}\left\{ J_1+J_2-J_3+2\alpha,J_1-J_2+J_3-2\beta,-J_1+J_2+J_3+2\beta,-J_1-J_2-J_3-2\alpha\right\}, \nn
\hat{M} &= \frac{1}{2}\left\{ \Delta-S_1 -S_2 +2\gamma ,\Delta + S_1 +S_2+2\delta,-\Delta-S_1 +S_2 -2\delta ,-\Delta+S_1 -S_2-2\gamma
\right\}.
\end{align}
The fact that at weak coupling we should find the $\mathfrak{sl}_2$ Bethe equations forces that $\beta=0$, as follows from the derivation in the next section. A final constraint from the $\mu_{12}$ asymptotics is that $\alpha=\gamma+\delta$.

We can fix the constants by comparison with the undeformed case. There the asymptotics are
\be
\label{eq:undefasymp}
\Pf_a \simeq A_a u^{-\tilde{M}^{\text{und}}_a}, \quad \Qf_i \simeq B_i u^{\hat{M}^{\text{und}}_i-1}\quad \Pf^a \simeq A^a u^{\tilde{M}^{\text{und}}_a-1}, \quad \Qf^i \simeq B^i u^{-\hat{M}^{\text{und}}_i},
\ee
with
\begin{align}
\tilde{M}^{\text{und}} &= \frac{1}{2}\left\{ J_1+J_2-J_3+2,J_1-J_2+J_3,-J_1+J_2+J_3+2,-J_1-J_2-J_3\right\}, \nn
\hat{M}^{\text{und}} &= \frac{1}{2}\left\{ \Delta-S_1 -S_2 +2 ,\Delta + S_1 +S_2,-\Delta-S_1 +S_2 +2 ,-\Delta+S_1 -S_2
\right\}.
\end{align}
In the undeformed limit our asymptotics for the $\Pf$s and $\Qf$s are proportional to some power of $\ad$ and therefore diverge or vanish. More precisely, for some $N_i>0$
\begin{align}
z^{-\tilde{M}_{1,2}} \rightarrow \bO{c^{-N_{1,2}}},\quad z^{\tilde{M}_{3,4}} \rightarrow \bO{c^{N_{3,4}}}.
\end{align}
In the left-right symmetric sector, where $A^1=A_4$, we find that the product of the coefficients $A_1A_4$ goes as $\bO{c}$, implying at least one of the coefficients vanishes in the undeformed limit. From these constraints we see that, if the undeformed limit at this level is regular, only the $\Pf_{1,2}$ asymptotics can have a finite undeformed limit, whereas the leading $\Pf_{3,4}$ asymptotics necessarily vanish. Hence the subleading term must become leading in the undeformed limit.\footnote{Since the $\Pf_{3,4}$ are auxiliary variables anyway, their asymptotics are fixed by consistency of the QSC, hence uniquely in terms of the well-defined undeformed limit of the $\Pf_{1,2}$.} From this reasoning we find the following comparison of the undeformed and deformed asymptotics:
\begin{align}
\tilde{M}_{1,2} = \tilde{M}^{\text{und}}_{1,2},\quad \tilde{M}_{3,4} - 2 = \tilde{M}^{\text{und}}_{3,4}.
\end{align}
This comparison holds precisely when $\alpha= 1$, and $\beta=0$ as found independently above. Performing a similar analysis for the $\Qf_i$ (where the roles of $\Qf_{1,2}$ and $\Qf_{3,4}$ are reversed) we find that
\begin{align}
\hat{M}_{1,2}-2 = \hat{M}^{\text{und}}_{1,2}-2,\quad \hat{M}_{3,4}  = \hat{M}^{\text{und}}_{3,4}-2,
\end{align}
which holds precisely when $\gamma = 1$ and $\delta=0$.

This completely fixes the asymptotics to be as in Eqs. \eqref{charges} and in particular leads to the following $\mu$ asymptotics:
\begin{align}
\mu_{12}&\simeq z^{\Delta-J_1},&\mu_{23}&\simeq z^{\Delta-J_3+1} \nn
\mu_{13}&\simeq z^{\Delta-J_2}, &\mu_{24}&\simeq z^{\Delta+J_2+2}\nn
\mu_{14}&\simeq z^{\Delta+J_3+1},&\mu_{34}&\simeq z^{\Delta+J_1+2}.
\end{align}
\subsubsection{Left-right symmetric sector}
As a concrete example, let us analyse what happens to the coefficients $A_i$ in the left-right symmetric case. Plugging in the asymptotics from Eqs. \eqref{charges} we find
\begin{align}
A^4A_4 &= A_1 A_4 =2 \frac{\prod_{j}\sinh\left(  \frac{ \tilde{M}_{4}- \hat{M}_{j}}{2} \ad \right)}{\prod_{b\neq 4} \sinh\left(  \frac{\tilde{M}_{4}- \tilde{M}_{b}}{2} \ad \right)} \nonumber \\
&=  -\frac{1}{2 \sinh (\ad /2)} \frac{\left( \cosh\left(  \frac{J+S+1}{2}\ad \right)-\cosh\left(  \frac{\Delta+1}{2}\ad \right)  \right)\left( \cosh\left(  \frac{J-S+3}{2}\ad \right)-\cosh\left( \frac{\Delta+1}{2}\ad \right)  \right) }{\sinh\left( \frac{J+1}{2} \ad \right)\sinh\left( \frac{J+2}{2}\ad \right) }.
\end{align}
This is almost a direct trigonometrisation of the undeformed Formula (16) of \cite{Marboe:2014gma}. However, as noted in the previous section, the undeformed limit of this expression vanishes. Moreover, the leading order of the expansion of this expression in $\ad$ is not the known undeformed expression. Concretely, $A_4$ must vanish to have a regular undeformed limit for the $\Pf_a$ functions, and we should consider the product of $A_1$ and the coefficient of the subleading term of $\Pf_4$ to find the product of coefficients that goes to the undeformed expression.

It is also interesting to note that if we expand the above formula in our branch point parameter $\theta$, assuming that $\Delta = J+S +\mathcal{O}\left(\theta^2\right)$ we find
\be
A_1 A_4 = \mathcal{O}\left(\theta^2\right),
\ee
analogous to the undeformed case where this is $\mathcal{O}\left(g^2\right)$. This indicates that it should be possible to adapt the analytic algorithm of \cite{Marboe:2014gma} to perturbatively solve our deformed QSC.

\subsection{Weak coupling Bethe equations}
\label{sec:weakcoupling}
We can check the derived quantum spectral curve and asymptotics by comparing to the Bethe-Yang equations of the $\eta$-deformed model as given in \cite{Arutynov:2014ota}. We will zoom in on the $\mathfrak{sl}_2$ sector of the theory. In this sector only three of the global charges ($J_1 = L$, $S_1=S$ and $\Delta = L+S +\mathcal{O}\left(\theta^2\right)$) are non-zero. The Bethe-Yang equations in this sector are
\be
\left(\frac{1}{q} \frac{x_j^++\xi}{x_j^-+\xi}\right)^L = \prod_{k\neq j}^{S}\left(\frac{x_j^+-x_k^-}{x_j^--x_k^+} \right)^{-1}  \frac{1-\frac{1}{x_j^+ x_k^-}}{1-\frac{1}{x_j^- x_k^+}}
\ee
and, defining $S(u)=\sin\frac{u}{2}$ and taking the $\theta \rightarrow 0$ limit, these reduce to
\be
\label{XXZBethe}
\left( \frac{S^+_i}{S^-_i}\right)^L= \prod_{k\neq i}^{S}\frac{S^{--}(u_j-u_k)}{S^{++}(u_j-u_k)},
\ee
which are the $\mathfrak{sl}_2$-XXZ Bethe equations. As in the undeformed case we expect to find these equations also from the QSC at zero coupling, by associating the roots $u_j$ with zeroes of $\mu_{12}^+$. These roots can then be associated to the exact Bethe roots in the TBA description as zeroes of $Y_{1,0}+1$, since zeroes of $\mu_{12}^+$ imply zeroes of $Y_{1,0}+1$.\footnote{See the discussion in Section 4.5 of \cite{Gromov:2014caa} or \cite{Gromov:2009zb}.} In particular we will see how this puts restrictions on the asymptotics defined in the previous section.

We start by expanding the $\Pf\mu$ system at lowest order: using $H$ symmetry we can set $A_1=\mathcal{O}\left( \theta^2\right)$, such that $\Pf_1$ vanishes at lowest order. This splits the $\Pf\mu$ system into two parts as in the undeformed case \cite{Marboe:2014gma} and using some algebra we obtain the following $TQ$-like equation for $Q=\mu_{12}^+$:
\be
-T Q +\frac{1}{\left(\Pf_2^-\right)^2}Q^{--}+\frac{1}{\left(\Pf_2^+\right)^2}Q^{++}=0,
\ee
where $T$ is given by the following rational function of $\Pf_a$s:
\be
\label{eq:defT}
T = \frac{\Pf_3^-}{\Pf_2^-}-\frac{\Pf_3^+}{\Pf_2^+}+\frac{1}{\left(\Pf_2^-\right)^2}+\frac{1}{\left(\Pf_2^+\right)^2}.
\ee
As long as we can make sure that $T$ is pole free, we can use the usual philosophy to obtain an equation on the zeroes of the function $Q$: at each zero $u_k$ of $Q$ we must have
\be
\label{wcbe1}
\left(\frac{\Pf_2^+}{\Pf_2^-}\right)^2(u_k) = - \frac{Q^{--}}{Q^{++}}(u_k).
\ee
To further analyse this, we first focus on $\mu_{12}$: as argued in Section \ref{sec:muperiodicity}, $\mu_{12}$ is a $4\pi$-periodic analytic function at lowest order. Taking into account the asymptotics from Eqs. \eqref{newconv} (which in this sector read $\mu_{12}\simeq z^S$) and the reality condition \eqref{mureality} we restrict the Fourier series on $\mu_{12}^+$, showing it is a real trigonometric polynomial, i.e. of the form
\be
Q(u) = \sum_{k=1}^S \left( a_k \sin(ku/2) + b_k \cos(k u/2)\right),
\ee
with real coefficients. Restricted to the complex strip with real part $[0,4\pi[$, $Q$ has $2S$ zeroes. Using Louiville's theorem we can prove easily that
\be
Q(u) \sim \prod_{k=1}^{2S} S \left( \frac{u-u_k}{2} \right),
\ee
where the $u_k$ are the zeroes of $Q$. The periodicity of $Q$, cf. Eqn. \eqref{eq:muperiod},
\be
\label{req1}
Q(u+2\pi) = (-1)^S Q(u),
\ee
relates the roots in a simple way, as it is equivalent to the following statement about the zeroes:
\be
\{u_k\}_{1\leq k \leq 2S} = \{u_k+2\pi \mod 4\pi \}_{1\leq k \leq 2S}.
\ee
This allows us to rewrite
\be
Q(u)  \sim \prod_{k=1}^{S} S(u-u_k),
\ee
giving us the right-hand side of Eqn. \eqref{wcbe1}. We can also analyse $\Pf_2$: at lowest order its branch cut vanishes and leaves a possible pole at zero. After factoring out this pole by multiplying with an appropriate power of the factor $S(u)$, $\Pf_2$ is a real analytic function with a convergent Fourier series. Using its asymptotics and reality we again restrict the series to be a trigonometric polynomial. In fact, assuming as in the undeformed case that the pole at the origin has order $L/2$, the trigonometric polynomial trivialises to a constant, leaving us with
\be
\Pf_2(u) \sim S(u)^{-L/2}.
\ee
This implies that $T$ of Eqn. \eqref{eq:defT} is indeed pole free, and if we combine this result with the above we find indeed that Eqn. \eqref{wcbe1} produces the $\mathfrak{sl}_2$-XXZ Bethe equations \eqref{XXZBethe}.

\section{Mirror duality and the undeformed mirror limit}
\label{sec:mirrordualityandlimit}

In the above we found the quantum spectral curve for the spectral problem of the $\eta$-deformed string, starting from the TBA equations describing the thermodynamics of the associated mirror models. As mentioned in the introduction, our family of $\eta$-deformed strings is actually closed under the associated double Wick rotation. Namely, taking $\theta \rightarrow \pi - \theta$ at fixed $\ad$ is equivalent to a double Wick rotation. To explicitly match the parametrisation this should be combined with a shift $u \rightarrow u + \pi$. To match the labelling of states and charges as in e.g. \cite{Arutyunov:2007tc}, we should moreover interchange the charges $J_{2,3} \leftrightarrow S_{1,2}$ and a re-identify the string circumference $J=J_1$, in terms of the mirror length $R=\ad J$ \cite{Arutynov:2014ota}. Note that the shift of $\theta$ and $u$ interchanges the $x$ functions $x_s$ and $x_m$ of Eqs. (\ref{eq:xs} \ref{eq:xm}) as it should.\footnote{To be precise, $x_s$ becomes $-x_m$ and vice versa. This relative sign is inconsequential and can be avoided by considering $\theta \rightarrow \theta + \pi$ instead \cite{Arutynov:2014ota} -- the model is invariant under $\theta \rightarrow - \theta$ -- but then the inequivalent models would be parametrised by $\theta \in [-\pi/2,0] \cup [\pi/2,\pi]$.} As such, our quantum spectral curve at $\theta=\theta_0$ not only describes the spectrum of our models at $\theta_0$, but also the thermodynamics of our models at $\pi-\theta_0$. Interestingly, this shift of theta actually exchanges the (analytic properties of) the $\Pf\mu$ and $\Qf\omega$ systems. Of course, to consider strict thermodynamics, in this second picture one does not want to add ``string excitations'' to the mirror model, meaning we would consider all charges except the energy $\Delta$ and the mirror length $R$ to be zero in the QSC asymptotics. We should also take into account that we started by computing Witten's index -- $\mbox{Tr}\left((-1)^F e^{-\beta H}\right)$ -- in the mirror theory, and to undo this and get back to the standard free energy, we should add $i\pi$ chemical potentials for the $y$ particles of the TBA. In the undeformed case it is well known that such chemical potentials introduce particular exponentially decay in the QSC asymptotics. It is an interesting question to understand the appropriate generalisation of these asymptotics in this context. Of course, once this is understood it should be simple to add general chemical potentials to the partition function.

In this context it is particularly interesting to consider the ``undeformed'' mirror limit $\theta\rightarrow\pi$,\footnote{At the level of the sigma model this is in some sense a maximal deformation limit, cf. Appendix \ref{App:parametereta}. Algebraically speaking, it is a contraction limit \cite{Pachol:2015mfa}.} because there we describe the spectral problem of the mirror model which is the light-cone gauge-fixed version of a string sigma model itself \cite{Arutyunov:2014cra,Arutyunov:2014jfa}, as well as the thermodynamics of the undeformed $\ads$ string which have recently been explicitly related to the Hagedorn temperature \cite{Harmark:2017yrv}. To concretely take this limit, we shift $u\rightarrow u + \pi$, rescale $u \rightarrow 2 \ad u$, and consider the limit $\ad \rightarrow 0^+$ with $\theta = 2 \arccos(2 g \sinh \ad$), obtained by shifting $\theta$ by $\pi$ in Eqs. \eqref{eq:parameters}. In this limit, the cut structure of the QSC is as expected: the $\Pf$ and $\mu$ functions now have \emph{long} cuts with branch points $\pm 2g$, while the $\Qf$ and $\omega$ functions have short cuts. Beyond this, the discussion of asymptotics immediately follows the one for the regular undeformed limit around Eqn. \eqref{eq:undefasymp}, up to the above mentioned interchange of charges.\footnote{Formally, we are still considering asymptotics around $u \rightarrow i \infty$, not crossing cuts. In the undeformed (mirror) limit we can move around at infinity and consider $u\rightarrow \infty$ instead, since the $\Pf$ and $\Qf$ functions have no obstructing cuts. More concretely, the analysis of for instance the $\mu$ asymptotics from the undeformed string and mirror TBA is unaffected by choice between $i \infty$ and $\infty$.} In other words, up to some state relabelling, the QSC for the spectrum of the mirror model is obtained by simply flipping the branch cut structure. This is formally equivalent to exchanging the undeformed $\Pf\mu$ and $\Qf\omega$ systems. These same QSC equations can also be used to efficiently compute the Hagedorn temperature in the setup of \cite{Harmark:2017yrv}. The only required modification is in the prescribed asymptotics, which should be by exponential decay of the form $e^{-\pi u}$ to account for the above-mentioned difference between Witten's index and the regular free energy, and a possible shift of the energy charge due to the finite difference between the $J$ charge and the classical scaling dimension which are taken to infinity in the spectral problem and Hagedorn temperature problem respectively. As this question is already under investigation by the authors of \cite{Harmark:2017yrv}, we will not pursue this in more detail here.

\section{Conclusions}

In this paper we have derived the quantum spectral curve for the $\eta$-deformed superstring, greatly simplifying the spectral problem for this model. Although the deformation does not affect the form of the equations, it shows up in the underlying analytic structure. As such, our QSC can be thought of as a trigonometrisation of the rational QSC describing the spectrum of $\mathcal{N}=4$ super Yang-Mills theory. Indeed, rather than living on a plane with cuts, our functions now live on a cylinder with cuts, or an appropriate cover in some cases. We rigorously derived this QSC starting from the $\eta$-deformed TBA equations via the associated analytic $Y$ and $T$ systems.

To describe arbitrary excited states, in analogy to the undeformed case we proposed asymptotics for the QSC functions depending on the quantum numbers of such states. In contrast to the undeformed model where power law asymptotics are specified for large real values of the spectral parameter, here this asymptotic direction is not available, and we instead considered exponential asymptotics for large imaginary spectral parameter. Our asymptotic prescription can nevertheless be smoothly linked to the power law prescription for the undeformed model. One important open question on this point is to construct the classical spectral curve for the $\eta$-deformed model, and contrast it with our QSC and asymptotics, similarly to how this was done in the undeformed case.

Next, it would be interesting to use the QSC to investigate the spectrum of the deformed theory, which we can think of as interpolating from the undeformed string spectrum to the undeformed mirror theory spectrum. Perturbatively we can try to approach this through the algorithm proposed in \cite{Marboe:2014gma,Marboe:2017dmb}. A particular challenge here is solving the trigonometric Bethe equations or Q system that would kick-start the procedure. We have started to analyse the solution for the $\eta$-deformed Konishi state, as a good test case from where to generalise.

As mentioned in the previous section, it would also be very interesting to use the trigonometric QSC to look at the thermodynamics of our models, in particular the undeformed string. Here it would be great to understand how to incorporate chemical potentials in our QSC, analogous to how this was done in the undeformed case \cite{Kazakov:2015efa} at least for purely imaginary chemical potentials corresponding to twists. In the undeformed case, a purely imaginary chemical potential $i \beta = \log \mbox{x}$ introduces exponential asymptotics of the form $\mbox{x}^{-i u}$. As power law behaviour naturally became exponential behaviour in our deformed case, it is not immediately clear how these twist exponentials should be modified. There are two reasons to believe these exponentials might not need modification at all. First, multiplying the $\Pf$ functions by similar exponentials -- $\mbox{x}^{-iu/2\ad}$ -- would precisely introduce the correct twists in the weak coupling Bethe equations, cf. Section \ref{sec:weakcoupling}. Second, from the perspective of the light-cone gauge-fixed string sigma model, certain twists do not translate to chemical potentials but instead affect the level matching condition, cf. e.g. \cite{deLeeuw:2012hp}. Setting $P=\beta$ in the discussion surrounding Eqn. \eqref{KPyexpansion}, would introduce exactly a term of the form $\mbox{x}^{-iu/2\ad}$ in $\mu_{12}$. Adding such exponentials raises one immediate concern however, as the frequency of this exponential, $\beta/2c$, is not a multiple of $\pi$ for generic $\beta$ or $c$ and it would therefore dramatically affect the surface on which our equations are defined. It would be great to fully develop our understanding of such chemical potentials or twists in the QSC. Once this is understood, it would be interesting to consider the Hagedorn temperature computation of \cite{Harmark:2017yrv} in our deformed setting, and interpolate from the string to the ``mirror'' Hagedorn temperature.

It is also relevant to note that the $\eta$-deformed $\ads$ string ``contains'' many other integrable deformations of the $\ads$ string, fitting into the class of homogeneous Yang-Baxter deformations \cite{Kawaguchi:2014qwa,Matsumoto:2014gwa,Matsumoto:2015uja,vanTongeren:2015soa,Osten:2016dvf,Hoare:2016wsk,Borsato:2016pas}. Namely, by considering particular coupled infinite boost and $q\rightarrow1$ limits, it is possible to extract many homogeneous Yang-Baxter deformations of $\ads$ at the level of the sigma model action \cite{Hoare:2016hwh}, including for instance the gravity dual of canonical non-commutative SYM. It would be very interesting to see whether such boosts can be implemented directly in the QSC, to thereby extract the QSC and in particular its asymptoticity data that describe these models which often have an interesting AdS/CFT interpretation \cite{vanTongeren:2015uha,vanTongeren:2016eeb}. Of course it would also be interesting to directly investigate the QSC description of homogeneous Yang-Baxter models, beyond the basic Cartan-twisted ones. The one loop spectral problem for the simplest non-Cartan twisted model -- a particular null dipole deformation -- has recently been investigated in \cite{Guica:2017mtd}, indicating non-trivial asymptotics for the QSC. Of course, it would be great to use the general algebraic structure of twisted models to formulate a general Yang-Baxter deformed QSC.

Another interesting direction to investigate is the model defined by the exact $q$-deformed S matrix for $q$ a root of unity instead of real.  The representation theory of quantum algebras is quite different when $q$ is a root of unity, which will presumably reflect itself in the QSC -- as it does in the TBA \cite{Arutyunov:2012zt,Arutyunov:2012ai} -- from both an algebraic as well as an analytical perspective. Namely, the number of $Y$ functions in this case is finite to begin with, and the additional overall real periodicity that we got in our real $q$ case becomes imaginary periodicity in the root of unity case, and this periodicity is generically not compatible with the periodicity we would expect for $\mu$.

Finally, one might wonder whether it is possible to find an elliptic deformation of the $\ads$ string, or at least its exact S matrix. The QSC as currently formulated makes heavy use of large $u$ asymptotics, which does not seem to allow for further compactification of the spectral plane. It thus seems that an elliptic QSC, if it exists, will have a considerably different structure.

\section*{Acknowledgements}

We would like to thank Gleb Arutyunov, Till Bargheer, Andrea Cavaglia, Nikolay Gromov, Lorenz Hilfiker, Vladimir Kazakov, F\"edor Levkovich-Maslyuk, Christian Marboe, Dmytro Volin, and Matthias Wilhelm for insightful discussions, and Gleb Arutyunov for valuable comments on the draft. R.K. is supported by the German Science Foundation (DFG) under the Collaborative Research Center (SFB) 676 ”Particles, Strings and the Early Universe” and the Research Training Group (RTG) 1670 ”Mathematics inspired by String Theory and Quantum Field Theory” . S.T. is supported by L.T. S.T. acknowledges further support by the SFB project 647  ``Space-Time-Matter: Analytic and Geometric Structures''.

\begin{appendices}
\section{Definitions and conventions}
\label{App:Definitions}

\subsection{Parameters of the $\eta$ model}
\label{App:parametereta}

The $\eta$-deformation of the $\ads$ superstring is a classically integrable sigma model. As for the undeformed string, its quantum spectrum can be described in an exact S-matrix approach. Namely, upon formal light-cone gauge fixing, the $\mathfrak{psu}_q(2,2|4)$ symmetry of the string breaks down to two copies of centrally extended $\mathfrak{psu}_q(2|2)$. Supposing that, as in the undeformed case \cite{Arutyunov:2006ak}, this algebra picks up a final central extension in the decompactification limit, the resulting algebra fixes the two body S-matrix of the model \cite{Beisert:2008tw}, see also \cite{Beisert:2017xqx}, up to a scalar factor constrained by crossing symmetry. The resulting dressing factor can be found similarly to the undeformed case \cite{Hoare:2011wr}, see also \cite{Arutynov:2014ota}, fixing the S matrix completely. This S-matrix, and hence the Bethe ansatz, depend on two parameters, $h$ and $q$. Meanwhile the $\eta$ model depends on the effective ``string'' tension $T$, and the deformation parameter $\varkappa$ appearing in the action.\footnote{The original deformation parameter in the action of the full model is called $\eta$, hence the model's name. In the bosonic action the natural deformation parameter is $\varkappa=2 \eta/(1-\eta)^2$, which turns out convenient.} The tree-level S matrix of the sigma model matches the expansion of the exact S-matrix perfectly provided we identify \cite{Arutyunov:2013ega,Arutyunov:2015qva}\footnote{The $\eta$-deformed string admits several formulations \cite{Delduc:2014kha} that look inequivalent at the geometric level but nevertheless have equivalent S matrices at least at tree level \cite{Hoare:2016ibq}.}
\begin{equation}
q = e^{-\varkappa/T}, \qquad \mbox{ and } \qquad h = \frac{T}{\sqrt{1+\varkappa^2}}.
\end{equation}
A priori it is not clear whether this identification holds beyond tree level. In fact, assuming $q$ to remain real, unitary of the exact S-matrix tells us it cannot, since it requires \cite{Arutynov:2014ota}
\begin{equation}
0 \leq \frac{h^2}{4} (q-1/q)^2 \leq 1,
\end{equation}
which would imply
\begin{equation}
0\leq \frac{T^2 \sinh^2{\frac{\varkappa}{T}}}{1+\varkappa^2}\leq 1.
\end{equation}
This is clearly violated for small $T$ but fine in the perturbative large $T$ regime. We will parametrise the unitarity-compatible space of couplings by $\theta$ and $a$
\begin{equation}
q = e^{-a}\qquad \mbox{ and } \qquad h \sinh a = \sin \frac{\theta}{2}
\end{equation}
which turn out natural for the Bethe ansatz and quantum spectral curve. Without loss of generality \cite{Arutynov:2014ota} we can restrict to positive real $a$ and $\theta \in [0,\pi]$.
\subsection{Kernels}
\label{App:kernelsandenergies}
The kernels that appear in the TBA equations \eqref{TBAeqns} are defined as
\begin{equation}
\begin{aligned}
K_M (u) &\,= \frac{1}{2\pi} \frac{ \sinh M\ad}{\cosh M\ad- \cos u},\\
K_{MN} (u)&\, = K_{M+N} (u)+ K_{|M-N|} (u)+2\sum_{j=1} ^{\min(M,N)-1} K_{|M-N|+2j} (u).
\end{aligned}
\end{equation}
The other kernels are defined directly from the scattering matrices
\begin{equation}
\begin{aligned}
S_\pm^{yQ}(u,v) &\,= q^{Q/2} \frac{(x(u))^{\mp1} - x^+(v) }{(x(u))^{\mp1} - x^-(v) }\sqrt{\frac{x^+(v)}{x^-(v)}}, \\
S_{xv}^{QM}(u,v) &\,= q^{Q} \frac{x^-(u) - x^{+M}(v) }{x^+(u) - x^{+M}(v) } \frac{x^-(u) - x^{-M}(v) }{x^+(u) - x^{-M}(v) }\frac{x^+(u)}{x^-(u)} \prod_{j=1}^{M-1} S_{Q+M-2j}(u-v) ,
\end{aligned}
\end{equation}
where $x^{\pm}(v) = x(v\pm i Q \ad)$, $x^{\pm M}(v) = x(v\pm i M \ad)$ and
\be
S_{Q}(u) = \frac{\sin \frac{1}{2} (u-i Q \ad)}{\sin \frac{1}{2} (u+i Q \ad)}.
\ee
They are given by
\begin{equation}
\begin{aligned}
K_{xv}^{QM}(u,v) &\, = \frac{1}{2\pi i} \frac{d}{du} \log S_{xv}^{QM} (u,v), \quad &
K_{vwx}^{QM}(u,v) &\, = -\frac{1}{2\pi i} \frac{d}{du} \log S_{xv}^{MQ} (v,u), \\
K_{\beta}^{Qy}(u,v) &\, = \frac{1}{2\pi i} \frac{d}{du} \log S_{\beta}^{Qy} (u,v), \quad &
K_{\beta}^{yQ}(u,v) &\, = \beta \frac{1}{2\pi i} \frac{d}{du} \log S_{\beta}^{Qy}(v,u).
\end{aligned}
\end{equation}
The driving term is defined as
\be
\tilde{\mathcal{E}}_Q = \log q^{Q} \frac{x(u-i Q a)+\xi}{x(u+i Q a)+\xi}
\ee
and finally for completeness the $\mathfrak{sl}(2)$ S-matrix kernel
\be
K^{PQ}_{\mathfrak{sl}(2)}(u,v) = \frac{1}{2\pi i} \frac{d}{du} \log S^{PQ}_{\mathfrak{sl}(2)}.
\ee
We will not need this kernel explicitly, except in its simplified form discussed in Appendix \ref{App:SimplifyingDressing} below. For this we do need some more detail on the dressing phase and associated kernel.

\paragraph{The dressing phase and kernel.} Similarly to the undeformed case, the dressing phase $\sigma$ is expressed in terms of $\chi$ functions as
\begin{equation}
\label{eq:dressingphasechidef}
\sigma(z_1,z_2) \equiv e^{i \theta_{12}}= e^{ i\left(\chi(x_1^+,x_2^+) - \chi(x_1^-,x_2^+) - \chi(x_1^+,x_2^-) + \chi(x_1^-,x_2^-)\right) }.
\end{equation}
Before Wick rotation to the mirror theory these $\chi$ functions are determined entirely in terms of a set of $\Phi$ functions
\begin{equation}
\chi(x_1,x_2) = \Phi(x_1,x_2),
\end{equation}
defined via a double contour integral as \cite{Hoare:2011wr}
\begin{equation}
\label{eq:phidef}
\Phi(x_1,x_2) = i \oint_{\mathcal{C}} \frac{dz}{2 \pi i} \frac{1}{z-x_1} \oint_{\mathcal{C}}\frac{dw}{2 \pi i}   \frac{1}{w-x_2}\log  \frac{\Gamma_{q^2}(1+\frac{i}{2\ad}(u(z)-u(w)))}{\Gamma_{q^2}(1-\frac{i}{2\ad}(u(z)-u(w)))},
\end{equation}
where $\Gamma_{q}$ denotes the $q$-gamma function. For our real $q$ case the integration contour is defined as \cite{Arutynov:2014ota}
\begin{equation}
\label{eq:dressingcontourdef}
\mathcal{C} \, : \,\,|y|^2-1+(y^*-y)\xi = 0,
\end{equation}
which are just the values $x_s(u)$ takes along the contour tracing its cut from $-\theta$ to $\theta$.

The natural dressing factor for the mirror model that appears in the TBA equations is
\begin{equation}
\Sigma_{12} = \frac{1-\frac{1}{x_1^+ x_2^-}}{1-\frac{1}{x_1^- x_2^+}} \sigma_{12},
\end{equation}
defined by appropriate analytic continuation of the above objects. Fused to describe bound state scattering we have \cite{Arutyunov:2012ai}\footnote{Note that the definition of $\Sigma$ in our real $q$ case \cite{Arutynov:2014ota} is slightly different from the one of \cite{Arutyunov:2012ai} for the $|q|=1$ case. The present definition is natural from the point of view of mirror duality. We effectively go back to the $|q|=1$ conventions in our definition of $K^{\Sigma}_{QM}$ below, to get kernels and TBA equations analogous to the undeformed model.}
\begin{align}
-i\log\Sigma^{QM}(y_1,y_2) =& \, \Phi(y_1^+,y_2^+)-\Phi(y_1^+,y_2^-)-\Phi(y_1^-,y_2^+)+\Phi(y_1^-,y_2^-) \label{eq:improveddressingphase} \\ &\, -\frac{1}{2}\left(\Psi(y_1^+,y_2^+)+\Psi(y_1^-,y_2^+)-\Psi(y_1^+,y_2^-)-\Psi(y_1^-,y_2^-)\right) \nonumber \\
&\, +\frac{1}{2}\left(\Psi(y_{2}^+,y_1^+)+\Psi(y_{2}^-,y_1^+)-\Psi(y_{2}^+,y_1^-)
-\Psi(y_{2}^-,y_1^-) \right) \nonumber \\
&\, - i \log \frac{i^Q \, \Gamma_{q^2}\left(M - \frac{i}{2a}(u(y_1^+) - v(y_2^+))\right)}{i^{M} \Gamma_{q^2}\left(Q+ \frac{i}{2a}(u(y_1^+) - v(y_2^+))\right)}\frac{1- \frac{1}{y_1^+y_2^-}}{1-\frac{1}{y_1^-y_2^+}}\sqrt{\frac{y_1^+ + \xi}{y_1^-+\xi} \frac{y_2^- + \xi}{y_2^+ + \xi}} \nonumber \\
& \, + \frac{i}{2} \log q^{Q-M} e^{-i(Q+M-2)(u-v)}\nonumber,
\end{align}
where
\begin{equation}
\label{eq:psidef}
\Psi(x_1,x_2) \equiv i \oint_{\mathcal{C}} \frac{dz}{2 \pi i} \frac{1}{z-x_2} \log  \frac{\Gamma_{q^2} (1+\frac{i}{2\ad}(u_1-u(z)))}{\Gamma_{q^2} (1-\frac{i}{2\ad}(u_1-u(z)))} \, .
\end{equation}

We define the associated integration kernel as
\begin{equation}
K^{\Sigma}_{QM}(u,v) = \frac{1}{2\pi i} \frac{d}{du} \log\Sigma^{QM}(y_1(u),y_2(v)) \sqrt{\tfrac{y_1^+}{y_1^-}\tfrac{y_1^-+\xi}{y_1^+ + \xi} \tfrac{y_2^-}{y_2^+}\tfrac{y_2^+ + \xi}{y_2^- + \xi}}.
\end{equation}
where we added a factor that reduces to one in the undeformed limit to simplify expressions below.

\paragraph{Energy and momentum.} The mirror energy and momentum are defined through the $x$ functions as
\be
\label{eq:mirrormomentum}
 \tilde{p}^Q = \frac{i}{\ad}\log \left( q^Q \frac{x^+}{x^-}\frac{x^-+\xi}{x^++\xi}\right), \quad \tilde{E}_Q= -\log \left( \frac{1}{q^Q}\frac{x^++\xi}{x^-+\xi}\right).
\ee

\paragraph{Identities.} Some important identities for the kernels are
\begin{equation}\label{kernelids}
\begin{aligned}
K_{Qy}(u,v) &\defeq& K_{-}^{Qy}(u,v)-K_{+}^{Qy}(u,v)  = K(u+ i Q \ad,v) -K(u- i Q\ad,v) \\
K_{yQ}(u,v) &\defeq& K_{-}^{yQ}(u,v)+K_{+}^{yQ}(u,v)  = K(u,v- i Q \ad) -K(u,v+ i Q \ad) \\
K_{-}^{yQ}(u,v) - K_{+}^{yQ}(u,v) &=& K_{-}^{Qy}(u,v)+K_{+}^{Qy}(u,v)  = K_Q(u-v),
\end{aligned}
\end{equation}
with
\be
K(u,v) = \frac{1}{2\pi i} \frac{d}{du} \log\frac{x(u) - 1/x(v) }{x(u) - x(v) }.
\ee
Let us also define the similar kernel
\be
\bar{K}(u,v) = \theta\left(|u|-\theta \right) \frac{1}{2\pi i} \frac{d}{du} \log\frac{x(u) - 1/x_s(v) }{x(u) - x_s(v) }
\ee
and the associated
\be
\widecheck{K}_{M}(u,v)  = \bar{K}(u+ i M \ad,v) +\bar{K}(u- i M \ad,v).
\ee

\section{Simplified TBA equations}
\label{App:simpTBA}

Acting with $(K+1)^{-1}$ defined in Eqn. \eqref{eq:K+1invdef} on the canonical TBA equations \eqref{TBAeqns} gives the \emph{simplified TBA equations}:
\begin{equation}
\begin{aligned}
\label{eq:simpTBA}
\log Y_1 = \, & \sum_\alpha L_{-}^{\alpha} \, \hat{\star}\, \s - L_2 \star \s -\ewD\,\check{\star}\, \s\, , \\
\log Y_Q = \, &  -(L_{Q-1} + L_{Q+1})\star \s +  \sum_\alpha L_{Q-1|vw}^{(\alpha)} \star \s, \qquad \qquad Q>1,\\
\log Y_{+}^{(\alpha)}/Y_{-}^{(\alpha)} = & \,\Lambda_Q \star K_{Qy}\,,\\
\log Y_{-}^{(\alpha)}Y_{+}^{(\alpha)}  = & \,\Lambda_Q \star (2K_{xv}^{Q1}\star \s - K_{Q})+
 2(\Lambda_{1|vw}^{(\alpha)}-\Lambda_{1|w}^{(\alpha)}) \star \s\,,\\
\log{Y_{M|vw}^{(\alpha)}} = & \,(\Lambda_{M+1|vw} + \Lambda_{M-1|vw}) \star  \s - \Lambda_{M+1} \star  \s + \delta_{M,1}(\Lambda_{-}^{(\alpha)}-\Lambda_{+}^{(\alpha)})  \, \hat{\star} \, \s\, ,\\
\log{Y_{M|w}^{(\alpha)}} = & \, (\Lambda_{M+1|w} + \Lambda_{M-1|w})\star  \s + \delta_{M,1} (L_{-}^{(\alpha)}-L_{+}^{(\alpha)}) \, \hat{\star} \, \s\, ,
\end{aligned}
\end{equation}
where $Y_{0|(v)w}=0$ and $\ewD$ is defined as
\begin{align}
\ewD = L \widecheck{\mathcal{E}} + \sum_{\alpha} \left( L^{(\alpha)}_{-} +  L^{(\alpha)}_{+} \right) \star \widecheck{K}  + 2 \Lambda_Q\star \widecheck{K}_Q^{\Sigma} + \sum_{\alpha} L_{M|vw}^{(\alpha)}\star \widecheck{K}_M.
\end{align}
This object plays a central role in the analysis of the $Y_Q$ TBA equations, see Appendix \ref{App:discYQ}.

\subsection{Contribution of the dressing phase}
\label{App:SimplifyingDressing}

The dressing phase largely drops out of the simplified TBA equations obtained by acting with $(K+1)^{-1}$. Stripping off an extra $\s$ by acting with $s^{-1}$ defined in eqn. \eqref{eq:sinvdef}, we are interested in
\begin{equation}
\check{K}^\Sigma_Q(u,v)  = \lim_{\epsilon\rightarrow 0^+}\left(K^{\Sigma}_{Q1} (u,v+i \ad - i \epsilon) + K^{\Sigma}_{Q1} (u,v-i \ad + i \epsilon)\right) - K^{\Sigma}_{Q2}(u,v),
\end{equation}
which vanishes for $|v|< \theta$ \cite{Arutyunov:2012ai}. Here we will need more detailed properties of this kernel for $|v|>\theta$. We can express this kernel in terms of simpler kernels already appearing in the TBA, similarly to how this was done in the undeformed case \cite{Arutyunov:2009ux}.

Let us work at the S matrix rather than kernel level, with $\check{\Sigma}_Q$ denoting the relevant S matrix. There are no particular subtleties in taking the $\epsilon \rightarrow 0$ limit and direct evaluation gives
\begin{align*}
-i \log \check{\Sigma}_Q = & \, \Phi(y_1^-,x)- \Phi(y_1^-,1/x)- \Phi(y_1^+,x)+ \Phi(y_1^+,1/x)\\
& \, + \frac{1}{2} \left( \Psi(y_1^-,x)- \Psi(y_1^-,1/x)+ \Psi(y_1^+,x)- \Psi(y_1^+,1/x)\right)\\
&\, + \Psi(x,y_1^+)-\Psi(x,y_1^-)\\
& \, - i \log i^Q \tfrac{\, \Gamma_{q^2}\left(\tfrac{Q}{2} - \frac{i}{2\ad}(u-v)\right)}{\Gamma_{q^2}\left(\tfrac{Q}{2}+ \frac{i}{2\ad}(u - v)\right)}\frac{1- \frac{1}{y_1^+ x}}{1-\frac{x}{y_1^-}}\sqrt{\tfrac{y_1^+}{y_1^-} \tfrac{1}{x^2}}\\
& \, + \frac{i}{2} \log q^{Q} e^{-i(Q-2)(u-v)}
\end{align*}
where $x=x(v - i 0^+)$.

Let us now rewrite the contour integrals in the $\Phi$ and $\Psi$ terms as integrals over the interval $[-\theta, \theta]$. The counterclockwise contour $\mathcal{C}$ can be described by $x(u)$ as $u$ runs from $\theta$ to $-\theta$ followed by $1/x(u)$ as it runs from $-\theta$ to $\theta$. This means that for any function $f(z)$ invariant under inverting its argument,
\begin{equation*}
\oint_{\mathcal{C}} \frac{dz}{2 \pi i} \frac{1}{z-y} f(z) = \int_{\hat{Z}_0}\frac{dt}{2\pi i} \frac{d x(t)}{dt} \left(\frac{1}{x(t)-y} + \frac{1}{x(t)^2} \frac{1}{\frac{1}{x(t)}-y} \right)f(x(t)),
\end{equation*}
with $\hat{Z}_0$ defined in Eq. \eqref{eq:Zs}. We recognise this combination of $x$ functions as
\begin{equation*}
\frac{\partial}{\partial t} \log S(s,t), \qquad S(s,t) = \frac{x(s)-x(t)}{x(s)-1/x(t)},
\end{equation*}
such that, with $v=u(y)$,
\begin{equation*}
\oint_{\mathcal{C}} \frac{dz}{2 \pi i} \frac{1}{z-y} f(z) = -\int_{\hat{Z}_0} \frac{dt}{2\pi i} \left(\frac{\partial}{\partial t} \log S(v,t)\right) f(x(t)).
\end{equation*}

With this identity, the first line above results in
\begin{align*}
&\partial_u \Delta \Phi \equiv \partial_u \left(\Phi(y_1^-,x)- \Phi(y_1^-,1/x)- \Phi(y_1^+,x)+ \Phi(y_1^+,1/x)\right)  \\
 & \qquad =  2 i \int_{\hat{Z}_0} \frac{dt_1}{2\pi i} \int_{\hat{Z}_0} \frac{dt_2}{2\pi i} \frac{\partial}{\partial u} \frac{\partial}{\partial t_1} \log S_{Qy}(u,t_1) \frac{\partial}{ \partial t_2} \log \bar{S}(t_2,v) \log S^{[2]}_{q\Gamma}(t_1,t_2),
\end{align*}

where\footnote{Note that we use the label $q\Gamma$ for readability, though on the right-hand side it is really $\Gamma_{q^2}$ which appears.}
\begin{equation*}
\log S^{[Q]}_{q\Gamma}(t_1,t_2) = \log   \frac{\Gamma_{q^2}(\frac{Q}{2}-\frac{i}{2\ad}(t_1-t_2))}{\Gamma_{q^2}(\frac{Q}{2}+\frac{i}{2\ad}(t_1-t_2))},
\end{equation*}
and we recall that $S_{Qy}(u,v) = S(u+ i \ad Q,v)/S(u - i \ad Q,v)$ which naturally comes out of the $y_1^\pm$ terms,\footnote{Our sign conventions for kernels differ from \cite{Arutyunov:2009ux} at this and other points.} and note that
\begin{equation*}
\frac{\partial}{\partial t} \log \frac{x(s)-x(t)}{x(s)-1/x(t)}\frac{1/x(s)-1/x(t)}{1/x(s)-x(t)} = 2\frac{\partial}{\partial t} \log \frac{x(t) - x(s)}{x(t) - 1/x(s)} = - 2 \frac{\partial}{ \partial t} \log \bar{S}(t,s), \qquad \mbox{Im}(s)<0
\end{equation*}
which naturally comes out of the $x$ terms. Integrating by parts in the $t_1$ integral, noting that $K_{Qy}$ vanishes at $\pm \theta$, and taking out factors of $2 \pi i$ to define kernels, we end up with
\begin{equation*}
\frac{1}{2\pi}\frac{\partial}{\partial u} \Delta \Phi =  2 K_{Qy} \,\hat{\star}\, K_{q\Gamma}^{[2]} \,\hat{\star}\, \bar{K},
\end{equation*}
where $K_{q\Gamma}^{[Q]}(u) = \tfrac{1}{2\pi i} \tfrac{d}{du} \log  S_{q\Gamma}^{[Q]}(u)$. For the first line of $\Psi$ terms we similarly find
\begin{align*}
& \Delta \Psi \equiv \frac{1}{2}\left(\Psi(y_1^-,x)- \Psi(y_1^-,1/x)+ \Psi(y_1^+,x)- \Psi(y_1^+,1/x) \right) \\
& \qquad = - i \int \frac{dt}{2\pi i}  \frac{\partial}{ \partial t} \log \bar{S}(t,v)  \left(\log S^{[2]}_{q\Gamma}(u-i \ad Q,t)+\log S^{[2]}_{q\Gamma}(u+i \ad Q,t)\right).
\end{align*}
We now notice that
\begin{align*}
S^{[2]}_{q\Gamma}(u-i \ad Q,t) S^{[2]}_{q\Gamma}(u+i \ad Q,t)  = &\, \frac{\Gamma_{q^2}(1+\frac{Q}{2}-\frac{i}{2\ad}(u-t))}{\Gamma_{q^2}(1+\frac{Q}{2}+\frac{i}{2\ad}(u-t))}\frac{\Gamma_{q^2}(1-\frac{Q}{2}-\frac{i}{2\ad}(u-t))}{\Gamma_{q^2}(1-\frac{Q}{2}+\frac{i}{2\ad}(u-t))}\\
= &\, (-1)^{Q-1} e^{i(Q-1)(u-t)} S^{[Q+2]}_{q\Gamma}(u,t) S^{[Q]}_{q\Gamma}(u,t),
\end{align*}
where in the second equality we used the defining property of the $q$-gamma function $Q-1$ times in the second term, the resulting product of ratios of trigonometric functions cancelling pairwise up to a phase. By the same property
\begin{equation*}
S^{[Q+2]}_{q\Gamma}(u,t) =S^{[Q]}_{q\Gamma}(u,t)/(e^{i(u-t)}S_Q(u-t))
\end{equation*}
We hence find
\begin{equation*}
\frac{1}{2\pi} \frac{\partial}{\partial u}  \Delta \Psi = (2 K_{q\Gamma}^{[Q]}-K_Q)\, \hat{\star}\,\bar{K} + \frac{Q-2}{2\pi} \star \bar{K}.
\end{equation*}
The next line gives, upon integration by parts again
\begin{equation*}
\frac{1}{2\pi} \frac{\partial}{\partial u} \left(\Psi(x,y_1^+) - \Psi(x,y_1^-)\right) = K_{Qy} \,\hat{\star}\, K_{\Gamma_{q^2}}^{[2]}.
\end{equation*}
For the second to last line, analogously to the undeformed case we directly find
\begin{equation*}
\frac{1}{2\pi i} \frac{\partial}{\partial u} \log i^Q \tfrac{\, \Gamma_{q^2}\left(\tfrac{Q}{2} - \frac{i}{2\ad}(u-v)\right)}{\Gamma_{q^2}\left(\tfrac{Q}{2}+ \frac{i}{2\ad}(u - v)\right)}\frac{y_1^+- \frac{1}{x}}{y_1^- - x}\sqrt{\tfrac{y_1^-}{y_1^+}x^2} = K_{q\Gamma}^{[Q]}(u-v) - \frac{1}{2} \bar{K}_Q(u,v) - \frac{1}{2} K_Q(u-v).
\end{equation*}
Finally, the last line precisely cancels the constant term in $\frac{1}{2\pi} \frac{\partial}{\partial u}  \Delta \Psi$ above, as
\begin{equation*}
1 \star \bar{K} = -\frac{1}{2}.
\end{equation*}
Putting everything together, we find
\begin{equation*}
\check{K}_Q^\Sigma =   2 K_{Qy} \,\hat{\star}\, K_\Gamma^{[2]} \,\hat{\star}\, \bar{K} + (2 K_{\Gamma_{q^2}}^{[Q]}-K_Q)\,\hat{\star}\, \bar{K} + K_{Qy} \,\hat{\star}\, K_{q\Gamma}^{[2]}+ K_{q\Gamma}^{[Q]} - \frac{1}{2} \bar{K}_Q - \frac{1}{2} K_Q
\end{equation*}

To simplify this, we note that the $\bar{K}$ and $K_Q$ kernel satisfy\footnote{To see this, extend the integration to run from $-\pi$ to $\pi$ a little above and a little below the real axis -- hence the factors of $1/2$ -- and then note that $K_Q(u-v)$ has poles at $u=v \pm Q i \ad$, and $\bar{K}(u,v)$ has ones at $u=v\pm i \epsilon$.}
\begin{equation*}
K_Q \,\hat{\star}\, \bar{K} = \frac{1}{2} \bar{K}_Q - \frac{1}{2} K_Q ,\qquad 1 \star \bar{K} = -\frac{1}{2}.
\end{equation*}
Similarly
\begin{equation}
\label{eq:Kqgammaidentity}
K^{[Q]}_{q\Gamma}\,\hat{\star}\, \bar{K}  = - \frac{1}{2} K^{[Q]}_{q\Gamma}+  \frac{1}{2}  \sum_{N=0}^\infty\bar{K}_{Q+2N},
\end{equation}
which will come back in Appendix \ref{App:discYQ} as well. Note that this expression is manifestly compatible with the relation $K^{[Q+2]}_{q\Gamma}  = K^{[Q]}_{q\Gamma} - K_Q -1$ encountered above. Using these identities we can finally simplify $\check{K}_Q^\Sigma$ to
\begin{align*}
\check{K}_Q^\Sigma & \, = \sum_{N=1}^{\infty} \left(K_{Qy}\,\hat{\star}\, \bar{K}_{2N} + \bar{K}_{Q+2N}\right).
\end{align*}

\paragraph{Contribution in the TBA.} The contribution of the improved mirror dressing factor to $\Delta$ appearing in the TBA equations is given by
\begin{equation}
\Delta^\Sigma(u) = 2 \sum_Q L_Q \star \check{K}^\Sigma_Q (u)
\end{equation}
This function has a short cut on the real line. We would like to continue it from the upper half-plane to have long cuts. Using the simplified expression for $\check{K}_Q^\Sigma$ discussed in the previous section, we can simplify the result by following Appendix C of \cite{Cavaglia:2010nm} with appropriately adapted integration contours, to arrive at
\begin{equation}
\begin{aligned}
\label{eq:simplifieddressing}
\Delta^\Sigma(u^*) & \, = \sum_\alpha \log Y^{(\alpha)}_- \star_{\gamma_x} \sum_{N=1}^{\infty} \bar{K}_{2N}(u^*) \\
& \, = \sum_\alpha \oint_{\gamma_x} \log Y^{(\alpha)}_-(z) \left(\sum_{N=1}^{\infty} K(z + 2 i N \ad, u) + K(z - 2 i N \ad, u)\right).
\end{aligned}
\end{equation}
This result is used in Appendix \ref{App:discYQ}.

\section{Simplifying the $\log Y_-/Y_+$ discontinuity relations}
\label{App:discYmin}
We show that, given the $Y$ system equation \eqref{Ysys:y}, we can equivalently impose two types of discontinuity equations on the $Y$ system such that the resulting analytic $Y$ system is equivalent to the TBA equations:\footnote{See \cite{Cavaglia:2010nm} for this derivation in the undeformed case.} either we impose that\footnote{As in the main text we will omit the index $\alpha$ since the derivation is identical for both values of the index.}
\be
\label{discnonlocal}
[\log Y_-]_0 = -\Lambda_P \star K_{Qy},
\ee
or we impose that
\be
\label{disclocal}
\left[\log \frac{Y_-}{Y_+} \right]_{\pm 2N}(u) = -\sum_{P=1}^{N} \left[\Lambda_P\right]_{\pm(2N-P)}(u) \text{ for } N\geq 1.
\ee
It is not difficult to see that the second set of discontinuities can be derived from the first one, noting that
\be
[\log Y_-]_0 = \log \frac{Y_-}{Y_+}.
\ee
We will show here that one can derive the first discontinuity equation from the second set. This allows us to use the local (and hence easier) discontinuity equations \eqref{disclocal}.

First, we need to define a function $G$ such that
\be
K(z,u) = \frac{1}{2\pi i} \frac{1}{2 \sin \frac{1}{2}(z-u)} \frac{G(z)}{G(u)}.
\ee
\label{GB}
This function exists uniquely up to normalisation as
\be
G(u) = \frac{1}{4\pi i} \frac{1}{\sin(1-u)K(1,u)},
\ee
where the choice to evaluate in $z=1$ is arbitrary.
Now consider the function
\be
T(z,u) = -G(u)2\pi iK(z,u)= -\frac{1}{2 \sin \frac{1}{2}(z-u)} G(z)
\ee
and
\bea
\label{Tproof}
& &-\sum_{Q=1}^{\infty} \int \frac{dz}{2\pi i} \Lambda_Q(z)\left(T(z-iQ \ad,u) -T(z+iQ\ad,u) \right) \nonumber \\
&=& -\sum_{Q=1}^{\infty} \int dz G(u) \Lambda_Q(z)\left(K(z+iQ\ad,u) -K(z-iQ\ad,u) \right) \nonumber \\
&=& -G(u)\sum_{Q=1}^{\infty} \int dz \Lambda_Q(z) K_{Qy}(z,u) = -G(u)\left(\Lambda_Q\star K_Qy(u)\right),
\eea
where the integrals all run over the interval $(-\pi,\pi]$, as in the rest of this appendix whenever the range is not specified.
This shows that the equation
\be
\label{Grel}
G(u)\log \frac{Y_-}{Y_+}(u) = -\sum_{Q=1}^{\infty} \int \frac{dz}{2\pi i} \Lambda_Q(z)\left(T(z-iQ\ad,u) -T(z+iQ\ad,u) \right)
\ee
is equivalent to the equation
\be
\log \frac{Y_-}{Y_+}(u) =  -\Lambda_P \star K_{Py}.
\ee
We will now derive the equation \eqref{Grel} from the new $2N$ discontinuities:

Let $\gamma$ be as in Fig \ref{fig:gammas}, then we can write
\be
G(u) \log \frac{Y_-}{Y_+}(u) = \oint_{\gamma} \frac{dz}{2\pi i} \frac{1}{2} \frac{\log \frac{Y_-}{Y_+}(z)}{\sin \frac{1}{2}(z-u)}G(z),
\ee
since the residue of $\frac{1}{\sin \frac{1}{2}(z-u)}$ is one and $\log \frac{Y_-}{Y_+}(z)G(z)$ is analytic on the inside of $\gamma$. We can deform $\gamma$ into $\Gamma$ (see Fig.\ref{fig:gammas}), which runs on both sides of the discontinuity lines at $\pm 2N i \ad $ all the way to infinity, using the $2\pi$-periodicity of the integrand to let the appearing vertical parts of the deformed contour vanish.

Define for simplicity the $2\pi$-periodic function
\be
B(z,u) =\frac{1}{2} \frac{G(z)}{\sin\frac{1}{2}(z-u)}
\ee
and notice that
\bea
&-&\sum_{N,\tau}\left(\int_{Z_{-2N\tau}+i\e}- \int_{Z_{-2N} -i\e} \right)\frac{dz}{2\pi i} \sum_{P=1}^{N}   \Lambda_P (z+i \tau P \ad)B(z,u) \nonumber \\
&=&-\sum_{N,\tau} \int\frac{dz}{2\pi i} \sum_{P=1}^{N} \left[\Lambda_P\right]_{-\tau (2N-P)}(z)  B(z-2i\tau N \ad,u) \nn
&=& \sum_{N,\tau}\int\frac{dz}{2\pi i} \left[ \log Y_-/Y_+ \right]_{-\tau 2 N}(z) B(z-2i\tau N \ad,u)\nonumber \\
&=& \sum_{N,\tau}\int\left(  \log Y_-/Y_+(z-i2N\tau\ad+i\e)-\log Y_-/Y_+(z-i2N\tau \ad-i\e) \right)B(z-2i\tau N \ad,u) \nonumber \\
&=&\oint_{\Gamma}\frac{dz}{2\pi i} \frac{1}{2}\frac{\log Y_-/Y_+}{\sin\frac{1}{2}(z-u)}G(z) =\oint_{\gamma}\frac{dz}{2\pi i} \frac{1}{2}\frac{\log Y_-/Y_+}{\sin\frac{1}{2}(z-u)}G(z),
\eea
where $\tau$ sums over $\pm 1$. This shows that
\bea
\label{discYy1}
\oint_{\gamma}\frac{dz}{2\pi i} \frac{1}{2}\frac{\log Y_-/Y_+}{\sin\frac{1}{2}(z-u)}G(z) = -\sum_{N,\tau}\left(\int_{Z_{-2N}+i\e}- \int_{Z_{-2N} -i\e} \right)\frac{dz}{2\pi i} \sum_{P=1}^{N}   \Lambda_P (z+i \tau P \ad)B(z,u).
\eea
Using the fact that $\Lambda_P$ and $B$ do not have any poles except at the branch points allows us to cancel integrals on the right-hand side by deforming the relevant contours:
\bea
&\phantom{=}& -\sum_{\tau} \sum_{N} \sum_{P=1}^{N} \left(\int_{Z_{-2N}+i\e}- \int_{Z_{-2N} -i\e} \right)\frac{dz}{2\pi i}  \Lambda_P (z+i \tau P \ad)B(z,u)\nonumber \\
&=& -\sum_{\tau} \sum_{P=1}^{\infty} \sum_{N=P}^{\infty}\left(\int_{Z_{-2N}+i\e}- \int_{Z_{-2N} -i\e} \right)\frac{dz}{2\pi i}  \Lambda_P (z+i \tau P \ad)B(z,u)\nonumber \\
&=& \sum_{\tau} \tau \sum_{P=1}^{\infty} \int_{Z_{-2N}+\tau i\e)} \frac{dz}{2\pi i}  \Lambda_P (z+i \tau P \ad)B(z,u) \nn
&=& \sum_{\tau} \tau \sum_{P=1}^{\infty} \int \frac{dz}{2\pi i}  \Lambda_P (z)\frac{1}{2} \frac{G(z-i \tau P\ad)}{\sin\frac{1}{2}(z-i \tau P\ad-u)}.
\eea
Now we see, using that
\be
T(z,u) =-\frac{1}{2 \sin \frac{1}{2}(z-u)} G(z)
\ee
that we can rewrite \eqref{discYy1} as follows:
\bea
\oint_{\gamma}\frac{dz}{2\pi i} \frac{1}{2}\frac{\log Y_-/Y_+}{\sin\frac{1}{2}(z-u)}G(z) &=&
-\sum_{\tau} \tau \sum_{P=1}^{\infty} \int \frac{dz}{2\pi i}  \Lambda_P (z-i \tau P a -i\e)T(z-i \tau Pa,u) \nonumber \\
&=&- \sum_{P=1}^{\infty} \int \frac{dz}{2\pi i} \Lambda_P (z) \left(  T(z-i Pa,u)-T(z+i Pa,u) \right),
\eea
which, as we have shown in \eqref{Tproof}, is equivalent to the statement
\be
\log \frac{Y_-}{Y_+}(u) =  -\Lambda_P \star K_{Py}.
\ee
In conclusion, this proves that the local discontinuities \eqref{disclocal} for $Y_-$ contain exactly the same amount of information as the non-local version \eqref{discnonlocal} and we can impose the local discontinuity relations on the $Y$ system.
\section{The $Y_Q$ discontinuity relation and reobtaining the $Y_Q$ TBA equation}
\label{App:discYQ}
By far the most complicated part of the transition from the TBA equations to the analytic $Y$ system is treating the $Y_Q$ functions. Most of this complication sits in the treatment of the dressing-phase kernel, but the general strategy of deriving the relevant analyticity conditions and subsequently proving the equivalence with the TBA equations is more convoluted than in the other three cases (for $Y_{\pm}$, $Y_{M|(v)w}$) as well.

The derivation consists of the following steps:
\begin{itemize}
\item Analyse and properly define the object $\widecheck{\Delta}$\footnote{Note that $\widecheck{\Delta}$ is a short-cutted function despite the presence of the check, which we keep to stay in line with the literature. In the rest of this paper a check is reserved for long-cutted functions, whereas we use hats for short-cutted functions.}
\item Define the long-cutted version of this object $\Delta$
\item Find a set of discontinuities to complement the $Y$ system
\item Prove that these discontinuities allow us to restore $\Delta$ completely
\item Use the $Y$ system to find expressions for the discontinuities $\left[ \log Y_Q\right]_{\pm(Q+2M)}$
\item Use the Cauchy integral theorem to restore the TBA equations from these discontinuities and $\Delta$.
\end{itemize}
The first step is straightforward: starting from the simplified TBA equations \eqref{eq:simpTBA} we derived the relevant $Y$ system equations \eqref{Ysys:Q}. Following the same prescription as in \cite{Arutyunov:2009ax} we find for $|$Re$(u)|>\theta$
\begin{align}
\left[ \log Y_1 \right]_{1} (u) =  \sum_{\alpha}L_-^{(\alpha)}(u+i\e)+\ewD(u) =: \cD.
\end{align}
The full discontinuity is the continuation of this function to the entire complex plane, where the term containing $L_-$ deserves special attention. We can write
\be
\label{deftD}
\cD \defeq  L \widecheck{\mathcal{E}}+ \sum_{\alpha}\widecheck{L}_y^{(\alpha)}  + \sum_{\alpha} \left( L^{(\alpha)}_{-} +  L^{(\alpha)}_{+} \right) \star \widecheck{K}  + 2 \Lambda_Q\star \widecheck{K}_Q^{\Sigma} + \sum_{\alpha} L_{M|vw}^{(\alpha)}\star \widecheck{K}_M,
\ee
where here the check indicates the short-cutted version of the respective kernels. In particular $\widecheck{L}_y^{(\alpha)}$ is the double-valued continuation of $L_-$ in the \uhp of the first sheet with a short cut. On the first sheet it can be characterised as
\be
\widecheck{L}_y^{(\alpha)}(u) =
\left\{
	\begin{array}{ll}
		L_-^{(\alpha)}(u)  & \mbox{if Im}(u)>0  \\
		L_+^{(\alpha)}(u)  & \mbox{if Im}(u)<0
	\end{array}
\right..
\ee
Also
\be
\widecheck{\mathcal{E}}(u) = \log \frac{x(u-i\e)+\xi}{x(u+i\e)+\xi}.
\ee
This defines $\cD$ as a function with short cuts. To change from short back to long cuts we define the following quantity $\Delta$:
\be
\label{delta2}
\Delta = -L \log \frac{x(u)+\xi}{1/x(u)+\xi} + \sum_{\alpha}L_-^{(\alpha)}  - \sum_{\alpha} \left( L^{(\alpha)}_{-} +  L^{(\alpha)}_{+} \right) \star K  - 2 \Lambda_Q\star K_Q^{\Sigma} - \sum_{\alpha} L_{M|vw}^{(\alpha)}\star K_M,
\ee
where we use the definition of $\widecheck{\mathcal{E}}$ to find a long-cutted version of this function. Then the function $\cD = [\log Y_1]_1$ with short cuts agrees with $\Delta$ on the strip $0<\text{Im}(u)<i\ad$ and can be continued to the rest of complex plane choosing short branch cuts. $\Delta$ on the other hand is defined using continuation choosing long branch cuts.
\subsection{Deriving discontinuities for $\cD$}
For this we will first derive the discontinuities of $\cD$. For most terms appearing in the definition of $\cD$ this uses the usual arguments, analysing the convolutions using the known properties of the appearing kernels. We find the discontinuities
\be
\left[ \cD \right]_{\pm2N} (u)= - \sum_{\alpha} \left( \mp \left[L^{(\alpha)}_{\mp}\right]_{\pm 2N} + \sum_{M=1}^N \left[   L_{M|vw}^{(\alpha)}\right]_{\pm(2N-M)} + \left[ \log Y_-^{(\alpha)}\right]_{0} \right).
\ee
The derivation of the term containing $L_{M|vw}$ functions follows the analysis illustrated in Section \ref{sec:derivingY1wdisc}. The $Y_{\pm}$-contribution is slightly more complicated:
there are two terms, the first one being the term not belonging to $\ewD$, which has as discontinuities
\be
\left[\sum_{\alpha}L_{\mp}^{(\alpha)}(u)\right]_{\pm2N},
\ee
using the representation of $L_y$ on the upper and lower half-plane. We can however combine them with the contribution coming from the term $\left(L^{(\alpha)}_{-} +L^{(\alpha)}_{+} \right) \star \widecheck{K}$. It is straightforward to find that its discontinuities are (only for positive $N$)
\be
-\left[L^{(\alpha)}_{-} +L^{(\alpha)}_{+}\right]_{-2N},
\ee
whereas the contributions for negative $N$ vanish. Adding both contributions together we see that the total contributions from $L_-$ sum up to be
\be
\pm \left[L^{(\alpha)}_{-}\right]_{\pm 2N},
\ee
with the usual implicit sum over $\alpha$. Finally the discontinuity of the dressing phase kernel follows directly from the simplification \eqref{eq:simplifieddressing} discussed in Appendix \ref{App:SimplifyingDressing} using the usual arguments. Changing now from short back to long cuts leads to the following discontinuity for $\Delta$:
\be
\left[\Delta\right]_{\pm 2N} (u) = \pm \sum_{\alpha} \left( \left[L^{(\alpha)}_{\mp}\right]_{\pm 2N} + \sum_{M=1}^N \left[   L_{M|vw}^{(\alpha)}\right]_{\pm(2N-M)} + \left[ \log Y_-^{(\alpha)}\right]_{0} \right),
\ee
which are in fact the same relations as were found for the undeformed case. We note that $\Delta$ in fact has one additional branch cut: analysing the energy term in its definition we find that $\Delta$, like in the undeformed case, has a logarithmic branch cut on the imaginary axis:
\be
\label{eq:logarithmicjump}
\Delta(iv +\e) - \Delta(iv -\e) = 2\pi i L,
\ee
using the principal branch description of the logarithm.
\subsection{Rederiving the TBA equation from the discontinuities}
Using the discontinuity relations for $\Delta$ we can start to rederive the TBA equation for the $Y_Q$-particles from the $Y$ system.To pursue the Cauchy integral method illustrated in the main text, we use the fact that $\Delta$ has square-root branch points at $\pm \theta$. Of course, we also assume that the relevant functions have no other poles than those at the branch points. A new aspect compared to the undeformed case is that we also have to deal with non-zero asymptotic behaviour of the $Y_Q$: to correctly rederive the TBA equation for $Y_Q$ we should consider a $Y$ system solution which asymptotically behaves as $\mp \ad QL$ as $u\rightarrow \pm i \infty$. This is consistent with the TBA equations \eqref{TBAeqns}, as the driving term $\mathcal{E}_Q$ behaves
as $\mp \ad QL$ as $u\rightarrow \pm i \infty$, whereas all the other terms on the right-hand side of the $Y_Q$ TBA equation vanish in those limits. In fact, analysing the other equations we find that only the $Y_Q$ are nonvanishing in those limits, since all kernels do vanish.
\subsubsection{Retrieving $\Delta$}
\begin{figure}[!t]
\centering
\begin{subfigure}{6cm}
\includegraphics[width=6cm]{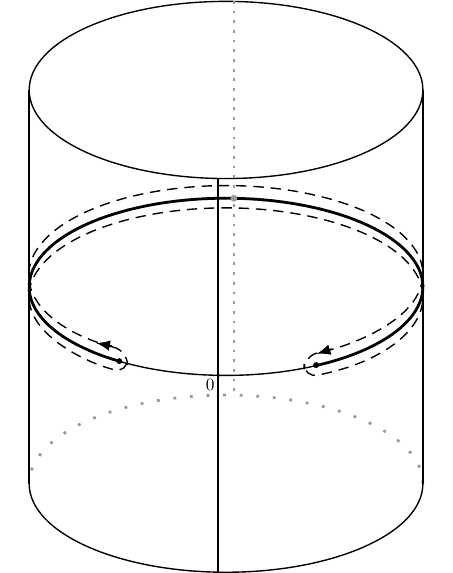}
\caption{}
\end{subfigure}
\qquad \qquad
\begin{subfigure}{6cm}
\includegraphics[width=6cm]{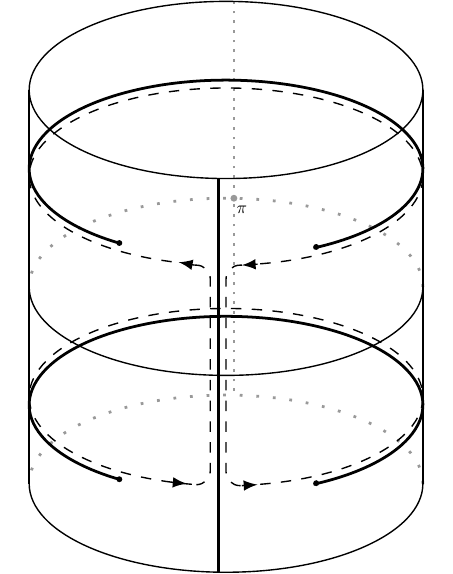}
\caption{}
\end{subfigure}
\caption{The contours $\gamma_x$ in (a) and $C_1$ in (b).}
\label{fig:gammaxC1}
\end{figure}

First we will find $\Delta$ from the local discontinuity relations given above. Using Cauchy we can write for $0<$Im$(u)<i \ad$
\begin{align}
G(u) \Delta(u) = \oint_{C_1} \frac{dz}{2\pi i}  \frac{\Delta(u)}{2\sin\frac{1}{2} (z-u)}G(z),
\end{align}
where the contour $C_1$ is depicted in Fig. \ref{fig:gammaxC1}. Deforming the contour now towards imaginary infinity to give it a $\Gamma$-like shape we reach the contour that encircles all horizontal branch cuts separately in the clock-wise direction (we will call this part $C_2$) and runs along the vertical branch cut on $i \R$ towards $+i\infty$ on the left and $-i\infty$ on the right. This shows that, using the definitions of $G$ and $B$ defined in Appendix \ref{GB}, we can write our original integral expression as
\begin{align}
\label{der1}
G(u) \Delta(u) &= \oint_{C_2} \frac{dz}{2\pi i}  \frac{\Delta(z)}{2 \sin\frac{1}{2} (z-u)}G(z) + \left(\int_{i\R -\e}-\int_{i\R +\e}\right) \frac{dz}{2\pi i}  \frac{\Delta(z)}{2\sin\frac{1}{2} (z-u)}G(z) \nn
&= I_{C_2} + L G(u)\log \frac{x(u)+\xi}{1/x(u)+\xi},
\end{align}
where we use the logarithmic jump of $\Delta$ \eqref{eq:logarithmicjump} and used the relation between $G$ and $K(z,u)$, as well as the definition of the kernel $K(z,u)$, to find the energy term hidden in $\Delta$.
Now we can work on massaging the integral $I_{C_2}$, defined as
\begin{align}
I_{C_2} &= \oint_{C_2} \frac{dz}{2\pi i}  \frac{\Delta(z)}{2 \sin\frac{1}{2} (z-u)}G(z) \nn
&= \sum_{N=1}^{\infty} \sum_{\tau} \oint_{\gamma_x} \frac{dz}{2\pi i}  \Delta(z+\tau i 2N \ad)B(z+\tau i 2N \ad,u),
\end{align}
with $\gamma_x$ as in \cite{Cavaglia:2010nm} being the contour around the line section connecting $\pm \theta$ through $\pi$ in the clockwise direction (see Fig. \ref{fig:gammaxC1}).
The summation variable $\tau$ in this appendix will always run over $\{\pm 1\}$. Now we can recognise the discontinuity of $\Delta$ and find
\begin{align}
I_{C_2} &= \sum_{N=1}^{\infty} \sum_{\tau} \oint_{\gamma_x} \frac{dz}{2\pi i}  \Delta(z+\tau i 2N \ad)B(z+\tau i 2N \ad,u) \nn
&= \sum_{N=1}^{\infty} \sum_{\tau}\int_{\hat{Z}_{0}}\frac{dz}{2\pi i}  \left[ \Delta(z)\right]_{\tau 2N} B(z+\tau i 2N \ad,u)\nn
&= \sum_{N=1}^{\infty} \sum_{\tau,\alpha}\tau G(u)\oint_{\gamma_x}dz
\left( \factornumbering{L^{(\alpha)}_{-\tau}(z+i\tau 2N\ad)}{$1$}
 + \factornumbering{\sum_{M=1}^N L_{M|vw}^{(\alpha)}(z+\tau(2N-M)i\ad)}{$2$}
 + \factornumbering{\log Y_-^{(\alpha)}(z)}{$3$} \right)\cdot\nn & \phantom{=} K(z+\tau i 2N \ad,u).
\end{align}
We have already extracted $G$ from the expression, but to continue we will split the contributions of the numbered factors. The first term can be treated as follows: we deform $\gamma_x$ using the analyticity of the integrand into two circles around the entire cylinder. Now we can rewrite the first term using the analyticity of $L_{\pm}$ in the $\substack{\mbox{\small lower}\\ \mbox{\small
upper}}$ half-plane and using that $K(z_*,u) =-K(z,u)$:
\begin{align}
& G(u) \sum_{N=1}^{\infty} \sum_{\tau,\alpha}\tau\oint_{\gamma_x}dz L^{(\alpha)}_{-\tau}(z+i\tau 2N\ad)K(z+\tau i 2N \ad,u)\nn
&= -G(u)\sum_{\alpha} \left(\int_{Z_{-2}+i\e}dz L^{(\alpha)}_{+}K(z,u)+\int_{Z_{2} -i\e}dz L^{(\alpha)}_{-}(z)K(z,u)\right),\nn
&= G(u)\sum_{\alpha} \left( L^{(\alpha)}_{-} -  \left(L_-^{(\alpha)}+L_+^{(\alpha)}\right)\star K (u) \right) ,
\end{align}
exactly giving us the relevant terms in the definition of $\Delta$ in \ref{delta2}.

The second term follows the standard procedure:
\begin{align}
& G(u) \sum_{N=1}^{\infty} \sum_{\tau,\alpha}\tau\oint_{\gamma_x}dz \sum_{M=1}^N L_{M|vw}^{(\alpha)}(z+\tau(2N-M)i\ad) K(z+\tau i 2N \ad,u) \nn
&= G(u) \sum_{M=1}^{\infty}\sum_{N=M}^{\infty} \sum_{\tau,\alpha}\tau\left(\int_{Z_{2N\tau}+i\e}-\int_{Z_{2N\tau} -i\e}\right)dz  L_{M|vw}^{(\alpha)}(z-\tau M i\ad) K(z,u)\nn
&=- G(u) \sum_{M=1}^{\infty} \sum_{\alpha}\int_{Z_{0}} dz L_{M|vw}^{(\alpha)}(z) \left( K_M(z,u)\right)  = -G(u) \sum_{M=1}^{\infty} \sum_{\alpha} L_{M|vw}^{(\alpha)}\star  K_M,
\end{align}
which also exactly matches the definition given in \ref{delta2}.
The third term can immediately be seen to give
\be
 G(u) \sum_{\alpha}\oint_{\gamma_x}dz \log Y_-^{(\alpha)}(z)\sum_{N=1}^{\infty} \left( K(z+ i 2N \ad,u)-K(z- i 2N \ad,u)\right),
\ee
which matches the expression \eqref{eq:simplifieddressing} for the dressing phase factor. So we see that we can indeed reconstruct $\Delta$ from its discontinuities.
\subsection{Reconstructing the $Y_Q$ TBA equation}
To rederive the TBA equation for $Y_Q$-particles we will follow the general strategy we developed in the previous sections. First we write $\log Y_Q$ as a Cauchy integral for $-i\ad Q<$Im$(u)<i\ad Q$ knowing it is regular in the strip between the lines at $\pm i\ad Q$:
\begin{align}
\log Y_Q (u) = \oint_{\gamma} \frac{dz}{2\pi i} \log Y_{Q}(z) H(z-u),
\end{align}
where we define $H(z) = \frac{1}{2} \cot\frac{u}{2}$ as before and $\gamma$ as in Fig. \ref{fig:gammas}. If we want to follow the strategy explained in the main text the next step is to deform $\gamma$ into $\Gamma$, which involves taking a limit. The result of this limit depends on the behaviour of $\log Y_Q$ as $u\rightarrow i \infty$. From the driving term in the TBA equations \eqref{TBAeqns} we see that we are interested in a solution of the $Y$ system that obeys
\be
\log Y_Q \rightarrow \mp \ad QL \text{ as } u \rightarrow \pm i \infty.
\ee
Writing $\gamma^{\pm}$ for the parts of $\gamma$ in the upper and lower half-plane respectively we now find
\begin{align}
\log Y_Q &= \oint_{\gamma} \frac{dz}{2\pi i} \log Y_{Q}(z) H(z-u) \nn
&= \int_{\gamma^+} \frac{dz}{2\pi i} \left( \log Y_{Q}(z) +\ad QL - \ad Q L \right)H(z-u)  + \int_{\gamma^-} \frac{dz}{2\pi i} \left( \log Y_{Q}(z) +\ad QL - \ad Q L \right)H(z-u) \nn
&= \ad Q L \oint_{z=u}\frac{dz}{2\pi i} H(z-u) +    \int_{\gamma^+} \frac{dz}{2\pi i} \left( \log Y_{Q}(z) +\ad QL  \right)H(z-u)  \nn
&+ \int_{\gamma^-} \frac{dz}{2\pi i} \left( \log Y_{Q}(z) - \ad Q L \right)H(z-u).
\end{align}
The contour integral can be done easily to give $\ad Q L$ and deforming $\gamma^{\pm}$ in the usual way to $\Gamma$, we get
{
\bea
\oint_{\gamma} \frac{dz}{2\pi i} \log Y_{Q}(z) H(z-u) &=& \ad Q L + \sum_{\tau}\sum_{l=0}^{\infty} \int_{Z_0} \left[ \log Y_{Q}(z) \right]_{\tau(Q+2l)} H(z-u).
\eea
}
So we will need discontinuities of the form
$$
\left[ \log Y_{Q} \right]_{\pm (Q+2l)},
$$
where $Q,l\in \mathbb{N}$. We will derive them from the $Y$ system. The derivation of these quantities is completely analogous to the one illustrated in \cite{Cavaglia:2010nm} and we will not treat it here. It uses the knowledge of the analyticity strips and branch cut locations of the $Y$ functions to combine discontinuities of the $Y$ system. The result is given in terms of $D$ functions, which are defined for $l\geq 0$ as
\begin{align}
D_{\tau (Q+2l)}^Q(u) &= \sum_{J=1}^{l+1}\sum_{M=J}^{Q+J-2} L_{vw|M}^{(\alpha)}(u+\tau(M+2l-2J+2)i\ad)  + L_-^{(\alpha)}(u+2\tau l i \ad) -  \sum_{J=1}^l  \Lambda_{J}(u+\tau(2l-J)i\ad)  \nn
&-\sum_{J=1}^l \left( 2 \sum_{M=1}^{Q-1}  \Lambda_{M+J}(u+\tau(M+2l-J)i\ad) + \Lambda_{Q+J}(u+\tau(Q+2l-J)i\ad)\right)\nn
&-\sum_{M=1}^{Q-1} \Lambda_{Q}(u+\tau(M+2l)i\ad),
\end{align}
and obey the following recursion relation \footnote{This is similar to the corresponding expression in \cite{Cavaglia:2010nm} except for the lower bound on $M$ for the $vw$ terms: it starts at $l+1$ instead of $l$.}:
\begin{align}
\label{recYQD}
D_{\tau (Q+2l)}^Q(u) - D_{\tau (Q+2l-2)}^Q(u +\tau 2 i\ad ) &=\sum_{M=l+1}^{Q+l-1} L_{vw|M}^{(\alpha)}(u+\tau M i\ad)-   \Lambda_{l}(u+\tau li\ad)  \nn
&- 2 \sum_{M=1+l}^{Q-1+l}  \Lambda_{M}(u+\tau Mi\ad) -\Lambda_{Q+l}(u+\tau (Q+l)).
\end{align}
The discontinuity of $Y_Q$ can now be written as follows for all $l\geq0$:
\begin{align}
\left[ \log Y_{Q} \right]_{\tau(Q+2l)}(u) &= \left[D_{\tau (Q+2l)}^Q\right]_0 (u) -\delta_{l,0}\left[\log Y_{1}\right]_{-\tau 1}(u).
\end{align}
Plugging this into our expression for $\log Y_Q$ leaves \begin{align}
&\oint_{\gamma} \frac{dz}{2\pi i} \log Y_{Q}(z) H(z-u)= \ad Q L
\nn &+ \sum_{\tau}\sum_{l=0}^{\infty} \int_{Z_0}\frac{dz}{2\pi i} \left(\left[D_{\tau (Q+2l)}^Q\right]_0 (u) -\delta_{l,0}\left[\log Y_{1}\right]_{-\tau 1}(u) \right) H(z+\tau(Q+2l)\ad i-u).
\end{align}
Let us massage these integrals in steps and consider the terms associated to different $Y$ functions separately. As a first step we simplify the contributions coming from the $D$ functions and $\log Y_1$ separately. The $D$ function contribution can be written as
\begin{align}
\label{eq:Dcontribution}
&\sum_{\tau}\sum_{l=0}^{\infty} \int_{Z_0}\frac{dz}{2\pi i} \left[D_{\tau (Q+2l)}^Q\right]_0 (u) H(z+\tau(Q+2l)\ad i-u) \nn
&=- \sum_{\tau}\sum_{l=1}^{\infty}\tau  \int_{Z_0 -i\tau \e } \left(D_{(Q+2l)\tau}^{Q}(z) -D_{(Q+2l-2)\tau}^{Q}(z+2\tau \ad)\right) H(z+\tau (Q+2l)\ad i-u) \nn
&-\sum_{\tau} \tau \int_{Z_0 -i\tau \e }\left(L_- + \sum_{M=1}^{Q-1} \Lambda_{M}(z+\tau i \ad M )+\sum_{M=2}^{Q} L_{vw|M-1}^{(\alpha)}(z+\tau (M-1) i\ad ) \right) H(z+\tau Q\ad i-u),
\end{align}
whereas the $\log Y_1$ contribution can be rewritten as
\begin{align}
&-\sum_{\tau}\int_{Z_0}\left[\log Y_{1}\right]_{-\tau 1}(u) H(z+\tau Q \ad i-u)\nn
&=-\int_{Z_0}\left[L_-^{(\alpha)} \right]_0H(z+ Q\ad i-u) -\oint_{\gamma_x}\log Y_{1}(u+i \ad) K_Q(z-u),
\end{align}
using the property
\be
\label{eq:Y1property}
\left[\log Y_1\right]_1+\left[\log Y_1\right]_{-1} = \left[L_-^{(\alpha)}\right]_0,
\ee
which follows directly from the $Y$-sytem. First we will treat the $Y_-$-terms, collecting all contributions in both the rewritten expressions:
\begin{align}
&\sum_{\tau}\sum_{l=0}^{\infty} \int_{Z_0}\frac{dz}{2\pi i} \left(\left[L_-^{(\alpha)}\right]_{\tau 2l}(u) -\delta_{l,0}\delta_{\tau,+1}\left[L_-^{(\alpha)} \right]_0(u) \right) H(z+\tau(Q+2l)\ad i-u)\nn
&=-\oint_{\gamma_x}\frac{dz}{2\pi i} L_-^{(\alpha)}(u)H(z+Q\ad i-u) + \sum_{\tau}\tau \int_{Z_0+i\tau \e }\frac{dz}{2\pi i} \L_-^{(\alpha)}(u) H(z-\tau Q\ad i-u)\nn
&= \int_{Z_0+i \e }  \L_-^{(\alpha)}(u) K_Q(z-u)
\end{align}
Now let us treat the remaining terms in \eqref{eq:Dcontribution}, including the bulk terms from the $D$s, applying the recursion relation \eqref{recYQD}:
\begin{align}
\label{derivYQ1}
&- \sum_{\tau}\sum_{l=1}^{\infty}\tau  \int_{Z_0 -i\tau \e } \left(D_{(Q+2l)\tau}^{Q}(z) -D_{(Q+2l-2)\tau}^{Q}(z+2\tau \ad)\right) H(z+\tau (Q+2l)\ad i-u) \nn
&-\sum_{\tau} \tau \int_{Z_0 -i\tau \e } \sum_{M=1}^{Q-1} \Lambda_{M}(z+\tau i \ad M) H(z+\tau Q\ad i-u)\nn
&=
- \sum_{\tau}\sum_{l=1}^{\infty}\tau  \int_{Z_0 -i\tau \e } \Bigg(-   \Lambda_{l}(z+\tau li\ad ) -\Lambda_{Q+l}(z+\tau (Q+l)) \nn
&\left.- 2 \sum_{M=1+l}^{Q-1+l}  \Lambda_{M}(z+\tau Mi\ad)+\sum_{M=l+1}^{Q+l-1} L_{vw|M}^{(\alpha)}(z+\tau M i\ad) \right) H(z+\tau (Q+2l)\ad i-u) \nn
&-\sum_{M=1}^{Q-1} \left(\Lambda_{M}-L_{vw|M}\right) \star K_{Q-M}.
\end{align}
The terms in the previous expression containing $Y_Q$ functions can be simplified to $-\sum_{M=1}^{\infty} \Lambda_M \star K_{M Q}$, whereas the $Y_{vw}$ functions can be simplified to the term
\begin{align}
&- \sum_{\tau}\sum_{l=1}^{\infty}\tau  \int_{Z_0 -i\tau \e } \sum_{M=1+l}^{Q+l-1} L_{vw|M}^{(\alpha)}(z+\tau M ia) H(z+\tau (Q+2l)\ad i-u)+\sum_{M=1}^{Q-1} L_{vw|M}\star K_{Q-M}\nn
&=\sum_{l=1}^{\infty}\sum_{M=1+l}^{Q+l-1} L_{vw|M}^{(\alpha)}(z)\star K_{Q+2l-M} +\sum_{M=1}^{Q-1} L_{vw|M}\star K_{Q-M} \nn
&=\sum_{l=1}^{\infty}\sum_{M=l}^{Q+l-2} L_{vw|M}^{(\alpha)}(z)\star K_{Q+2l-M-2}
=\sum_{M=1}^{\infty}\sum_{l=\max(0,M-Q+1)}^{M} L_{vw|M}^{(\alpha)}(z)\star K_{Q+2l-M} \nn
&=\sum_{M=1}^{\infty}\sum_{l=0}^{M} L_{vw|M}^{(\alpha)}(z)\star K_{Q+2l-M},
\end{align}
using that the terms we introduce in the last line sum up to zero due to the antisymmetry of $K_M$ in $M$. Let us summarise what our results thus far are, by writing the simplest expression we have for the right-hand side of the $Y_Q$ TBA equation.
\begin{align}
\label{prelres}
\log Y_Q(u) &= -\oint_{\gamma_x}\log Y_{1}(u+i\ad) K_Q(z-u)-\sum_{M=1}^{\infty} \Lambda_M \star K_{MQ}(u)+\int_{Z_0+i \e }  \L_-^{(\alpha)}(u) K_Q(z-u)\nn
&+\sum_{M=1}^{\infty} \sum_{l=0}^{M} L_{M|vw}\star K_{Q+2l-M}(u)+ \ad Q L.
\end{align}
The next step is using our knowledge of $\Delta$ to rewrite the first contour integral:
\begin{align}
\label{eq:y1expl}
&\oint_{\gamma_x}\log Y_{1}(u+i\ad) K_Q(z-u) = \int_{\check{Z}_0}\cD(u) K_Q(z-u)
\nn
 &= \frac{1}{2}\oint_{\gamma_x} \left(\Delta-\sum_{\alpha}L_-^{(\alpha)}\right)(u) K_Q(z-u)
+ \int_{\check{Z}_0})\left( \sum_{\alpha}\check{L}_-^{(\alpha)}(u)\right) K_Q(z-u)
\nn &= \frac{1}{2}\oint_{\gamma_x} \left(-L \log \frac{x(u)+\xi}{1/x(u)+\xi} - \sum_{\alpha} \left( L^{(\alpha)}_{-} +  L^{(\alpha)}_{+} \right) \hat{\star} K  - 2 \Lambda_M\star K_M^{\Sigma} - \sum_{\alpha} L_{M|vw}^{(\alpha)}\star K_M\right)K_Q(z-u)
\nn &+ \int_{\check{Z}_0}\left( \sum_{\alpha}\check{L}_-^{(\alpha)}(z)\right) K_Q(z-u)
\end{align}
We can now look at all the separate terms to recognise the TBA-contributions.
\subsubsection{Driving term $\mathcal{E}_Q$}
First we treat the energy contribution: it is retrieved from
\begin{align}
- \frac{L}{2}\oint_{\gamma_x}  \log \frac{x(u)+\xi}{1/x(u)+\xi}K_Q(z-u).
\end{align}
It is interesting to note that because our functions are real-periodic the integration contour $\gamma_x$ is actually closed on the cylinder. Therefore, considering the slightly more general integral
\begin{align}
\oint_{\gamma_x}dz f(z) K_Q(z-u)
\end{align}
we can directly apply Cauchy's theorem to the domain on the outside of this contour, picking up residues at all the poles. $K_M$ only has two poles, so we get for a function $f$ which is analytic and regular on on this domain the result
\begin{align}
\label{fsimple}
\oint_{\gamma_x}dz f(z) K_Q(z-u) =  f(u+i Q \ad) - f(u-i Q \ad)
\end{align}
We cannot directly apply this formula to the energy term, since that has a branch cut at the imaginary axis. However, incorporating the branch cuts is quite straightforward:
\begin{align}
\label{eq:rederiveenergy}
&\oint_{\gamma_x}dz \log \frac{x(u)+\xi}{1/x(u)+\xi} K_Q(z-u)= \log \frac{x(u+i Q \ad)+\xi}{1/x(u+i Q \ad)+\xi}- \log \frac{x(u-i Q \ad)+\xi}{1/x(u-i Q \ad)+\xi} \nn
&+ \left( \int_{i \R-\epsilon}-\int_{i \R+\epsilon}\right) dz \log \frac{x(u)+\xi}{1/x(u)+\xi}K_Q(z-u) \nn
&=2 \log q^{-Q} \frac{x(u+i Q a)+\xi}{x(u-i Q a)+\xi}
+\int_{i \R}2\pi i K_Q(z-u)dz \nn
&= -2\tilde{\mathcal{E}}_Q- 2 \ad Q,
\end{align}
where the second term is due to the branch cut along the imaginary axis. After correctly accounting for prefactors this term exactly cancels the extra term $\ad Q L$ we picked up from the boundary condition on $\log Y_Q$. This is a feature not present in the undeformed case: there both the integral over $K_Q$ and $\log Y_Q$ at infinity vanish already separately.
\subsubsection{$L_{M|vw}$}
For rewriting the contribution of the $L_{M|vw}$ we can explicitly write it as
\be
-\frac{1}{2}\oint_{\gamma_x}dz \sum_{\alpha} L_{M|vw}^{(\alpha)}\star\left(K(t+i M\ad,z)+K(t-i M \ad,z) \right)K_Q(z-u).
\ee
This can further be simplified by contour deformation for some of the terms involved of $\gamma_x$, which governs the integration over $z$:
by moving both halves of the contour onto the other sheet we have formed a contour that looks exactly like $\gamma_x$, but it runs in the other direction and on the second sheet. This leads to
\begin{align}
&-\frac{1}{2}\oint_{\gamma_x}dz \sum_{\alpha} L_{M|vw}^{(\alpha)}\star\left(K(t+i M\ad,z)+K(t-i M \ad,z) \right)K_Q(z-u)
\nn
&= \frac{1}{2\pi i}\oint_{\gamma_x}dz \sum_{\alpha} L_{M|vw}^{(\alpha)}\star_t
\left(\frac{d}{dt}  \log \left(x(t+i M \ad) -x(z)\right)\left(x(t-i M \ad) -x(z)\right)  \right)K_Q(z-u),
\end{align}
which can be simplified even further by deforming $\gamma_x$ again using an extension of \eqref{fsimple}: if $f$ has poles itself, the right-hand side of \eqref{fsimple} will also feature the poles of $f$. In this case
\be
f(z) = \frac{d}{dt}  \log \left(x(t+i M \ad) -x(z)\right)\left(x(t-i M \ad) -x(z)\right),
\ee
which gives for the $L_{M|vw}$ contribution
\begin{align}
& \frac{1}{2\pi i}\oint_{\gamma_x}dz \sum_{\alpha} L_{M|vw}^{(\alpha)}\star_t
\left(\frac{d}{dt}  \log \left(x(t+i M \ad) -x(z)\right)\left(x(t-i M \ad) -x(z)\right)  \right)K_Q(z-u)
\nn &= \sum_{\alpha} L_{M|vw}^{(\alpha)}\star \frac{1}{2\pi i} \frac{d}{dt}\Bigg( \log \left(
\frac{\sin 1/2( t+i M \ad-u+i Q \ad)}{\sin 1/2( t+i M \ad-u-i Q \ad)}
\frac{\sin 1/2( t-i M \ad-u+i Q \ad)}{\sin 1/2( t-i M \ad-u-i Q \ad)} \right)
\nn &+ \log \left(\frac{x(t+i M \ad) -x(u+iQ\ad)}{x(t+i M\ad) -x(u-iQ\ad)}\frac{x(t-i M\ad) -x(u+iQ\ad)}{x(t-i M\ad) -x(u-iQ\ad)}
\right) \Bigg)\nn
&= -\sum_{\alpha} L_{M|vw}^{(\alpha)}\star \left( K^{MQ}_{vwx}(u) -  \sum_{i=0}^{M} K_{M+Q-2i} \right)  = -\sum_{\alpha} L_{M|vw}^{(\alpha)}\star \left( K^{MQ}_{vwx}(u) -  \sum_{i=0}^{M} K_{Q-M+2i} \right).
\end{align}
If we now combine all the $Y_{vw}$-contributions on the right-hand side of the $Y_Q$ TBA equation (see also \eqref{prelres}) we get
\begin{align}
\sum_{l=0}^{M} L_{M|vw}\star K_{Q+2l-M}(u) -\frac{1}{2}\oint_{\gamma_x}dz \sum_{\alpha} L_{M|vw}^{(\alpha)}\star K_M(z)K_Q(z-u) = \sum_{\alpha} \sum_{M=1}^{\infty} L_{M|vw}^{(\alpha)}\star K^{MQ}_{vwx}(u)
\end{align}
which is the actual TBA-contribution due to the $vw$ functions.
\subsubsection{$L_-$}
The contributions of $L_-$ in \eqref{eq:y1expl} to the TBA equation can be summed up as
\begin{align}
& \frac{1}{2}\oint_{\gamma_x}\sum_{\alpha} \left( L^{(\alpha)}_{-} +  L^{(\alpha)}_{+} \right) \hat{\star} K(z) K_Q(z-u) \nn
&- \int_{\check{Z}_0} \left( \sum_{\alpha}\check{L}_-^{(\alpha)}(z)\right) K_Q(z-u)  + \int_{Z_0+i \e }  \L_-^{(\alpha)}(u) K_Q(z-u).
\end{align}
We can rewrite the first integral by contracting the integral contour. For this, we need to continue the integral through $(-\theta,\theta)$, resulting in a residue being picked up due to the convolution on this interval. Therefore, the first integral reads
\begin{align}
&\frac{1}{2}\oint_{\gamma_x}\sum_{\alpha} \left( L^{(\alpha)}_{-} +  L^{(\alpha)}_{+} \right) \hat{\star} K(z) K_Q(z-u)  \nn
&= -\left( L^{(\alpha)}_{-} \hat{\star}K_-^{yQ} - L^{(\alpha)}_{+} \hat{\star}K_+^{yQ}\right)
-\int_{Z_0+i \e }  \L_-^{(\alpha)}(u) K_Q(z-u)\nn
&+ \int_{\check{Z}_0})\left( \sum_{\alpha}\check{L}_-^{(\alpha)}(z)\right) K_Q(z-u)
\end{align}
such that after combining them we find as the total $L_{\pm}$ contribution
\be
L_{\beta} \hat{\star} K_{\beta}^{yQ}.
\ee
\subsubsection{Restoring the dressing phase contribution}
The dressing phase kernel in the TBA equations gets restored by the term
\be
\oint_{\gamma_x} \Lambda_P\star K_P^{\Sigma}(z) K_Q(z-u).
\ee
We can first rewrite the product $\Delta^{\Sigma}= \Lambda_P\star K_P^{\Sigma}$ using the result \eqref{eq:simplifieddressing} from Appendix \ref{App:SimplifyingDressing}. Using the kernel $K_{q\Gamma}^{[Q]}$ and the identity \eqref{eq:Kqgammaidentity}
 we replace the infinite sum containing $K$ kernels in \eqref{eq:simplifieddressing} by the terms containing $K_{q\Gamma}^{[N]}$. Using the TBA equation for $Y_-$ we then obtain
\be
\Delta^{\Sigma}(u) = 2 \Lambda_P \star \oint_{\gamma_x} ds K^{Py}_-(s) \left( \oint_{\gamma_x} dt K_{q\Gamma}^{[2]}(s-t)K(t,u)-K_{q\Gamma}^{[2]}(s-u) \right) \text{ for } u \not\in \check{Z}_0.
\ee
We can recognise the right-hand side of this equation as the discontinuity of the function
\be
2 \Lambda_P \star \oint_{\gamma_x} ds K^{Py}_-(s) \oint_{\gamma_x} dt K_{q\Gamma}^{[2]}(s-t)\frac{1}{2\pi i} \frac{d}{dt} \log \left( x(t) -x(u) \right),
\ee
such that we can immediately rewrite
\be
\oint_{\gamma_x}dz \, \Delta^{\Sigma}(z) K_Q(z-u) =
- 2 \Lambda_{P} \star \oint_{\gamma_x} ds \, K^{Py}_-(s) \oint_{\gamma_x} dt \, K_{q\Gamma}^{[2]}(s-t)K^{yQ}_-(t,u).
\ee
Following \cite{Cavaglia:2010nm} and in particular using equation (4.17) in \cite{Arutynov:2014ota} we find that indeed
\be
\oint_{\gamma_x}dz \, \Delta^{\Sigma}(z) K_Q(z-u) =
- 2 \Lambda_{P} \star K_{P  Q}^{\Sigma}.
\ee
Together with the already present $\Lambda_P \star K_{PQ}$ term this forms the dressing phase term in the TBA equations.

\subsection*{Combining all partial results}
Combining all the partial results in the previous subsections and plugging them into Eqn. \eqref{prelres} we get
\be
\log Y_Q(u) = -L \tilde{\mathcal{E}}_Q + \sum_{\alpha}\left( \sum_{M=1}^{\infty} L_{M|vw}^{(\alpha)}\star K^{MQ}_{vwx}(u)+ L_{\beta}^{\alpha} \hat{\star} K_{\beta}^{yQ}\right)   + \Lambda_P \star K_{\mathfrak{sl}(2)}^{PQ},
\ee
which indeed coincides with the TBA equation \eqref{TBAeqns} for $Y_Q$.

\section{Defining the $T$ gauges}
\label{App:Tgauges}
The derivation of the QSC from the TBA equations requires us to pass through the analytic $T$ system as described in Section \ref{sec:T-system}. This part of the analysis is highly technical, since, as in the undeformed case, we have to deal with the fact that there is no gauge in which the $T$s have nice analyticity properties everywhere on the $T$ hook. Even though the derivation has many similarities to the undeformed case we opt to present a self-contained derivation, to aid the reader and simultaneously provide a convenient summary of the results for the undeformed case presented in \cite{Gromov:2011cx} (in particular Sections 3.3-3.4 and Appendix C.2-C.3) and \cite{Gromov:2014caa} (Appendix B.1).

Our aim is to derive two sets of $T$s from the analytic $Y$ system, the $\bfT$ gauge and the $\mbT$ gauge, which have nice properties as listed in Section \ref{sec:T-system}. Additionally, we will prove that the $\hbT$ gauge, obtained from the $\mbT$ gauge, is $\Z_4$-symmetric, which will allow us to derive the $\Pf\mu$ parametrisation of the $T$ system (see Appendix \ref{App:hbT-gauge}).

In this appendix we will distinguish different types of conjugation: we will write $\overline{f(u)}$ for the conjugation of the image of $u$ under $f$, as usual. $\bar{f}(u)$ indicates the image of $u$ under $\bar{f}$, being the conjugation of the function only. For example, for an analytic function $f$, $\bar{f}$ is obtained from the converging Taylor series of $f$ by conjugating all its coefficients.

The functions which are analytic in the strip
\be
\left\{ u \in \C\, | \, |\text{Im}(u)| < M\ad\right\},
\ee
possibly with the exception of a finite number of poles form the set $\mathcal{A}_M$. If we want $f\in \mA_M$ to also be pole free, i.e. be analytic, on the strip, we write $f \in \mAp_M$.

Our starting point is a set of $Y$s satisfying the ground-state TBA equations, and therefore also the analytic $Y$ system. The derivation is split up into the following steps:
\begin{enumerate}
\item Construct a gauge $\JT$ in the upper band with the correct analyticity properties
\item Use the gauge freedom of the $T$ system to define from $\JT$ the $\bfT$ gauge
\item Define the $\mbT$ gauge directly in terms of the $\bfT$ gauge
\item Define gauges $\mcTR$ and $\mcTL$ which have the correct analyticity properties in the right and left band respectively and are $\Z_4$-symmetric
\item Prove that the $\mbT$ gauge is related to the $\mcTR$ and $\mcTL$ gauges by a very simple gauge transformation, from which we can conclude that the $\mbT$ also have the correct analyticity properties and are $\Z_4$-symmetric.
\end{enumerate}
\subsection{Constructing $\JT$}
We start building the gauge $\JT$ for the upper band $a\geq s$, adapting the method outlined in \cite{Balog:2011nm}. First we set $\JT_{0,0}=1=\JT_{a,\pm2}= 1$ for $a\geq 2$. From the parametrisation of the $Y$ system we find
\be
\label{YsinJTs}
1+Y_{a,0} = \frac{\JT_{a,0}^+\JT_{a,0}^-}{\JT_{a+1,0}\JT_{a-1,0}} \text{ for } a>0.
\ee
By a trigonometrisation of the chain lemma in \cite{Balog:2011nm} we can solve this equation for $\JT_{a,0}$s:
\be
\label{rand1}
\sigma_a = \JT_{a,0} = \exp \left(\ssum{j} \Lambda_j \star l_a^j \right),
\ee
with
\be
l_a^j = \sum_{n=0}^{j-1} K_{a+1-j+2n}.
\ee
The application of this lemma requires:
\begin{itemize}
\item No poles and zeroes in the strip with $|$Im$(u)<(a-1)\ad$, and $1+Y_{a,0} \in \mA_a$, for $a\geq 2$ .
\item No poles and zeroes in the physical strip for $1+Y_1$.
\item That $1+Y_a$ does not attain negative real values to avoid unwanted branch cut behaviour.
\end{itemize}
These properties can be analysed in the asymptotic large-volume solution of the $Y$ system and appear to be respected. We will assume these properties remain satisfied at finite volume. The cut structure of all the $Y$s of course follows rigorously from the TBA equations. In addition to the list above, we find that $Y_1$ does not have a $Z_0$ cut and that all $Y$ functions are real on the real line. This will imply that also the $\JT$ are real. The solution \eqref{rand1} solves the Eqn. \eqref{YsinJTs} at least in a neighbourhood of the physical strip, but we can in most cases extend this: the solution already implies that the $\JT_{a,0}$ do not have a cut at $Z_{\pm1}$. For the rest we use induction: Suppose we have shown that for all $a\leq k$ $\JT_{a,0}$ has no cuts until possibly $Z_{\pm (a+1)}$. Using the defining relation \eqref{YsinJTs} and the knowledge that $Y_a \in \mathcal{A}_a$ we can now derive that $\JT_{k+1,0}$ has no cuts until possibly $Z_{\pm (k+2)}$: from the defining relation for $a=k$ we immediately see that $\JT_{k+1,0}$ has no cuts until possibly $Z_{\pm k}$. If it has a cut at $Z_{\pm k}$, the $a=k+1$ equation tells us that $\JT_{k+2,0}$ has to have a cut at $Z_{\pm(k-1)}$. Continuing up the chain of equations we find now that there is a $\JT_{m,0}$ with a cut at $Z_{\pm 1}$, which we know is not possible. So $\JT_{k+1,0}$ has no cut at $Z_{\pm k}$. Completely analogously one proves that also the cuts at $Z_{\pm (k+1)}$ are not possible.

Secondly, we look at poles. We know that $1+Y_a \in \mAp_{a-1}$. Moreover, $1+Y_1,\JT_{a,0} \in \mAp_1$.  Using the same arguments as for the cut structure we find that $\JT_1 \in \mAp_2$ and $\JT_a \in \mAp_{a-1}$ for $a\geq 2$.

Now we have completely defined and analysed the central $\JT$s. We can find the remaining $\JT$s in the upper band using the Hirota equation and
\be
\label{YsinTs2}
1+Y_{a,s}^{-1} = \frac{\JT_{a,s}^+\JT_{a,s}^-}{\JT_{a,s+1}\JT_{a,s-1}}.
\ee
First we fill the band for $a\geq 2$ solving the finite-difference equation
\be
\JT_{a,1}^+\JT_{a,1}^- = \JT_{a,0}(1+Y_{a,1}^{-1}),
\ee
which also follows from the $T$ parametrisation of the $Y$ functions. Moreover, from the analyticity of the right-hand side we immediately see that  $\JT_{a,1} \in \mathcal{A}_a$. To complete the upper band we only need to define $\JT_{1,\pm 1}$, which we do using the Hirota equation. It follows immediately that $\JT_{1, \pm 1} \in \mathcal{A}_{1}$. This finishes the construction of the analytic gauge $\JT$. Note that from the construction it follows that the $\JT$ gauge is real.
\subsubsection{Constructing $\bfT$}
From the $\JT$ gauge we can construct a gauge with even better properties, known as the $\bfT$ gauge. In addition to the analyticity properties listed in Table \ref{Tsanalyticity} the $\bfT$s satisfy the following identities, which were dubbed ``group-theoretical'' in \cite{Gromov:2011cx}:
\be
\bfT_{3,\pm 2} =\bfT_{2,\pm 3}, \quad \bfT_{0,0}^+=\bfT_{0,0}^-.
\ee
We start from the discontinuity relation for $Y_{\pm}^{(\alpha)}$, which on the $Y$ hook reads
\be
\left[ \log Y_{1,\pm 1} Y_{2 \pm,2} \right]_{2n} = - \sum_{a=1}^n \left[ \log 1+Y_{a,0} \right]_{2n-a} \text{ for } n>0.
\ee
Notice that the right-hand side of this expression is independent of $\pm$ ($\alpha=l,r$ in our TBA notation). This is due to the fact that the product $Y_{1,\pm 1} Y_{2,\pm2}$ is the same on both sides of the $Y$ hook, so
\be
\label{qmodlity}
Y_{1, 1} Y_{2 ,2} = Y_{1, -1} Y_{2 ,-2}
\ee
as follows from the TBA equations. Replacing the $Y$s by their parametrisation in $\JT$s leads to a telescoping cancellation of terms and we are left with
\be
\left[ \log \frac{1}{Y_{1,\pm 1} Y_{2,\pm 2}} \frac{\JT_{1,0}}{\JT_{0,0}^-} \right]_{2n} = 0 \text{ for } n>0.
\ee
Thus it follows that for a solution of the $Y$ system
the function $\bfB$ defined by
\be
\textbf{B} = \frac{1}{Y_{1,\pm 1} Y_{2,\pm 2}} \frac{\JT_{1,0}}{\JT_{0,0}^-}
\ee
is analytic in the upper half-plane.

Now we are ready to define the $\bfT$ gauge: to modify the $\JT$ gauge we consider the most general gauge transformation
\be
\label{bfT}
\textbf{T}_{a,s} = f_1^{[a+s]}f_2^{[a-s]}\bar{f_1}^{[-a-s]}\bar{f_2}^{[-a+s]} \JT_{a,s}
\ee
for two unknown functions $f_1,f_2$ such that $f_1^-,f_2^-$ are analytic in the upper half-plane. This gauge does not destroy the reality nor the analyticity of the $\bfT$s. On $f_1,f_2$ we will impose that
\be
\mathbf{B} = \frac{\left(f_1 f_2\right)^-}{\left(f_1 f_2\right)^+}.
\ee
If we can find $f_1,f_2$ satisfying this constraint, then the $\bfT$s satisfy the identities
\bea
\label{eq:bft1}
\frac{\bfT_{3,\pm 2}\bfT_{0,\pm 1}}{\bfT_{2,\pm 3}\bfT_{0,0}^-} = 1 =
\frac{\bfT_{3,\pm 2}\bfT_{0,\pm 1}}{\bfT_{2,\pm 3}\bfT_{0,0}^+},
\eea
where the second equality is just complex conjugation of the first equality. We have more freedom in choosing $f_1,f_2$ still. Imposing
\be
\bfT_{0,1}=\bfT_{0,-1}
\ee
leads to the equation
\be
\label{realdiffff}
\frac{\JT_{0,-1}}{\JT_{0,1}}= \frac{\left(f_1/f_2\right)^+}{\left(f_1/f_2\right)^-}\frac{\left(\bar{f}_1/\bar{f}_2\right)^-}{\left(\bar{f}_1/\bar{f}_2\right)^+},
\ee
where we notice that the two fractions containing the $f_i$ are complex conjugate functions. The left hand side is real and can be decomposed as
\be
\frac{\JT_{0,-1}}{\JT_{0,1}} = \bfH \overline{\bfH},
\ee
with $\bfH$ analytic in the upper half-plane, implying we should solve
\be
\mathbf{H} = \frac{\left(f_1/f_2\right)^+}{\left(f_1/f_2\right)^-}.
\ee
To find $f_1,f_2$ we now have to solve
\be
\mathbf{B} = \frac{\left(f_1 f_2\right)^-}{\left(f_1 f_2\right)^+}, \quad \mathbf{H} = \frac{\left(f_1/f_2\right)^+}{\left(f_1/f_2\right)^-},
\ee
which are two finite-difference equations of the exact same form. Their solution can be found as we will discuss in the next subsection, solving the logarithmic version of this equations. With these solutions we can find $f_1,f_2$ and define the $\bfT$ gauge. The group-theoretical properties follow from Eqn. \eqref{eq:bft1} and show in particular that $\mu$, defined as the square root of $\bfT_{0,1} = \bfT_{0,0}^+$ is $2i\ad$-periodic in the mirror kinematics.
\subsubsection{Solving periodic difference equations: constructing $f$}
To find the gauge transformation in the previous section we are supposed to solve the finite-difference equations
\begin{align}
\label{ffromB}
\log \textbf{B} &= \log \left(f^-\right)-\log \left(f^+\right) \nn
- \log \textbf{H} &= \log \left(g^-\right)-\log \left(g^+\right),
\end{align}
where $f=f_1 f_2$ and $g= f_1/f_2$. A formal solution of the first equation in \eqref{ffromB} (the second follows completely analogously) can be written down immediately as the infinite sum
\be
\label{formalsol1}
 \log \left(f\right) = \ssum{n} \log \textbf{B}^{[2n-1]},
\ee
but this will in general not be convergent. We can regularise this sum by considering a spectral representation of $\bfB$ in the \uhp. To formalise this we prove the following lemma.
\\[5mm]
\textbf{Lemma} Let $\textbf{f}:\C\rightarrow \C$ be a $2\pi$-periodic function regular in the upper half-plane obeying
\be
\lim_{u\rightarrow i \infty} \textbf{f}(u) \in \R,
\ee
and which converges uniformly to an $L^1$ function on $[-\pi,\pi]$. Then it admits a spectral representation
\be
\bff(u) = \int_{Z_0} \mathcal{K}(u-v)\rho_{\small \bff}(v)dv \text{ for Im}(u)>0,
\ee
where $\mathcal{K}(u) = -\frac{1}{2 \pi i} \cot(u)$ and where
\be
\rho_{\small \bff}(u) = 2 \lim_{\e \rightarrow 0} \text{Re}\left(\bff(u+i\e)\right) = \lim_{\e \rightarrow 0} \bff(u+i\e)+\overline{\bff}(u-i\e)	.
\ee

The proof is an application of Cauchy's theorem after splitting the integral into two parts using the definition of $\rho$ and hinges on the fact that high in the upper half-plane the real part of the cotangent becomes negligible, causing the two integrals to cancel each other, except for a contribution of the pole of the cotangent which yields $\bff$.

We can define the kernel
\be
\psi_{\ad}(u) = \mathcal{K}(u) + \frac{1}{2\pi i} \cot(2i\ad ) + \sum_{n=1}^{\infty} \left( \mathcal{K}^{[2n]}(u) + \frac{1}{2\pi i} \cot(2(n+1) i \ad )\right),
\ee
where the sum gives a combination of $q$-polygamma functions. Now it is easy to check that
\bea
\log \left( f\right) &=& \psi_{\ad}^+ \star \rho_{\log \bfB}, \nn
\log \left( g\right) &=& \psi_{\ad}^+ \star \rho_{-\log \bfH},
\eea
solve the finite-difference equations we started with.\footnote{It is easy to check numerically in the asymptotic finite-volume solution that $\log \bfB$ and $\log \bfH$ satisfy the requirements of the lemma above. We assume this extends beyond the asymptotic solution.}
\subsection{Constructing $\mbT$}
We have now constructed the $\bfT$ gauge, which has nice properties in the upper band. This gauge however does not have nice properties in the right band, which is where we would like to start our $\Pf\mu$ parametrisation. We can define such a nice right-band gauge directly from the $\bfT$ gauge through the transformation
\be
\label{mbt}
\mbT_{a,s} = (-1)^{as}  \bfT_{a,s} \left(\bfT_{0,0}^{[a+s]}\right)^{\frac{a-2}{2}}.
\ee
One can check directly that this gauge satisfies the Hirota equation on the $T$ hook. However, whether it has nice properties is far from obvious from this definition. We will spend the rest of this appendix proving its properties. In order to do this, we first define two gauges $\mcTR$ and $\mcTL$ which behave nicely in the right and left band respectively, from a solution of the analytic $Y$ system.
\subsection{Constructing $\mcTR$ and $\mcTL$}
The construction of the $\mcTR$ and $\mcTL$ gauges follows Section 3.3 of \cite{Gromov:2011cx}. We will focus on the $\mcTR$ gauge and simply denote it by $\mcT$, the construction of the $\mcTL$ gauge is analogous.

Consider a set of $\mcT$ functions in the right band ($s\geq a$) parametrised by resolvents $G$ and $\bar{G}$ as follows:
\begin{align}
\mcT_{0,s} &= 1, \nn
\mcT_{1,s} &= s + G^{[+s]}+\bar{G}^{[-s]}, \nn
\mcT_{2,s} &= \left(1 + G^{[s+1]}-G^{[s-1]}\right)\left(1 + \bar{G}^{[-s-1]}-\bar{G}^{[-s+1]}\right),
\end{align}
The resolvents $G,\bar{G}$ are parametrised by $\rho$ through the definitions
\begin{align}
G(u) &= \int_{Z_0} dv\, \mK(u-v) \rho(v) \text{ for Im}(u)>0,\nn
\bar{G}(u) &= \int_{Z_0} dv\, \bar{\mK}(u-v) \rho(v) \text{ for Im}(u)<0,
\end{align}
such that we have $\rho = G^{[+\e]}+\bar{G}^{[-\e]}$. This gauge has the required analyticity properties and solves the Hirota equation on the right band, so we only need to check the presence of $\Z_4$ symmetry. Ultimately, this follows from the fact that
\be
\label{YplusYmin}
Y_{1,1}(u+i\e)=1/Y_{2,2}(u-i\e) \text{ for } u \in \check{Z}_{0}.
\ee
Indeed, the ratio
\be
\label{rY1Y2}
r= \frac{1+1/Y_{2,2}}{1+Y_{1,1}} = \frac{\mcT_{2,2}^+\mcT_{2,2}^-\mcT_{0,1}}{\mcT_{1,1}^+\mcT_{1,1}^-\mcT_{2,3}}
\ee
is expressed in right-band $\mcT$ functions only and satisfies $r(u+i\e)=1/r(u-i\e)$ for $ u \in \check{Z}_{0}$ due to Eqn. \eqref{YplusYmin}. Using our parametrisation this implies $G^{[\e]}=-\bar{G}^{[-\e]}$, which implies that the function $\hat{G}$ defined by
\be
\hat{G}(u) = \int_{\hat{Z}_0} dv \, \mK(u-v) \rho(v)
\ee
has only one short $Z_0$ cut and coincides with $G$ ($\bar{G}$) on the upper (lower) half-plane.
This allows us to consider short-cutted versions of the $\mcT$ gauge we have defined before. These functions can be found by continuation and are given by
\begin{align}
\hmcT_{0,s} &= 1, \nn
\hmcT_{1,s} &= s + \hat{G}^{[+s]}-\hat{G}^{[-s]}, \nn
\hmcT_{2,s} &= \left(1 + \hat{G}^{[s+1]}-\hat{G}^{[s-1]}\right)\left(1 + \hat{G}^{[-s-1]}-\hat{G}^{[-s+1]}\right) = \hmcT_{1,1}^+\hmcT_{1,1}^-.
\end{align}
These $\hmcT$s coincide with the $\mcT$s just above the real axis, which can be extended to any domain where $\mcT$s are cut free. Moreover, these $\hmcT$s define a solution of the Hirota equation on an infinite band by analytic continuation in $s$, which is possible because the dependence on $s$ is analytic. From the definitions of the $\hmcT$s it is straightforward to check that they satisfy
\be
\label{z4hmct}
\hmcT_{a,s} = (-1)^a \hmcT_{a,-s},
\ee
in other words the $\hmcT$s are $\Z_4$-symmetric. Note that this short-cutted gauge only coincides with the long-cutted one on the right band and only just above the real line. Also, for consistency of the derivation we should require that the second equality in \eqref{rY1Y2} holds, leading to the following nonlinear integral equation for the density $\rho$:
\be
\label{nonlinearr}
\frac{1+1/Y_{2,2}}{1+Y_{1,1}}= \frac{\left(1+\mK_1^+\slashconv \rho -\rho/2 \right)\left(1+\mK_1^-\slashconv \rho -\rho/2 \right)}{\left(1+\mK_1^+\slashconv \rho +\rho/2 \right)\left(1+\mK_1^-\slashconv \rho +\rho/2 \right)} \text{ for } u\in \hat{Z}_0,
\ee
where
\begin{align}
\mK(u) &= -\frac{1}{2\pi i}\cot(u), \nn
\mK_s(u) &= \mK^{[s]}(u)-\mK^{[-s]}(u), \nn
K\slashconv f (u) &= \text{P.V.} \int_{\hat{Z}_0} dv\, K(u-v)f(v).
\end{align}
The undeformed limit of this equation was numerically shown to have a solution \cite{Gromov:2011cx}. We have not proven that this equation always has a solution in our case, but will assume this is the case.
\subsubsection{The $\mbT$ gauge is analytic}
Having defined the $\mcT$ gauge we can see how it relates to the $\mbT$ gauge defined before, since two gauges parametrising the same set of $Y$ functions are related by a gauge transformation, as in Eqn. \eqref{eq:Tgaugetransformation}. This will show all the wanted properties of the $\mbT$ gauge.

The first thing to check is that the $\mbT$ gauge has the right analyticity properties. In order for both gauges to be real we have to restrict the gauge transformation \eqref{eq:Tgaugetransformation} to
\be
\mbT_{a,s} = f_1^{[a+s]}f_2^{[a-s]}\bar{f_1}^{[-a-s]}\bar{f_2}^{[-a+s]} \mcT_{a,s}.
\ee
From $\mcT_{0,s}= \mbT_{0,s} =1 $ we get $f_1f_2^{++}=1$. From the definition of $\mbT_{2,s}$ and the analyticity of the $\bfT$ gauge it follows directly that $\mbT_{2,s}\in \mA_{s-1}$ for $s\geq 2$. Since also $\mcT_{2,s}\in \mA_{s-1}$ for $s\geq 2$ we find that $f_1^{++}/f_1^{--}$ is analytic in the upper half-plane. The last thing to check is the analyticity of the $\mbT_{1,s}$ functions. For this we use the discontinuity relation \eqref{res2}. Parametrising it using the $\bfT$ and $\mcT$ gauge and using that $Y_{2,2} = \frac{\bfT_{2,1}}{\bfT_{1,2}}$ and that
\be
\label{Y11Y22}
Y_{1,\pm 1}Y_{2,\pm 2} = \frac{\bfT_{0,0}^-}{\bfT_{1,0}}
\ee
we find that
\be
\frac{\bfT_{1,2} \mcT_{1,1}^-}{\bfT_{1,1}^- \mcT_{1,2}} =  -\frac{\mbT_{1,2} \mcT_{1,1}^-}{\mbT_{1,1}^- \mcT_{1,2}}
\ee
is also analytic in the upper half-plane. Plugging in our gauge transformation this implies that $f_1^{++}f_1^{--}/f_1^2$ is analytic in the upper half-plane. All of these constraints now imply that $f_1/f_1^{--}$ is analytic in the upper half-plane. For the remaining $\mbT_{1,s}$ the gauge transformation now takes the form
\be
\label{gaugetrsfmmcTR}
\mbT_{1,s} = \left(  \left(\frac{f_1}{f_1^{--}}\right)^{[s]} \times \left(\frac{\overline{f_1}}{\overline{f_1}^{++}}\right)^{[-s]}   \right) \mcT_{1,s} \text{ for } s\geq 1,	
\ee
which shows that $\mbT_{1,s} \in \mA_s$, as desired. So we find that the $\mbT$ gauge indeed has the right analyticity properties. In the next section we will show that $\mbT$ is actually $\Z_4$-symmetric.
\subsubsection{So is $\mbT$ really $\Z_4$-symmetric?}
The real difficulty is to prove that the $\mbT$ gauge is actually $\Z_4$-symmetric. We will go through the following steps: first we will prove that the product $\hat{\mbT}_{1,1}\hat{\mbT}_{1,-1}$ has only two cuts at $Z_{\pm1}$. Second we will show that we can force each of these functions to have only two cuts. We can then use this to put more constraints on the gauge transformation in \eqref{gaugetrsfmmcTR} and conclude $\Z_4$ symmetry from there.

\textbf{Step 1:} the derivation starts with the discontinuity relation belonging to the $Y_Q$ functions (see \eqref{discs1}) and is a generalisation of Appendix C.3 of \cite{Gromov:2011cx} to nonsymmetric states. On the $Y$ hook it reads for $N\geq 1$
\be
\left[ \left[\log Y_{1,0}^+  \right]_0^{[2N]} \right]_0 = \sum_{\pm} \left( \left[\log\left(1+Y_{1,\pm 1}^{[2N]} \right) + \sum_{m=1}^N \log\left(1+Y_{m+1,\pm 1}  \right)^{[2N-m]}  \right]_0 - \log Y_{1,\pm 1}Y_{2,\pm 2} \right).
\ee
Using the $\bfT$ gauge we can compute $\left[\log Y_{1,0}^+  \right]_0$:
\be
Y_{1,0} = \frac{\bfT_{1,1}\bfT_{1,-1}}{\bfT_{0,0}\bfT_{2,0}} = \frac{\mbT_{1,1}\mbT_{1,-1}}{\bfT_{2,0}}
\ee
and using that $\bfT_{2,0} \in \mA_{3}$ we find
\be
\left[\log Y_{1,0}^+  \right]_0 = \left[\log \mbT_{1,1}^+\mbT_{1,-1}^+  \right]_0 = \log \mbT_{1,1}^+/\tilde{\mbT}_{1,1}^+\log \mbT_{1,-1}^+/\tilde{\mbT}_{1,-1}^+,
\ee	
where the $\tilde{\mbT}_{1,\pm 1}$ coincide with the $\mbT_{1,\pm 1}$ below $Z_1$ and have a short cut $\hat{Z_1}$. All other cuts in the upper half-plane are long. We use the analyticity of the $\bfT$ and $\mbT$ gauges to telescope the sums in the expression above:
\be
\label{telescoped1}
\left[\log \left(\frac{\mbT_{1,1}^+ \mbT_{1,-1}^+ }{\tilde{\mbT}_{1,1}^+ \tilde{\mbT_{1,-1}}^+}  \frac{\bfT_{0,1}\bfT_{0,-1}}{\bfT_{1,1}^+\bfT_{1,-1}^+}\right)  \right]_{2N} = -2 \log Y_{1,1}Y_{2,2},
\ee
where we use that $Y_{1,1}Y_{2,2}= Y_{1,-1}Y_{2,-2}$. We know that $\mbT_{1,\pm1} = -\bfT_{1,\pm1} \bfT_{0,0}^{-1/2}$, so the argument of the log on the left-hand side becomes
\be
\frac{\bfT_{0,0}^+}{\tilde{\mbT}_{1,1}^+ \tilde{\mbT}_{1,-1}^+}	.
\ee
Now we can use the property that
\be
\left[ \log \bfT_{0,0} \right]_{2N+1} = - 2 \log Y_{1,\pm 1}Y_{2, \pm 2} \text{ for } N\in \Z,
\ee
which follows directly from $Y_{1,\pm 1}Y_{2,\pm 2} = \bfT_{1,0}/\bfT_{0,0}^+$, the periodicity of $\bfT_{0,0}$ and the fact that $\bfT_{1,0} \in \mA_1$. With the previous discontinuity relation fitting perfectly, we find that what remains in \eqref{telescoped1} is just
\be
\left[\log \tilde{\mbT}_{1,1} \tilde{\mbT}_{1,-1} \right]_{2N+1} = 0 \text{ for } |N| \geq 1,
\ee
after repeating the argument for negative $N$ as well. Using that the potential cuts are located at $Z_{2N+1}$ with $N\in \Z$ we find that $\tilde{\mbT}_{1,1} \tilde{\mbT}_{1,-1}$ has just $Z_{\pm 1}$ cuts, finishing step one of our derivation as this directly implies that the product $\hat{\mbT}_{1,1}\hat{\mbT}_{1,-1}$ has only two short $Z_{\pm1}$ cuts.

\textbf{Step 2:} Next we want to prove that we can force each of the functions $\hat{\mbT}_{1,1}$ and $\hat{\mbT}_{1,-1}$ to have only two cuts. To do this we use a gauge freedom of $\bfT$ that we have not used yet and will alter the $\mbT$ since this last gauge is based on the former. Going through the properties of $\bfT$ one finds that nothing changes after transforming them as
\be
\bfT_{a,s} \rightarrow \left(\prod_{k=-(|s|-1)/2}^{(|s|-1)/2} \frac{e^{i \phi^{[a+2k]}}}{e^{i \phi^{[-a+2k]}}}  \right)^{\text{sgn}(s)} \bfT_{a,s},
\ee
as long as $\phi$ is real with one long $Z_0$ cut. To change the function $\mbT_{1,1}$ we see how this gauge transformation changes the definition of $\mbT$: as a function with short cuts we find
\be
\hbT_{1,1}^{\text{new}} = e^{i \hat{\phi}^+-\bar{\hat{\phi}}^-}\hbT_{1,1}.
\ee
Now, since we want $\hbT_{1,1}^{\text{new}}$ to have only 2 $Z_{\pm1}$-cuts we should cancel any extra cuts in $\hbT_{1,1}$ through $\phi$:
\be
i \left[\,\hat{\phi} \, \right]_{-2N}= - \left[ \log \hbT_{1,1}\right]_{-1-2N},\quad
i \left[\,\bar{\hat{\phi}}  \, \right]_{2N}= \left[ \log \hbT_{1,1}\right]_{1+2N} \text{ for } N\in \N.
\ee
Constructing a $\phi$ that obeys these rules can be done as follows: we can firstly find a short-cutted function $\psi$ that obeys
\be
i \left[\,\hat{\psi} \, \right]_{-2N}= - \left[ \log \hbT_{1,1}\right]_{-1-2N},\quad
i \left[\,\hat{\psi}  \, \right]_{2N}= \left[ \log \hbT_{1,1}\right]_{1+2N} \text{ for } N\in \N,
\ee
and define $\phi$ as solution to the Riemann-Hilbert problem
\be
\label{eq:RH1}
\phi(u+i\e)+\phi(u-i\e)=  \psi \text{ for } u \in \check{Z}_0.
\ee
One can solve this equation using the same method outlined in Section \ref{sec:pmusystem} to find the auxiliary $\mu$ elements, after multiplying the unknown function $\phi$ with a pure square root on the cylinder and dividing $\psi$ by the same function, turning the relative sign between the shifts of $\phi$ in \eqref{eq:RH1} into a minus. Any solution is real analytic on the mirror sheet with a long $Z_0$ cut, so we can use any to force $\hbT_{1,1}$ to have only $2$ cuts. It follows immediately from this and the fact that the product $ \hbT_{1,1} \hbT_{1,-1}$ has only two cuts that now both functions have exactly two cuts at $Z_{\pm1}$.
\subsubsection{Actually proving $\hbT$ is $\Z_4$-symmetric}
The gauge transformation \eqref{gaugetrsfmmcTR} can be written like
\be
\mbT_{1,s} = h^{[+s]}\bar{h}^{[-s]}\mcT_{1,s}
\ee
where $h$ is a function analytic in the upper half-plane. Translating this equation for $s=1$ to the short-cutted sheet reads
\be
\hbT_{1,1} = \hat{h}^{+}\bar{\hat{h}}^{-}\hmcT_{1,1}.
\ee
Knowing that the $T$ functions on both sides of this equation have only two short cuts at $\hat{Z}_{\pm1}$ we find that $\hat{h}$ has only one $\hat{Z}_0$ cut. The only remaining thing to prove is that $\overline{\hat{h}}/\hat{h}$ is a constant to conclude that the $\hbT$ gauge is also $\Z_4$-symmetric. Indeed, in that case we find
\be
\label{hbtsimhmct}
\hbT_{0,s}=1,\quad \hbT_{1,s}\sim \hat{h}^{[+s]}\hat{h}^{[-s]}\hmcT_{1,s}, \quad \hbT_{2,s}\sim \hat{h}^{[s+1]}\hat{h}^{[s-1]}\hat{h}^{[-s+1]}\hat{h}^{[-s-1]}\hmcT_{2,s},
\ee
which directly implies $\Z_4$ symmetry using Eqn. \eqref{z4hmct}.

The argument starts off with the Hirota equation for $\mbT_{1,1}$ and $\mbT_{2,2}$:
\begin{align}
\label{Hirota1122}
\mbT_{1,1}^+\mbT_{1,1}^- &= \mbT_{1,0}\mbT_{1,2}+\mbT_{0,1}\mbT_{2,1} \nn
\mbT_{2,2}^+\mbT_{2,2}^- &= \mbT_{2,1}\mbT_{2,3}+\mbT_{2,1}\mbT_{2,3}.
\end{align}
Now, defining $\mF = \sqrt{\bfT_{0,0}}$ we can use the properties
\begin{equation}
\mbT_{0,1} =1, \quad \mbT_{1,1}, \mbT_{3,2} = - \mF^+ \mbT_{2,3},\quad \mbT_{1,0} = -\mF^+Y_{1,1}Y_{2,2},
\end{equation}
where the last two properties follow from the $\bfT$ gauge and the gauge transformation \eqref{mbt}. In particular we have used Eqn. \eqref{Y11Y22} to obtain the last equation. Consider $u\in \hat{Z}_0$, then we have
\be
\mbT_{2,3}=\hbT_{2,3},\quad \mbT_{2,2}(u\pm i \ad )=\hbT_{2,2}(u\pm i \ad \mp i\e ),\quad \mbT_{1,1}(u\pm i \ad )=\hbT_{1,1}(u\pm i \ad \mp i\e ).
\ee
Additionally we find from the gauge transformation that
\be
\hbT_{2,s} = \frac{\hat{h}^{[s-1]}\bar{\hat{h}}^{[-s+1]}}{\hat{h}^{[-s+1]}\bar{\hat{h}}^{[s-1]}} \hbT_{1,1}^{[+s]} \hbT_{1,1}^{[-s]},
\ee
which for real $h$ simplifies to $\hbT_{2,s} =\hbT_{1,1}^{[+s]} \hbT_{1,1}^{[-s]}$. Excluding $\hbT_{2,1}$ from Eqn. \eqref{Hirota1122} and then using these properties we get
\be
\hat{h}^{[\e]}\bar{\hat{h}}^{[-\e]} = \frac{\left(1-Y_{1,1}Y_{2,2}\right) \mF^+}{\rho},
\ee
where we have used the parametrisation of $\mcT$ in terms of $\rho$ to simplify the right-hand side. The right-hand side is analytic on a neighbourhood of the real line, which implies the left hand side is that too, leading to
\be
\hat{h}^{[\e]}\bar{\hat{h}}^{[-\e]} = \hat{h}^{[-\e]}\overline{\hat{h}}^{[\e]},
\ee
which in turn tells us that $\bar{\hat{h}}/\hat{h}$ is cut free on the complex plane. Using the regularity requirement that our $\hbT$s do not have any poles except possibly at the branch points we want $\hat{h}$ to be pole and zero free. Assuming in analogy to the undeformed case that the Fourier series of $\bar{\hat{h}}/\hat{h}$ is a finite sum it follows directly that this series can only be a constant, hence leading to Eqn. \eqref{hbtsimhmct} and therefore implying the $\Z_4$ symmetry of the $\hbT$ gauge.

So now we have proven the existence of a $\Z_4$-symmetric gauge $\mbT$ related to the $\bfT$ gauge by Eqn. \eqref{bfT}. It has the analyticity domains we hoped for and we were able to choose $\hbT_{1,1}$ to have exactly two cuts, at $Z_{\pm1}$. This gauge is an excellent starting point to continue our simplification.
\section{The Wronskian parametrisation of $\hbT$}
\label{App:hbT-gauge}

In this appendix we show that we can decompose the $\hbT$ into more basic $q$ functions, which each have only one cut.

We start by using the general Wronskian parametrisation for an infinite band of $T$ functions satisfying the Hirota equation as in \cite{Gromov:2011cx,Krichever:1996qd}. Given that the $\hbT$ gauge is a solution of the Hirota equation on an infinite band of width three (using continuation in $s$) it admits a Wronskian parametrisation
\begin{align}
\hat{\mathbb{T}}_{0,s} = 1, \qquad \hat{\mathbb{T}}_{1,s} = \hat{q}_1^{[+s]}\hat{q}_2^{[-s]}-\hat{q}_1^{[-s]}\hat{q}_2^{[+s]},\qquad \hat{\mathbb{T}}_{2,s} = \hat{\mathbb{T}}_{1,1}^{[+s]}\hat{\mathbb{T}}_{1,1}^{[-s]}.
\end{align}
for two unknown functions $q_1,2$, where we have used its $\Z_4$ symmetry and the fact that $\hbT_{0,s}=1$ to reduce the number of independent functions from six to two. This parametrisation is symmetric under the action of $u$-dependent doubly-periodic (with halfperiods $\pi$ and $i\ad$) $\mathfrak{sl}_2$ matrices $H$ which act as $q_i \rightarrow H_{ij}q_j$.

The further analysis of the right band mostly relies on the cut structure of the functions present, but requires solving one particular Riemann-Hilbert problem for $H$, which rotates our solution into a solution with good analyticity properties. We see that these $q$s should satisfy a system of Baxter equations
\begin{align}
\label{Be1}
\hat{q}^{[2r-1]}\hat{\mathbb{T}}_{1,1} & = \hat{q}^+\hat{\mathbb{T}}_{1,r}^{[r-1]}   - \hat{q}^-\hat{\mathbb{T}}_{1,r-1}^{[r]} \nn
\hat{q}^{[-2r+1]}\hat{\mathbb{T}}_{1,1} & = \hat{q}^-\hat{\mathbb{T}}_{1,r}^{[-r+1]}   - \hat{q}^+\hat{\mathbb{T}}_{1,r-1}^{[-r]},
\end{align}
which can be used to analyse the solutions: using that $\hat{\mathbb{T}}_{1,1}$ is regular at $3i\ad$ we find that
\be
\label{49}
\hat{\mathbb{T}}_{1,1}^{[-3]} [\hat{q}]_{2r-4} = \hat{\mathbb{T}}_{1,r}^{[r-4]}[\hat{q}]_{-2}-\hat{\mathbb{T}}_{1,r-1}^{[r-4]}[\hat{q}]_{-4}, \quad \text{for } r>2,
\ee
from which one can derive the Riemann-Hilbert problem
\be
\label{RH1}
[H_i^j]_0 \hat{q}_j^{[-2n]}+H_i^j \left[\hat{q}_j^{[-2n]}\right]_0 = 0 \quad \text{ for } j,n=1,2
\ee
for a symmetry transformation $H$ that will regularise $\hat{q}$ to have only one short cut on the real axis. This follows from the fact that \eqref{49} relates the cuts of $\hat{q}$ in the upper half-plane to the first two cuts above the real line. Taking the conjugate of this equation does the analogous thing in the lower half-plane. Therefore, by making sure that $\hat{q}$ is regular at those two cuts ensures that it has only one cut on the real line. This is achieved by finding a symmetry transformation $H$ that satisfies Eqn. \eqref{RH1}. Whether this equation has a solution can be seen by analysing the related integral equation
\be
\label{integralH}
H_i^j = P_i^j + K_{\ad} \, \hat{\star}\left( H_i^k [ A_k^n]_0 (A^{-1})_n^j \right),
\ee
with $A_i^n = \hat{q}_i^{[-2n]}$ an invertible $2\times2$-matrix (the $q_i$ are independent solutions of a Baxter equation), $P_i^j$ a branch-cut free elliptic function with half-periods $(2\pi, i\ad)$ and $K_{\ad}$ an elliptic integration kernel since it should reflect the real periodicity of $\hbT$s. The simplest choice for this kernel was introduced in Eqn. \eqref{zetakernel} as
\be
K_\ad (u) = \zeta(u) -\zeta(u+2\pi) +\zeta(2\pi),
\ee
with $\zeta$ the quasi-elliptic Weierstra\ss\, function with quasi-halfperiods $(2\pi, i\ad)$. This kernel is elliptic and has exactly two simple poles in its fundamental parallellogram with residues $1$ and $-1$. If we take discontinuities on both sides of Eqn. \eqref{integralH} we find for Im$(u)>0$:
\begin{align}
\left[ H_i^j \right]_0(u) &= H_i^k [ A_k^n]_0 (A^{-1})_n^j,
\end{align}
showing that a solution to \eqref{integralH} also solves the Riemann-Hilbert problem \eqref{RH1}. One can view the integral equation \eqref{integralH} as an eigenvalue equation for the integral operator $L$ defined by
\be
L(T) = P^j + K_{\ad} \,\hat{\star} \left( T^k [ A_k^n]_0 (A^{-1})_n^j \right).
\ee
Whether an appropriate eigenfunction with eigenvalue one exists can in principle be determined using Fredholm theory. We will assume this is the case. Forgetting for now the condition that det$(H) = 1$ we find a linear combination $\hat{q}_i'$ of the old $\hat{q}_i$ that has only one cut on the real axis. At this stage one can use the fact that the $\hat{\mathbb{T}}_{1,s}$ have only $2$ cuts at $\pm i\ad s$ to conclude that det$H$ is cut free and can be absorbed in $\hat{q}_i'$ without spoiling their cut structure. This shows that $H$ is indeed a symmetry transformation such that the new $\hat{q}_i'$ have exactly one short cut, as asserted.
\section{Deriving the $\Qf\omega$ system}
\label{App:qomega}
In this appendix we construct the $\Qf\omega$ system from the $\Pf\mu$ system, in the upper half-plane conventions.

Define the $4\times4$-matrix $U$ as
\be
U^b_a = \delta^b_a + \Pf_a \Pf^b
\ee
and consider the finite-difference equation
\be
X_a^- = U^b_a X_b^+
\ee
for the unknown functions $X_a$, $a=1,\cdots 4$. As in the undeformed case, we can construct formal solutions which are analytic in the upper half-plane by the infinite product $X_a = \left(U^{[+1]}U^{[+3]}\cdots \right)^b_a C_b$ with $C_b$ an arbitrary $2i\ad$-periodic vector. Note that $X_a$ inherits periodicity properties of the $\Pf_a$ and is periodic by construction. Taking $4$ independent vectors $C_{b|i}$ labelled by $i$ we can build the matrix $M_{bi}$ and thus find $4$ independent solutions
\be
\Qm_{a|i} = \left(U^{[+1]}U^{[+3]}\cdots \right)^j_a M_{ji}.
\ee
We can now define the $\Qf$ functions in the same way:
\be
\Qf_i = -P^a \Qm_{a|i}^+ \text{ for Im}(u)>0,
\ee
since we are now working in the upper half-plane-conventions (see Section \ref{sec:conventions}). An immediate consequence of these definitions is
\be
\label{QPF1}
\Qm_{a|i}^+-\Qm_{a|i}^- = \Pf_a Q_i.
\ee
Viewed as functions with long cuts the $\Qf$s only have one cut on the real axis, as follows by following the proof in the undeformed case \cite{Gromov:2014caa} verbatim. We moreover see that the $\Qf$s inherit the exact periodicity properties of the $\Pf$s.

Defining $\omega$ is also straightforward. Consider the continuation of $\Qf$ through the long cut on the real axis,
\be
\label{defomega1}
\Qft_i = -\Pft^a \Qm_{a|i}^{\pm} = P_b\mu^{ba}\Qm_{a|i}^{\pm},
\ee
where the shift direction in $\Qm_{a|i}^{\pm}$ is again irrelevant as a simple application of the orthogonality condition $\Pf_a\Pf^a = 0$ will show. Define the $\Qf$ with upper indices as
\be
\Qf^i = P_a \left(\Qm^{a|i}\right)^+,\quad \text{with } \Qm^{a|i} = \left(\Qm_{a|i}\right)^{-t},
\ee
with $M^{-t}$ indicating the transpose of the inverse of $M$. We can find that
\be
\left(\Qm^{a|i}\right)^- = V_b^a\left(\Qm^{a|i}\right)^+
\ee
where $V = U^{-1}$ is simply given by $V_b^a = \delta_b^a -P_b P^a$. This shows directly that $\Qf^i$ also has only one long cut on the real axis and we can write
\be
\Pf_a = -\Qf^i \Qm_{a|i}^{\pm}.
\ee
This allows us to get rid of $\Pf$ in Eqn. \eqref{defomega1} and find
\begin{align}
\Qft_i = -\Qf^j \Qm_{a|j}^{\pm} \mu_{ab}\Qm_{b|j}^{\pm} = -\Qf^j \omega_{ji}, \quad \text{where we define } \omega_{ji}=\Qm_{a|j}^{-}\mu_{ab}\Qm_{b|i}^{-},
\end{align}
such that $\omega$ is an antisymmetric matrix with long cuts playing the role analogous to $\mu$ in the $\Pf\mu$ system. Analogously to the undeformed case we find that $\omega$ is $2i\ad$-periodic when viewed as a function with short cuts and that its discontinuity relation is
\be
\tilde{\omega}_{ij}-\omega_{ij} = \Qf_i \Qft_j -\Qft_i \Qf_j.
\ee
Theoretically one could expect problems in setting the Pfaffian of $\omega$ to one due to the restrictions of the trigonometric case, but this is not the case: det$(U) = 1$, so from the defining equation of $\Qm_{a|i}$ we find that $\det \Qm$ is an $2i\ad$- periodic function. Since $\Qm$ is analytic in the upper half-plane, its $2i\ad$ periodicity implies that det$(\Qm)$ does not have any cuts. So we can rescale it without introducing new branch cuts in our construction, by rescaling $\Qm_{a|i}$. This gives us det$(\omega)=1$ and combined with antisymmetry this implies Pf$(\omega)=1$ up to a sign, which we can fix by rescaling by $-1$ if necessary. Finally, we can introduce the inverse of $\omega$ to complete the $\Qf\omega$ system with upper indices. This completes the construction of the $\Qf\omega$ system.
\end{appendices}

\bibliographystyle{nb}

\bibliography{Stijnsbibfile}

\begin{thebibliography}{10}
\ifx\href\asklfhas\newcommand{\href}[2]{#2}\fi
\ifx\arxivref\asklfhas\newcommand{\arxivref}[2]{\href{http://arxiv.org/abs/#1}{#2}}\fi
\ifx\doiref\asklfhas\newcommand{\doiref}[2]{\href{http://dx.doi.org/#1}{#2}}\fi
\raggedright
\small
\parskip 0pt

\bibitem{Arutyunov:2009ga}
G.~Arutyunov and S.~Frolov,
\textit{``{Foundations of the $\ads$ Superstring. Part I}''},
\textsf{\doiref{10.1088/1751-8113/42/25/254003}{J.Phys.~A42,~254003~(2009)}},
\texttt{\arxivref{0901.4937}{arxiv:0901.4937}}.

\bibitem{Beisert:2010jr}
N.~Beisert, C.~Ahn, L.~F.~Alday, Z.~Bajnok, J.~M.~Drummond et~al.,
\textit{``{Review of AdS/CFT Integrability: An Overview}''},
\textsf{\doiref{10.1007/s11005-011-0529-2}{Lett.Math.Phys.~99,~3~(2012)}},
\texttt{\arxivref{1012.3982}{arxiv:1012.3982}}.

\bibitem{Bombardelli:2016rwb}
D.~Bombardelli, A.~Cagnazzo, R.~Frassek, F.~Levkovich-Maslyuk, F.~Loebbert,
  S.~Negro, I.~M.~Sz\'{e}cs\'{e}nyi, A.~Sfondrini, S.~J.~van~Tongeren and
  A.~Torrielli,
\textit{``An Integrability Primer for the Gauge-Gravity Correspondence: an
  Introduction''},
\textsf{\doiref{10.1088/1751-8113/49/32/320301}{J.Phys.A~49,~320301~(2016)}},
\texttt{\arxivref{1606.02945}{arxiv:1606.02945}}.

\bibitem{Gromov:2013pga}
N.~Gromov, V.~Kazakov, S.~Leurent and D.~Volin,
\textit{``{Quantum Spectral Curve for Planar $\mathcal{N} =$ Super-Yang-Mills
  Theory}''},
\textsf{\doiref{10.1103/PhysRevLett.112.011602}{Phys.Rev.Lett.~112,~011602~(2014)}},
\texttt{\arxivref{1305.1939}{arxiv:1305.1939}}.

\bibitem{Arutyunov:2007tc}
G.~Arutyunov and S.~Frolov,
\textit{``{On String S-matrix, Bound States and TBA}''},
\textsf{\doiref{10.1088/1126-6708/2007/12/024}{JHEP~0712,~024~(2007)}},
\texttt{\arxivref{0710.1568}{arxiv:0710.1568}}.

\bibitem{Arutyunov:2009zu}
G.~Arutyunov and S.~Frolov,
\textit{``{String hypothesis for the $\ads$ mirror}''},
\textsf{\doiref{10.1088/1126-6708/2009/03/152}{JHEP~0903,~152~(2009)}},
\texttt{\arxivref{0901.1417}{arxiv:0901.1417}}.

\bibitem{Bombardelli:2009ns}
D.~Bombardelli, D.~Fioravanti and R.~Tateo,
\textit{``{Thermodynamic Bethe Ansatz for planar AdS/CFT: A Proposal}''},
\textsf{\doiref{10.1088/1751-8113/42/37/375401,
  10.1088/1751-8113/42/37/375401}{J.Phys.A~A42,~375401~(2009)}},
\texttt{\arxivref{0902.3930}{arxiv:0902.3930}}.

\bibitem{Arutyunov:2009ur}
G.~Arutyunov and S.~Frolov,
\textit{``{Thermodynamic Bethe Ansatz for the $\ads$ Mirror Model}''},
\textsf{\doiref{10.1088/1126-6708/2009/05/068}{JHEP~0905,~068~(2009)}},
\texttt{\arxivref{0903.0141}{arxiv:0903.0141}}.

\bibitem{Gromov:2009bc}
N.~Gromov, V.~Kazakov, A.~Kozak and P.~Vieira,
\textit{``{Exact Spectrum of Anomalous Dimensions of Planar N = 4
  Supersymmetric Yang-Mills Theory: TBA and excited states}''},
\textsf{\doiref{10.1007/s11005-010-0374-8}{Lett.Math.Phys.~91,~265~(2010)}},
\texttt{\arxivref{0902.4458}{arxiv:0902.4458}}.

\bibitem{Arutyunov:2009ux}
G.~Arutyunov and S.~Frolov,
\textit{``{Simplified TBA equations of the $\ads$ mirror model}''},
\textsf{\doiref{10.1088/1126-6708/2009/11/019}{JHEP~0911,~019~(2009)}},
\texttt{\arxivref{0907.2647}{arxiv:0907.2647}}.

\bibitem{Cavaglia:2010nm}
A.~Cavaglia, D.~Fioravanti and R.~Tateo,
\textit{``{Extended Y-system for the $AdS_5/CFT_4$ correspondence}''},
\textsf{\doiref{10.1016/j.nuclphysb.2010.09.015}{Nucl.Phys.~B843,~302~(2011)}},
\texttt{\arxivref{1005.3016}{arxiv:1005.3016}}.

\bibitem{Balog:2011nm}
J.~Balog and A.~Hegedus,
\textit{``{$\ads$ mirror TBA equations from Y-system and discontinuity
  relations}''},
\textsf{\doiref{10.1007/JHEP08(2011)095}{JHEP~1108,~095~(2011)}},
\texttt{\arxivref{1104.4054}{arxiv:1104.4054}}.

\bibitem{Gromov:2011cx}
N.~Gromov, V.~Kazakov, S.~Leurent and D.~Volin,
\textit{``{Solving the AdS/CFT Y-system}''},
\textsf{\doiref{10.1007/JHEP07(2012)023}{JHEP~1207,~023~(2012)}},
\texttt{\arxivref{1110.0562}{arxiv:1110.0562}}.

\bibitem{Balog:2012zt}
J.~Balog and A.~Hegedus,
\textit{``{Hybrid-NLIE for the AdS/CFT spectral problem}''},
\textsf{\doiref{10.1007/JHEP08(2012)022}{JHEP~1208,~022~(2012)}},
\texttt{\arxivref{1202.3244}{arxiv:1202.3244}}.

\bibitem{Bombardelli:2017vhk}
D.~Bombardelli, A.~Cavaglià, D.~Fioravanti, N.~Gromov and R.~Tateo,
\textit{``{The full Quantum Spectral Curve for AdS$_4$/CFT$_3$}''},
\texttt{\arxivref{1701.00473}{arxiv:1701.00473}}.

\bibitem{Cavaglia:2015nta}
A.~Cavaglià, M.~Cornagliotto, M.~Mattelliano and R.~Tateo,
\textit{``{A Riemann-Hilbert formulation for the finite temperature Hubbard
  model}''},
\textsf{\doiref{10.1007/JHEP06(2015)015}{JHEP~1506,~015~(2015)}},
\texttt{\arxivref{1501.04651}{arxiv:1501.04651}}.

\bibitem{Gromov:2015dfa}
N.~Gromov and F.~Levkovich-Maslyuk,
\textit{``{Quantum Spectral Curve for a cusped Wilson line in $ \mathcal{N}=4 $
  SYM}''},
\textsf{\doiref{10.1007/JHEP04(2016)134}{JHEP~1604,~134~(2016)}},
\texttt{\arxivref{1510.02098}{arxiv:1510.02098}}.

\bibitem{Kazakov:2015efa}
V.~Kazakov, S.~Leurent and D.~Volin,
\textit{``{T-system on T-hook: Grassmannian Solution and Twisted Quantum
  Spectral Curve}''},
\textsf{\doiref{10.1007/JHEP12(2016)044}{JHEP~1612,~044~(2016)}},
\texttt{\arxivref{1510.02100}{arxiv:1510.02100}}.

\bibitem{Lunin:2005jy}
O.~Lunin and J.~M.~Maldacena,
\textit{``{Deforming field theories with $U(1)\times U(1)$ global symmetry and
  their gravity duals}''},
\textsf{\doiref{10.1088/1126-6708/2005/05/033}{JHEP~0505,~033~(2005)}},
\texttt{\arxivref{hep-th/0502086}{hep-th/0502086}}.

\bibitem{Frolov:2005ty}
S.~Frolov, R.~Roiban and A.~A.~Tseytlin,
\textit{``{Gauge-string duality for superconformal deformations of N=4 super
  Yang-Mills theory}''},
\textsf{\doiref{10.1088/1126-6708/2005/07/045}{JHEP~0507,~045~(2005)}},
\texttt{\arxivref{hep-th/0503192}{hep-th/0503192}}.

\bibitem{Frolov:2005dj}
S.~Frolov,
\textit{``{Lax pair for strings in Lunin-Maldacena background}''},
\textsf{\doiref{10.1088/1126-6708/2005/05/069}{JHEP~0505,~069~(2005)}},
\texttt{\arxivref{hep-th/0503201}{hep-th/0503201}}.

\bibitem{Ahn:2017mff}
C.~Ahn, J.~Balog and F.~Ravanini,
\textit{``{NLIE for the Sausage model}''},
\textsf{\doiref{10.1088/1751-8121/aa7780}{J.~Phys.~A50,~314005~(2017)}},
\texttt{\arxivref{1701.08933}{arxiv:1701.08933}}.

\bibitem{Delduc:2013qra}
F.~Delduc, M.~Magro and B.~Vicedo,
\textit{``{An integrable deformation of the $\ads$ superstring action}''},
\textsf{\doiref{10.1103/PhysRevLett.112.051601}{Phys.Rev.Lett.~112,~051601~(2014)}},
\texttt{\arxivref{1309.5850}{arxiv:1309.5850}}.

\bibitem{Arutyunov:2013ega}
G.~Arutyunov, R.~Borsato and S.~Frolov,
\textit{``{S-matrix for strings on $\eta$-deformed $AdS_{5} \times S^5$}''},
\textsf{\doiref{10.1007/JHEP04(2014)002}{JHEP~1404,~002~(2014)}},
\texttt{\arxivref{1312.3542}{arxiv:1312.3542}}.

\bibitem{Delduc:2014kha}
F.~Delduc, M.~Magro and B.~Vicedo,
\textit{``{Derivation of the action and symmetries of the $q$-deformed $\ads$
  superstring}''},
\textsf{\doiref{10.1007/JHEP10(2014)132}{JHEP~1410,~132~(2014)}},
\texttt{\arxivref{1406.6286}{arxiv:1406.6286}}.

\bibitem{Klimcik:2002zj}
C.~Klimcik,
\textit{``{Yang-Baxter sigma models and dS/AdS T duality}''},
\textsf{\doiref{10.1088/1126-6708/2002/12/051}{JHEP~0212,~051~(2002)}},
\texttt{\arxivref{hep-th/0210095}{hep-th/0210095}}.

\bibitem{Klimcik:2008eq}
C.~Klimcik,
\textit{``{On integrability of the Yang-Baxter sigma-model}''},
\textsf{\doiref{10.1063/1.3116242}{J.Math.Phys.~50,~043508~(2009)}},
\texttt{\arxivref{0802.3518}{arxiv:0802.3518}}.

\bibitem{Matsumoto:2014nra}
T.~Matsumoto and K.~Yoshida,
\textit{``{Lunin-Maldacena backgrounds from the classical Yang-Baxter equation
  - towards the gravity/CYBE correspondence}''},
\textsf{\doiref{10.1007/JHEP06(2014)135}{JHEP~1406,~135~(2014)}},
\texttt{\arxivref{1404.1838}{arxiv:1404.1838}}.

\bibitem{vanTongeren:2015uha}
S.~J.~van~Tongeren,
\textit{``{Yang–Baxter deformations, AdS/CFT, and twist-noncommutative gauge
  theory}''},
\textsf{\doiref{10.1016/j.nuclphysb.2016.01.012}{Nucl.~Phys.~B904,~148~(2016)}},
\texttt{\arxivref{1506.01023}{arxiv:1506.01023}}.

\bibitem{vanTongeren:2016eeb}
S.~J.~van~Tongeren,
\textit{``{Almost abelian twists and AdS/CFT}''},
\textsf{\doiref{10.1016/j.physletb.2016.12.002}{Phys.~Lett.~B765,~344~(2017)}},
\texttt{\arxivref{1610.05677}{arxiv:1610.05677}}.

\bibitem{Arutyunov:2015qva}
G.~Arutyunov, R.~Borsato and S.~Frolov,
\textit{``{Puzzles of $\eta$-deformed AdS$_5 \times$ S$^5$}''},
\textsf{\doiref{10.1007/JHEP12(2015)049}{JHEP~1512,~049~(2015)}},
\texttt{\arxivref{1507.04239}{arxiv:1507.04239}}.

\bibitem{Hoare:2015wia}
B.~Hoare and A.~A.~Tseytlin,
\textit{``{Type IIB supergravity solution for the T-dual of the $\eta$-deformed
  AdS$_{5} \times$ S$^{5}$ superstring}''},
\textsf{\doiref{10.1007/JHEP10(2015)060}{JHEP~1510,~060~(2015)}},
\texttt{\arxivref{1508.01150}{arxiv:1508.01150}}.

\bibitem{Wulff:2016tju}
L.~Wulff and A.~A.~Tseytlin,
\textit{``{Kappa-symmetry of superstring sigma model and generalized 10d
  supergravity equations}''},
\textsf{\doiref{10.1007/JHEP06(2016)174}{JHEP~1606,~174~(2016)}},
\texttt{\arxivref{1605.04884}{arxiv:1605.04884}}.

\bibitem{Borsato:2016ose}
R.~Borsato and L.~Wulff,
\textit{``{Target space supergeometry of $\eta$ and $\lambda$-deformed
  strings}''},
\textsf{\doiref{10.1007/JHEP10(2016)045}{JHEP~1610,~045~(2016)}},
\texttt{\arxivref{1608.03570}{arxiv:1608.03570}}.

\bibitem{Arutyunov:2015mqj}
G.~Arutyunov, S.~Frolov, B.~Hoare, R.~Roiban and A.~A.~Tseytlin,
\textit{``{Scale invariance of the $\eta$-deformed $AdS_5\times S^5$
  superstring, T-duality and modified type II equations}''},
\textsf{\doiref{10.1016/j.nuclphysb.2015.12.012}{Nucl.~Phys.~B903,~262~(2016)}},
\texttt{\arxivref{1511.05795}{arxiv:1511.05795}}.

\bibitem{Baguet:2016prz}
A.~Baguet, M.~Magro and H.~Samtleben,
\textit{``{Generalized IIB supergravity from exceptional field theory}''},
\textsf{\doiref{10.1007/JHEP03(2017)100}{JHEP~1703,~100~(2017)}},
\texttt{\arxivref{1612.07210}{arxiv:1612.07210}}.

\bibitem{Sakamoto:2017wor}
J.-i.~Sakamoto, Y.~Sakatani and K.~Yoshida,
\textit{``{Weyl invariance for generalized supergravity backgrounds from the
  doubled formalism}''},
\textsf{\doiref{10.1093/ptep/ptx067}{PTEP~2017,~053B07~(2017)}},
\texttt{\arxivref{1703.09213}{arxiv:1703.09213}}.

\bibitem{Sfetsos:2013wia}
K.~Sfetsos,
\textit{``{Integrable interpolations: From exact CFTs to non-Abelian
  T-duals}''},
\textsf{\doiref{10.1016/j.nuclphysb.2014.01.004}{Nucl.Phys.~B880,~225~(2014)}},
\texttt{\arxivref{1312.4560}{arxiv:1312.4560}}.

\bibitem{Hollowood:2014qma}
T.~J.~Hollowood, J.~L.~Miramontes and D.~M.~Schmidtt,
\textit{``{An Integrable Deformation of the $\ads$ Superstring}''},
\textsf{\doiref{10.1088/1751-8113/47/49/495402}{J.Phys.~A47,~495402~(2014)}},
\texttt{\arxivref{1409.1538}{arxiv:1409.1538}}.

\bibitem{Demulder:2015lva}
S.~Demulder, K.~Sfetsos and D.~C.~Thompson,
\textit{``{Integrable $\lambda$-deformations: Squashing Coset CFTs and
  $AdS_5\times S^5$}''},
\textsf{\doiref{10.1007/JHEP07(2015)019}{JHEP~1507,~019~(2015)}},
\texttt{\arxivref{1504.02781}{arxiv:1504.02781}}.

\bibitem{Arutynov:2014ota}
G.~Arutyunov, M.~de~Leeuw and S.~J.~van~Tongeren,
\textit{``{The exact spectrum and mirror duality of the
  $(\text{AdS}_5{\times}S^5)_\eta$ superstring}''},
\textsf{\doiref{10.1007/s11232-015-0243-9}{Theor.~Math.~Phys.~182,~23~(2015)}},
\texttt{\arxivref{1403.6104}{arxiv:1403.6104}},
[Teor. Mat. Fiz.182,no.1,28(2014)].

\bibitem{Arutyunov:2014cra}
G.~Arutyunov and S.~J.~van~Tongeren,
\textit{``{$\mathrm{AdS}_5 \times \mathrm{S}^5$ mirror model as a string sigma
  model}''},
\textsf{\doiref{10.1103/PhysRevLett.113.261605}{Phys.~Rev.~Lett.~113,~261605~(2014)}},
\texttt{\arxivref{1406.2304}{arxiv:1406.2304}}.

\bibitem{Arutyunov:2014jfa}
G.~Arutyunov and S.~J.~van~Tongeren,
\textit{``{Double Wick rotating Green-Schwarz strings}''},
\textsf{\doiref{10.1007/JHEP05(2015)027}{JHEP~1505,~027~(2015)}},
\texttt{\arxivref{1412.5137}{arxiv:1412.5137}}.

\bibitem{Arutyunov:2014cda}
G.~Arutyunov and D.~Medina-Rincon,
\textit{``{Deformed Neumann model from spinning strings on ($\ads$)$_\eta$}''},
\textsf{\doiref{10.1007/JHEP10(2014)050}{JHEP~1410,~50~(2014)}},
\texttt{\arxivref{1406.2536}{arxiv:1406.2536}}.

\bibitem{Banerjee:2014bca}
A.~Banerjee and K.~L.~Panigrahi,
\textit{``{On the rotating and oscillating strings in (AdS$_{3}$ x
  S$^{3}$)$_{\kappa}$}''},
\textsf{\doiref{10.1007/JHEP09(2014)048}{JHEP~1409,~048~(2014)}},
\texttt{\arxivref{1406.3642}{arxiv:1406.3642}}.

\bibitem{Kameyama:2014vma}
T.~Kameyama and K.~Yoshida,
\textit{``{A new coordinate system for $q$-deformed AdS$_{5} \times$ S$^5$ and
  classical string solutions}''},
\textsf{\doiref{10.1088/1751-8113/48/7/075401}{J.Phys.~A48,~075401~(2015)}},
\texttt{\arxivref{1408.2189}{arxiv:1408.2189}}.

\bibitem{Khouchen:2015jfa}
M.~Khouchen and J.~Klusoň,
\textit{``{D-brane on deformed AdS$_{3} \times$ S$^{3}$}''},
\textsf{\doiref{10.1007/JHEP08(2015)046}{JHEP~1508,~046~(2015)}},
\texttt{\arxivref{1505.04946}{arxiv:1505.04946}}.

\bibitem{Roychowdhury:2016bsv}
D.~Roychowdhury,
\textit{``{Multispin magnons on deformed $ AdS_{3}\times S^{3} $}''},
\textsf{\doiref{10.1103/PhysRevD.95.086009}{Phys.~Rev.~D95,~086009~(2017)}},
\texttt{\arxivref{1612.06217}{arxiv:1612.06217}}.

\bibitem{Hernandez:2017raj}
R.~Hernandez and J.~M.~Nieto,
\textit{``{Spinning strings in the $\eta$-deformed Neumann-Rosochatius
  system}''},
\texttt{\arxivref{1707.08032}{arxiv:1707.08032}}.

\bibitem{Harmark:2017yrv}
T.~Harmark and M.~Wilhelm,
\textit{``{The Hagedorn temperature of AdS5/CFT4 via integrability}''},
\texttt{\arxivref{1706.03074}{arxiv:1706.03074}}.

\bibitem{Gromov:2014caa}
N.~Gromov, V.~Kazakov, S.~Leurent and D.~Volin,
\textit{``{Quantum spectral curve for arbitrary state/operator in
  AdS$_{5}$/CFT$_{4}$}''},
\textsf{\doiref{10.1007/JHEP09(2015)187}{JHEP~1509,~187~(2015)}},
\texttt{\arxivref{1405.4857}{arxiv:1405.4857}}.

\bibitem{Arutyunov:2009ax}
G.~Arutyunov, S.~Frolov and R.~Suzuki,
\textit{``{Exploring the mirror TBA}''},
\textsf{\doiref{10.1007/JHEP05(2010)031}{JHEP~1005,~031~(2010)}},
\texttt{\arxivref{0911.2224}{arxiv:0911.2224}}.

\bibitem{Dorey:1996re}
P.~Dorey and R.~Tateo,
\textit{``{Excited states by analytic continuation of TBA equations}''},
\textsf{\doiref{10.1016/S0550-3213(96)00516-0}{Nucl.Phys.~B482,~639~(1996)}},
\texttt{\arxivref{hep-th/9607167}{hep-th/9607167}}.

\bibitem{Arutyunov:2012zt}
G.~Arutyunov, M.~de~Leeuw and S.~J.~van~Tongeren,
\textit{``{The Quantum Deformed Mirror TBA I}''},
\textsf{\doiref{10.1007/JHEP10(2012)090}{JHEP~1210,~090~(2012)}},
\texttt{\arxivref{1208.3478}{arxiv:1208.3478}}.

\bibitem{Arutyunov:2012ai}
G.~Arutyunov, M.~de~Leeuw and S.~J.~van~Tongeren,
\textit{``{The Quantum Deformed Mirror TBA II}''},
\textsf{\doiref{10.1007/JHEP02(2013)012}{JHEP~1302,~012~(2013)}},
\texttt{\arxivref{1210.8185}{arxiv:1210.8185}}.

\bibitem{Krichever:1996qd}
I.~Krichever, O.~Lipan, P.~Wiegmann and A.~Zabrodin,
\textit{``{Quantum integrable systems and elliptic solutions of classical
  discrete nonlinear equations}''},
\textsf{\doiref{10.1007/s002200050165}{Commun.Math.Phys.~188,~267~(1997)}},
\texttt{\arxivref{hep-th/9604080}{hep-th/9604080}}.

\bibitem{Marboe:2014gma}
C.~Marboe and D.~Volin,
\textit{``{Quantum spectral curve as a tool for a perturbative quantum field
  theory}''},
\textsf{\doiref{10.1016/j.nuclphysb.2015.08.021}{Nucl.~Phys.~B899,~810~(2015)}},
\texttt{\arxivref{1411.4758}{arxiv:1411.4758}}.

\bibitem{Gromov:2009zb}
N.~Gromov, V.~Kazakov and P.~Vieira,
\textit{``{Exact Spectrum of Planar ${\cal N}=4$ Supersymmetric Yang-Mills
  Theory: Konishi Dimension at Any Coupling}''},
\textsf{\doiref{10.1103/PhysRevLett.104.211601}{Phys.Rev.Lett.~104,~211601~(2010)}},
\texttt{\arxivref{0906.4240}{arxiv:0906.4240}}.

\bibitem{Pachol:2015mfa}
A.~Pachoł and S.~J.~van~Tongeren,
\textit{``{Quantum deformations of the flat space superstring}''},
\textsf{\doiref{10.1103/PhysRevD.93.026008}{Phys.~Rev.~D93,~026008~(2016)}},
\texttt{\arxivref{1510.02389}{arxiv:1510.02389}}.

\bibitem{Marboe:2017dmb}
C.~Marboe and D.~Volin,
\textit{``{The full spectrum of AdS5/CFT4 I: Representation theory and one-loop
  Q-system}''},
\texttt{\arxivref{1701.03704}{arxiv:1701.03704}}.

\bibitem{deLeeuw:2012hp}
M.~de~Leeuw and S.~J.~van~Tongeren,
\textit{``{The spectral problem for strings on twisted AdS$_5\times$S$^5$}''},
\textsf{\doiref{10.1016/j.nuclphysb.2012.03.004}{Nucl.~Phys.~B860,~339~(2012)}},
\texttt{\arxivref{1201.1451}{arxiv:1201.1451}}.

\bibitem{Kawaguchi:2014qwa}
I.~Kawaguchi, T.~Matsumoto and K.~Yoshida,
\textit{``{Jordanian deformations of the $\ads$ superstring}''},
\textsf{\doiref{10.1007/JHEP04(2014)153}{JHEP~1404,~153~(2014)}},
\texttt{\arxivref{1401.4855}{arxiv:1401.4855}}.

\bibitem{Matsumoto:2014gwa}
T.~Matsumoto and K.~Yoshida,
\textit{``{Integrability of classical strings dual for noncommutative gauge
  theories}''},
\textsf{\doiref{10.1007/JHEP06(2014)163}{JHEP~1406,~163~(2014)}},
\texttt{\arxivref{1404.3657}{arxiv:1404.3657}}.

\bibitem{Matsumoto:2015uja}
T.~Matsumoto and K.~Yoshida,
\textit{``{Schr\"odinger geometries arising from Yang-Baxter deformations}''},
\textsf{\doiref{10.1007/JHEP04(2015)180}{JHEP~1504,~180~(2015)}},
\texttt{\arxivref{1502.00740}{arxiv:1502.00740}}.

\bibitem{vanTongeren:2015soa}
S.~J.~van~Tongeren,
\textit{``{On classical Yang-Baxter based deformations of the
  AdS$_{5}\times$S$^{5}$ superstring}''},
\textsf{\doiref{10.1007/JHEP06(2015)048}{JHEP~1506,~048~(2015)}},
\texttt{\arxivref{1504.05516}{arxiv:1504.05516}}.

\bibitem{Osten:2016dvf}
D.~Osten and S.~J.~van~Tongeren,
\textit{``{Abelian Yang–Baxter deformations and TsT transformations}''},
\textsf{\doiref{10.1016/j.nuclphysb.2016.12.007}{Nucl.~Phys.~B915,~184~(2017)}},
\texttt{\arxivref{1608.08504}{arxiv:1608.08504}}.

\bibitem{Hoare:2016wsk}
B.~Hoare and A.~A.~Tseytlin,
\textit{``{Homogeneous Yang-Baxter deformations as non-abelian duals of the
  AdS$_5$ sigma-model}''},
\textsf{\doiref{10.1088/1751-8113/49/49/494001}{J.~Phys.~A49,~494001~(2016)}},
\texttt{\arxivref{1609.02550}{arxiv:1609.02550}}.

\bibitem{Borsato:2016pas}
R.~Borsato and L.~Wulff,
\textit{``{Integrable Deformations of $T$-Dual $\sigma$ Models}''},
\textsf{\doiref{10.1103/PhysRevLett.117.251602}{Phys.~Rev.~Lett.~117,~251602~(2016)}},
\texttt{\arxivref{1609.09834}{arxiv:1609.09834}}.

\bibitem{Hoare:2016hwh}
B.~Hoare and S.~J.~van~Tongeren,
\textit{``{On jordanian deformations of AdS$_5$ and supergravity}''},
\textsf{\doiref{10.1088/1751-8113/49/43/434006}{J.~Phys.~A49,~434006~(2016)}},
\texttt{\arxivref{1605.03554}{arxiv:1605.03554}}.

\bibitem{Guica:2017mtd}
M.~Guica, F.~Levkovich-Maslyuk and K.~Zarembo,
\textit{``{Integrability in dipole-deformed N=4 super Yang-Mills}''},
\texttt{\arxivref{1706.07957}{arxiv:1706.07957}}.

\bibitem{Arutyunov:2006ak}
G.~Arutyunov, S.~Frolov, J.~Plefka and M.~Zamaklar,
\textit{``The off-shell symmetry algebra of the light-cone
  $\mathit{AdS}_{5}\times \mathit{S}^5$ superstring''},
\textsf{J.~Phys.~A40,~3583~(2007)},
\texttt{\arxivref{hep-th/0609157}{hep-th/0609157}}.

\bibitem{Beisert:2008tw}
N.~Beisert and P.~Koroteev,
\textit{``{Quantum Deformations of the One-Dimensional Hubbard Model}''},
\textsf{\doiref{10.1088/1751-8113/41/25/255204}{J.~Phys.~A41,~255204~(2008)}},
\texttt{\arxivref{0802.0777}{arxiv:0802.0777}}.

\bibitem{Beisert:2017xqx}
N.~Beisert, R.~Hecht and B.~Hoare,
\textit{``{Maximally extended sl(2|2), q-deformed d(2,1;epsilon) and 3D
  kappa-Poincar\'e}''},
\texttt{\arxivref{1704.05093}{arxiv:1704.05093}}.

\bibitem{Hoare:2011wr}
B.~Hoare, T.~J.~Hollowood and J.~L.~Miramontes,
\textit{``{q-Deformation of the $\ads$ Superstring S-matrix and its
  Relativistic Limit}''},
\textsf{\doiref{10.1007/JHEP03(2012)015}{JHEP~1203,~015~(2012)}},
\texttt{\arxivref{1112.4485}{arxiv:1112.4485}}.

\bibitem{Hoare:2016ibq}
B.~Hoare and S.~J.~van~Tongeren,
\textit{``{Non-split and split deformations of ${\mathrm{AdS}}_{5}$}''},
\textsf{\doiref{10.1088/1751-8113/49/48/484003}{J.~Phys.~A49,~484003~(2016)}},
\texttt{\arxivref{1605.03552}{arxiv:1605.03552}}.

\end{thebibliography}
\end{document}